\documentclass[twocolumn]{article}
\usepackage{geometry} 
\usepackage[linktoc=all]{hyperref}
\hypersetup{
	bookmarksopen=true,
	bookmarksdepth=3,
	colorlinks=true,
	citecolor=blue,
	linkcolor=[rgb]{0.6,0,0}
}
\usepackage{graphicx}
\usepackage{cuted}
\usepackage{subcaption}
\usepackage{amssymb,amsmath}
\usepackage{listings}
\usepackage{xcolor}
\usepackage{array}
\newcolumntype{L}[1]{>{\raggedright\arraybackslash}m{#1}}
\newcolumntype{C}[1]{>{\centering\arraybackslash}m{#1}}
\newcolumntype{R}[1]{>{\raggedleft\arraybackslash}m{#1}}

\newcommand{\scalar}{\,_{^{^\circ}}}

\usepackage[style=numeric-comp, sorting=none, backend=bibtex]{biblatex}
\bibliography{bibliography}

\usepackage{fancyhdr}
\begin{document}

\title{On the accuracy and performance of the lattice Boltzmann method with 64-bit, 32-bit and novel 16-bit number formats}

\author{Moritz Lehmann$^{1*}$, Mathias J. Krause$^2$, Giorgio Amati$^3$, Marcello Sega$^4$, Jens Harting$^{4,5}$\\and Stephan Gekle$^1$}
\date{\today}
\maketitle
\newpage
*Correspondence: moritz.lehmann@uni-bayreuth.de\\
$^1$Biofluid Simulation and Modeling -- Theorethische Physik VI, University of Bayreuth\\
$^2$Institute of Mechanical Process Engineering and Mechanics, Karlsruhe Institute of Technology\\
$^3$SCAI -- SuperComputing Applications and Innovation Department, CINECA\\
$^4$Helmholtz Institute Erlangen-Nürnberg for Renewable Energy, Forschungszentrum Jülich\\
$^5$Department of Chemical and Biological Engineering and Department of Physics, Friedrich-Alexander-Universität Erlangen-Nürnberg

\section*{Abstract}
Fluid dynamics simulations with the lattice Boltzmann method (LBM) are very memory-intensive. Alongside reduction in memory footprint, significant performance benefits can be achieved by using FP32 (single) precision compared to FP64 (double) precision, especially on GPUs. 
Here, we evaluate the possibility to use even FP16 and Posit16 (half) precision for storing fluid populations, while still carrying arithmetic operations in FP32. 
For this, we first show that the commonly occurring number range in the LBM is a lot smaller than the FP16 number range. 
Based on this observation, we develop novel 16-bit formats -- based on a modified IEEE-754 and on a modified Posit standard -- that are specifically tailored to the needs of the LBM. 
We then carry out an in-depth characterization of LBM accuracy for six different test systems with increasing complexity: Poiseuille flow, Taylor-Green vortices, Karman vortex streets, lid-driven cavity, a microcapsule in shear flow (utilizing the immersed-boundary method) and finally the impact of a raindrop (based on a Volume-of-Fluid approach). 
We find that the difference in accuracy between FP64 and FP32 is negligible in almost all cases, and that for a large number of cases even 16-bit is sufficient. Finally, we provide a detailed performance analysis of all precision levels on a large number of hardware microarchitectures and show that significant speedup is achieved with mixed FP32/16-bit.

\paragraph{Keywords:} LBM, floating-point, FP16, Posit, mixed precision, customized precision, GPU, OpenCL

\section{Introduction}
The lattice Boltzmann method (LBM) \cite{kruger2017lattice, chapman1990mathematical, purqon2017accuracy, benzi1992lattice} is a powerful tool to simulate fluid flow. The parallel nature of the underlying algorithm has led to (multi-)GPU implementations \cite{tekic2012implementation, haeusl2019mpi, lehmann2020analytic, lehmann2021ejection, laermanns2021tracing, lehmann2019high, schreiber2011free, holzer2020highly, takavc2021cross, ho2017improving, habich2013performance, riesinger2017holistic, aksnes2010porous, kummerlander2021implicit, geveler2010lattice, beny2019toward, boroni2017full, griebel2005meshfree, limbach2006espresso, institute2016espresso, walsh2009accelerating, ZSH21, mawson2014memory, tolke2008teraflop, herschlag2018gpu, delbosc2014optimized, bailey2009accelerating, obrecht2013multi, de2019performance, obrecht2011new, tran2017performance, rinaldi2012lattice, rinaldifluid, beny2019efficient, ames2020multi, xiong2012efficient, zhu2020efficient, duchateau2015accelerating, janssen2015validation, habich2011performance, calore2015optimizing, hong2015scalable, xian2011multi, obrecht2010global, kuznik2010lbm, feichtinger2011flexible, calore2016massively, horgalattice, onodera2020locally, falcucci2021extreme, zitz2019lattice, mohrhard:19} becoming a popular choice as speedup can be up to two orders of magnitude compared to CPUs at similar power consumption. 
However, most GPUs have only poor FP64 (double precision) arithmetic capabilities and thus the vast majority of GPU codes has been implemented in FP32 (single precision), while most CPU codes are written in FP64. 
This difference, and in particular whether FP32 is sufficient for LBM simulations compared to FP64, has been a point of persistent discussion within the LBM community \cite{takavc2021cross, ho2017improving, habich2013performance, riesinger2017holistic, aksnes2010porous, kummerlander2021implicit, tolke2008teraflop, herschlag2018gpu, delbosc2014optimized, bailey2009accelerating, obrecht2013multi, de2019performance, kuznik2010lbm, feichtinger2011flexible, calore2016massively, horgalattice, onodera2020locally, falcucci2021extreme, zitz2019lattice, gray2016enhancing, wittmann2013comparison, wittmann2016hardware, bonaccorso2020lbsoft}.
Nevertheless, only few papers \cite{skordos1993initial, gray2016enhancing, aksnes2010porous, obrecht2013multi, de2019performance, kuznik2010lbm} provide some comparison on how floating-point formats affect the accuracy of the LBM and mostly find only insignificant differences between FP64 and FP32 except at very low velocity and where floating-point round-off leads to spontaneous symmetry breaking. 
Besides the question of accuracy, a quantitative performance comparison across different hardware microarchitectures is missing as the vast majority of LBM software is either written only for CPUs \cite{wellein2006towards, krause2021openlb, olb03, krause:08, olb14, latt2021palabos, min2013performance, mountrakis2015parallel, kotsalos2019bridging, kotsalos2020palabos, wellein2006single, lintermann2020lattice, SSFKSHC17} or only for Nvidia GPUs \cite{mawson2014memory, tolke2008teraflop, herschlag2018gpu, delbosc2014optimized, bailey2009accelerating, obrecht2013multi, de2019performance, obrecht2011new, tran2017performance, rinaldi2012lattice, rinaldifluid, beny2019efficient, ames2020multi, xiong2012efficient, zhu2020efficient, duchateau2015accelerating, janssen2015validation, habich2011performance, calore2015optimizing, hong2015scalable, xian2011multi, obrecht2010global, kuznik2010lbm, feichtinger2011flexible, calore2016massively, horgalattice, onodera2020locally} or CPUs and Nvidia GPUs \cite{riesinger2017holistic, aksnes2010porous, kummerlander2021implicit, geveler2010lattice, beny2019toward, boroni2017full, griebel2005meshfree, limbach2006espresso, institute2016espresso, walsh2009accelerating, ZSH21}. 

A second point of concern has been the amount of video memory on GPUs, which is in general smaller than standard memory on CPU systems and can thus lead to restrictions in domain size. 
LBM solely works on density distribution functions (DDFs) $f_i$ (also called fluid populations) -- floating-point numbers \cite{ieee1985ieee, goldberg1991every, is2008754, kahan1996ieee} -- that need to be loaded from and written to video memory in every time step. 
These DDFs take up the majority of the consumed memory. 
If wanting to reduce the memory footprint of LBM with reduced floating-point precision, it comes to mind to store the DDFs in a lower precision number format (streaming step) while doing arithmetic in higher (floating-point) precision (collision step). This is equivalent to decoupling arithmetic precision and memory precision \cite{grutzmacher2018modular, anzt2019toward}. 
As a desirable side effect, since the limiting factor regarding compute time is memory bandwidth \cite{lehmann2019high, schreiber2011free, holzer2020highly, takavc2021cross, ho2017improving, habich2013performance, riesinger2017holistic, aksnes2010porous, kummerlander2021implicit, geveler2010lattice, mawson2014memory, tolke2008teraflop, herschlag2018gpu, delbosc2014optimized, bailey2009accelerating, obrecht2013multi, de2019performance, obrecht2011new, tran2017performance, rinaldi2012lattice, rinaldifluid, beny2019efficient, ames2020multi, xiong2012efficient, zhu2020efficient, duchateau2015accelerating, kuznik2010lbm, feichtinger2011flexible, calore2016massively, horgalattice, mohrhard:19, wellein2006towards, wittmann2013comparison, wittmann2016hardware, krause:10b, succi2019towards, d2002multiple}, lower precision DDFs also vastly increase performance.
Such a mixed precision variant, where arithmetic is done in FP64 and DDF storage in FP32, is already used by \cite{obrecht2013multi} and \cite{falcucci2021extreme}. 
Using FP32 arithmetic and FP16 DDF storage would be even better, but has not yet been attempted due to concerns about possibly insufficient accuracy. Lower 16-bit precision has already been successfully applied to other fluid solvers \cite{klower2021compressing, klower2020number, hatfield2019accelerating} and to a lot of other high performance computing software \cite{langou2006exploiting, haidar2020mixed}.\\

The purpose of this work is thus two-fold: 
Firstly, in order to render mixed FP32/16-bit precisions feasible for LBM, we introduce novel 16-bit number formats that turn out to be superior to standard IEEE-754 FP16 in LBM applications and in many cases perform as accurately as FP32. 
Therein we leverage that some of the FP32 bits do not contain physical information or are entirely unused, similar to \cite{klower2021compressing}. 
This approach requires minimal code interventions and can be easily combined with any velocity set, collision operator, swap algorithm or LBM extension. 
In addition to using custom-built floating-point formats, we show that shifting the DDFs by subtracting the lattice weights and computing the equilibrium DDFs in a specific order of operations as originally proposed by Skordos \cite{skordos1993initial} and further investigated by He and Luo \cite{he1997lattice} and Gray and Boek \cite{gray2016enhancing} -- an optimization beneficial across all floating-point formats and already widely used \cite{lehmann2021ejection, laermanns2021tracing, lehmann2019high, haeusl2019mpi, lehmann2020analytic, krause2021openlb, olb03, krause:08, olb14, latt2021palabos, min2013performance, mountrakis2015parallel, kotsalos2019bridging, kotsalos2020palabos, griebel2005meshfree, limbach2006espresso, institute2016espresso, tolke2008teraflop, skordos1993initial, gray2016enhancing, he1997lattice, feichtinger2011flexible, obrecht2013multi, d2002multiple} -- turns out absolutely crucial for the 16-bit compression.

Secondly, we present an extensive comparison of FP64, FP32, FP16, shifted Posit16 as well as our novel custom formats. 
Regarding LBM accuracy, we study Poiseuille flow through a cylinder \cite{kruger2011unit}, Taylor-Green vortex energy dissipation \cite{taylor1937mechanism, skordos1993initial}, Karman vortices \cite{karman1911ueber} from flow around a cylinder, lid-driven cavity \cite{rinaldi2012lattice, obrecht2011new, kuznik2010lbm, delbosc2014optimized, mawson2014memory, hong2015scalable, cattenone2020basis, alim2009lattice, neumann2013dynamic, ghia1982high, jiang1994large, yang1998implicit}, deformation of a microcapsule in shear flow \cite{BarthesBiesel_2016, guckenberger2016bending, guckenberger2017theory} with the immersed-boundary method extension and microplastic particle transport during a raindrop impact \cite{lehmann2021ejection} with the Volume-of-Fluid and immersed-boundary method extensions. 
Regarding performance, we exploit the capability of our \textsl{FluidX3D} LBM implementation written in OpenCL \cite{lehmann2021ejection, laermanns2021tracing, lehmann2019high, haeusl2019mpi, lehmann2020analytic} to provide benchmarks for all floating-point variants on a large variety of hardware, from the worlds fastest datacenter GPU over various consumer GPUs and CPUs from different vendors to even a mobile phone ARM System-on-a-Chip (SoC), and show roofline analysis \cite{wittmann2016hardware, williams2009roofline, succi2019towards} for one hardware example.

\section{Lattice Boltzmann algorithm}

\subsection{LBM -- overview}
The lattice Boltzmann method is a Navier-Stokes flow solver that discretizes space into a Cartesian lattice and time into discrete time steps \cite{kruger2017lattice, chapman1990mathematical, purqon2017accuracy, benzi1992lattice}. 
For each point on the lattice, density $\rho$ and velocity $\vec{u}$ of the flow are computed from so-called density distribution functions (DDFs, also called fluid populations) $f_i$. The DDFs are floating-point numbers and represent how many fluid molecules move between neighboring lattice points in each time step. Because of the lattice, only certain directions are possible for this exchange and there are various levels of this directional discretization, in 3D typically $19$ (including the center point) where space-diagonal directions are left out. After exchange of DDFs from and to neighboring lattice points (streaming), the DDFs are redistributed locally on each lattice point (collision). For the collision, there are various approaches, the most common being the single-relaxation-time (SRT), two-relaxation-time (TRT) and multi-relaxation-time (MRT) collision operators \cite{kruger2017lattice, lehmann2019high}.\\
The computation of the streaming part is done by copying the DDFs in memory to their new location. The algorithm is provided in appendix \ref{sec:appendix:lbm:without} with notation as in appendix \ref{sec-list-of-physical-quantities}.

\subsection{DDF-shifting}
To achieve maximum accuracy, it is essential not to work with the density distribution functions (DDFs) $f_i$ directly, but with shifted $f_i^\text{shifted}:=f_i-w_i$ instead \cite{skordos1993initial, he1997lattice, gray2016enhancing, feichtinger2011flexible, d2002multiple}. $w_i=f_i^\text{eq}(\rho=1,\vec{u}=0)$ are the lattice weights and $\rho$ and $\vec{u}$ are the local fluid density and velocity. 
This requires a small change in the equilibrium DDF computation
\begin{align}
&f_i^\text{eq-shifted}(\rho,\vec{u}):=f_i^\text{eq}(\rho,\vec{u})-w_i=\\
&=w_i\,\rho\,\cdot\left(\frac{(\vec{u}\scalar\vec{c}_i)^2}{2\,c^4}+\frac{\vec{u}\scalar\vec{c}_i}{c^2}+1-\frac{\vec{u}\scalar\vec{u}}{2\,c^2}\right)-w_i=\\
&=w_i\,\rho\,\cdot\left(\frac{(\vec{u}\scalar\vec{c}_i)^2}{2\,c^4}-\frac{\vec{u}\scalar\vec{u}}{2\,c^2}+\frac{\vec{u}\scalar\vec{c}_i}{c^2}\right)+w_i\,(\rho-1)\label{eq:lbm-equilibrium-shifted}
\end{align}
and density summation:
\begin{equation} \label{eq:rho-shifted}
\rho=\sum_i\left(f_i^\text{shifted}+w_i\right)=\left(\sum_i f_i^\text{shifted}\right)+1
\end{equation}
We emphasize that it is key to choose equation \eqref{eq:lbm-equilibrium-shifted} exactly as presented without changing the order of operations\footnote{To minimize the overall number of floating-point operations, terms should be pre-computed such that $f_i^\text{eq-shifted}=A\,\cdot(\frac{1}{2}\,(B^2+C)\pm B)+D$ requires only $3$ fused-multiply-add (FMA) operations.}, otherwise the accuracy may not be enhanced at all \cite{skordos1993initial, he1997lattice, gray2016enhancing}. With this exact order, the round-off error due to different sums is minimized. 
This offers a large benefit, most prominently on FP16 accuracy, by substantially reducing numerical loss of significance at no additional computational cost. Since it is also beneficial for regular FP32 accuracy, it is already widely used in LBM codes such as our \textsl{FluidX3D} \cite{lehmann2021ejection, laermanns2021tracing, lehmann2019high, haeusl2019mpi, lehmann2020analytic}, OpenLB \cite{krause2021openlb, olb03, krause:08, olb14}, ESPResSo \cite{griebel2005meshfree, limbach2006espresso, institute2016espresso}, Palabos \cite{latt2021palabos, min2013performance, mountrakis2015parallel, kotsalos2019bridging, kotsalos2020palabos} and some versions of waLBerla \cite{feichtinger2011flexible}. In the appendix in section \ref{sec:appendix:lbm} we provide the entire algorithm without and with DDF-shifting for comparison and in section \ref{sec-list-of-physical-quantities} we clarify our notation.\\
We also recommend doing the summation of the DDFs in alternating '$+$' and '$-$' order during computation of the velocity $\vec{u}$ to further reduce numerical loss of significance, for example $u_x=(f_{ 1}-f_{ 2}+f_{ 7}-f_{ 8}+f_{ 9}-f_{10}+f_{13}-f_{14}+f_{15}-f_{16})/\rho$ for the $x$-component in D3Q19.\\
Gray and Boek \cite{gray2016enhancing} also propose to compute $(\rho-1)=\sum_i f_i^\text{shifted}$ as a separate variable and directly insert it into eq.~\eqref{eq:lbm-equilibrium-shifted}; while we do not advise against this, we found its benefit to be insignificant at any floating-point precision while increasing complexity of the code and thus omit it in our implementation.

\subsection{Which range of numbers does the LBM use?}
\begin{figure}[!htb]
\centering \includegraphics[width=8cm]{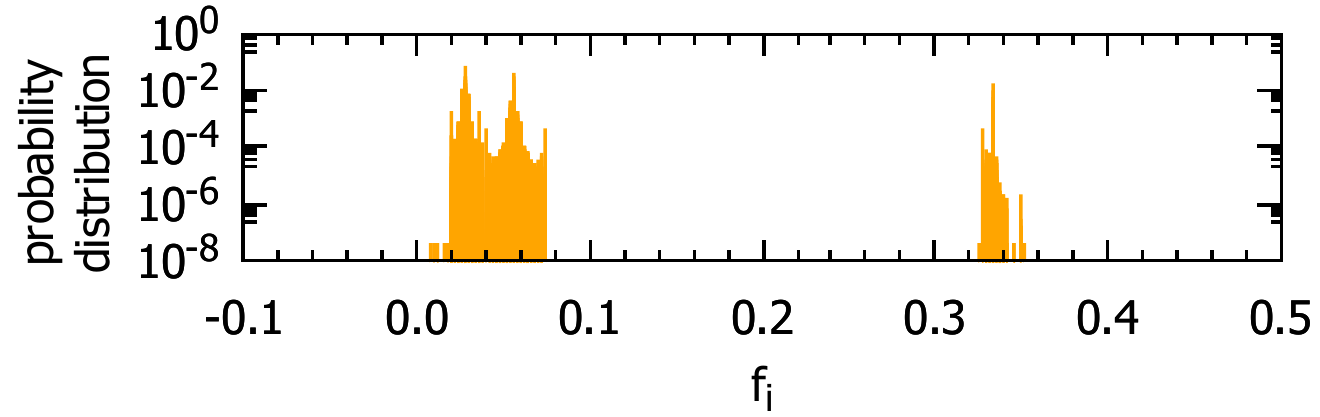}\\
\centering \includegraphics[width=8cm]{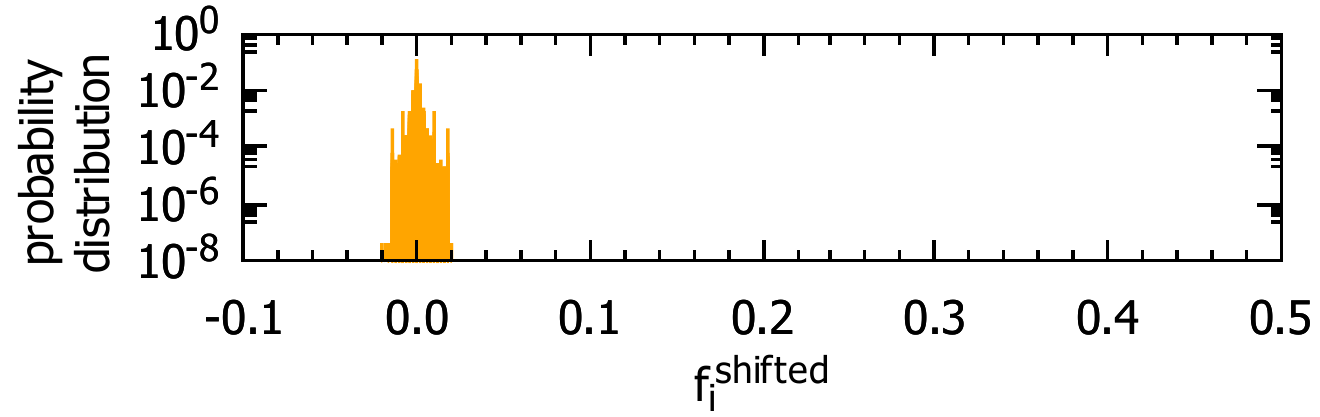}\vspace*{0.1cm}
\caption{Histogram of the DDFs for the lid-driven cavity simulation from section \ref{sec:ldc} after $100000$ LBM time steps. The simulation is performed without the DDF-shifting (top) and with DDF-shifting (bottom), both times in FP32/FP32.} \label{fig:fi}
\end{figure}

In figure \ref{fig:fi} we present the distribution of $f_i$ and $f_i^\text{shifted}$ for the example system of the lid-driven cavity from section \ref{sec:ldc}.
Similar data for the remaining setups are given in the appendix in figure \ref{fig:setups:fi}. 
It is quite remarkable how the number range in all cases is very limited. 
The $f_i$ accumulate around the LBM lattice weights (for D3Q19 $w_i\in\{\frac{1}{36},\frac{1}{18},\frac{1}{3}\}$) and the $f_i^\text{shifted}$ accumulate around $0$, where floating-point accuracy is best. So for FP32 not only are the trailing bits of the mantissa expected to be nonphysical, numerical noise \cite{klower2021compressing}, but also some bits of the exponent are entirely unused, meaning one can waive these bits without loosing accuracy.\\
To find the number range of $f_i$ and $f_i^\text{shifted}$, we insert $\vec{u}_j=c\,\frac{\vec{c}_j}{|\vec{c}_j|}$ in eq. \eqref{eq:lbm-equilibrium} and eq. \eqref{eq:lbm-equilibrium-shifted} and find that $\frac{1}{\rho}\,|f_i^\text{eq}|\leq\delta$ or $\frac{1}{\rho}\,|f_i^\text{eq-shifted}|\leq\delta^\text{shifted}$ respectively, with the values of $\delta$ and $\delta^\text{shifted}$ depending on the velocity set in use (table \ref{tab:delta}).
\begin{table}[!htb] \begin{center} \begin{tabular}{l|c c c c c c}
\hline
 				& \rotatebox[origin=c]{90}{ D2Q9 }	& \rotatebox[origin=c]{90}{ D3Q7 }	& \rotatebox[origin=c]{90}{ D3Q13 }	& \rotatebox[origin=c]{90}{ D3Q15 }	& \rotatebox[origin=c]{90}{ D3Q19 }	& \rotatebox[origin=c]{90}{ D3Q27 }\\
\hline
$\delta$				& $0.45$	& $0.47$	& $0.50$	& $0.42$	& $0.34$	& $0.30$\\
$\delta^\text{shifted}$	& $0.31$	& $0.35$	& $0.25$	& $0.31$	& $0.17$	& $0.21$\\
\hline
\end{tabular} \end{center}
\vspace*{-0.2cm}
\caption{The numerical value of $\delta$ and $\delta^\text{shifted}$ depending on the used velocity set.} \label{tab:delta}
\end{table}\\
With $\tau>0.5$, through eq. \eqref{eq:lbm-collision}, we get in the worst-case
\begin{align}
|f_i|&\lessapprox|2\,f_i^\text{eq}|\lessapprox2\,\rho\,\delta\\
|f_i^\text{shifted}|&\lessapprox|2\,f_i^\text{eq-shifted}|\lessapprox2\,\rho\,\delta^\text{shifted}
\end{align}
respectively, because the DDFs in stable simulations are expected to follow the equilibrium DDFs. The density $\rho$ typically deviates only little from $\rho\approx1$. Assuming $\rho<2$ leads to $|f_i|\lessapprox2$ being the worst-case maximum number range (D3Q13, no DDF-shifting). With the more typical D3Q19 and DDF-shifting, the same number range $|f_i^\text{shifted}|\lessapprox2$ restricts the density to a less strict $\rho<6$. 
Keeping the sign is required, because $f_i^\text{shifted}$ (and also $f_i$) can be negative.

\section{Number representation models}
\begin{figure}[!htb]
\centering \includegraphics[width=8cm]{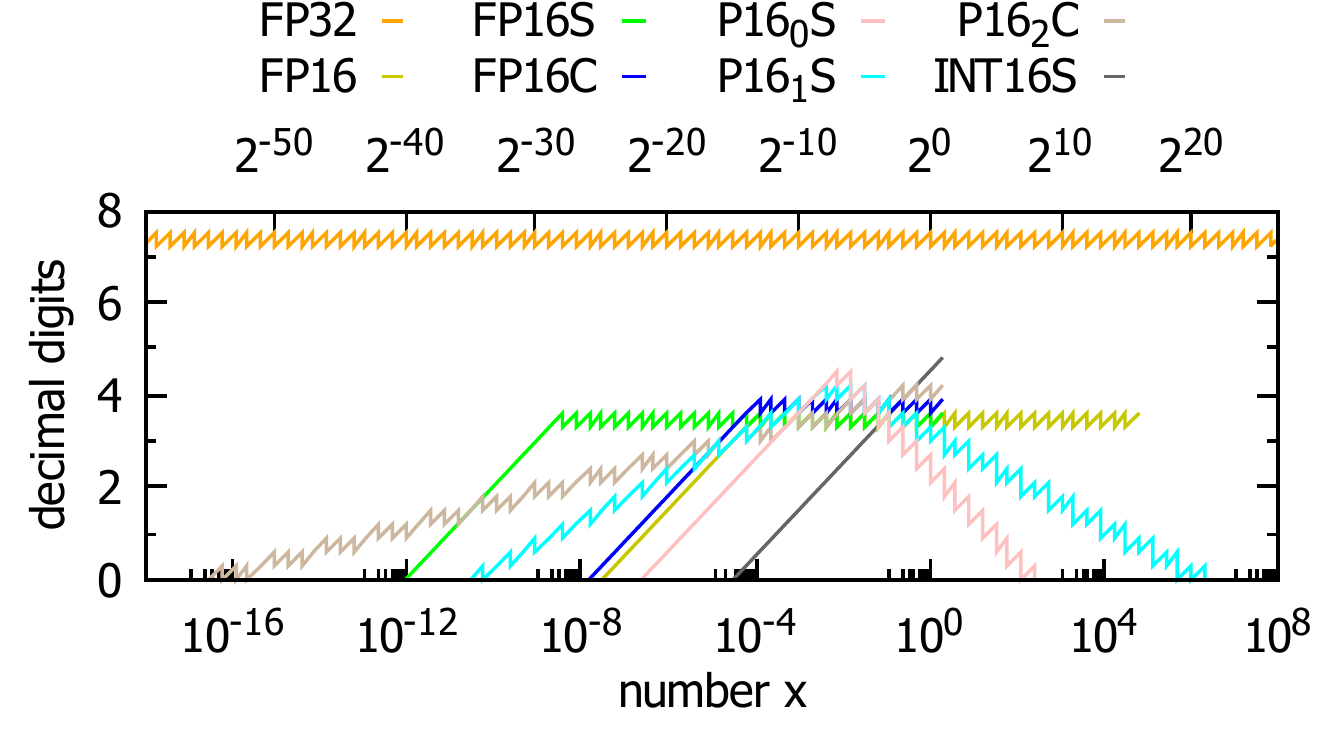}\vspace*{0.1cm}
\caption{Accuracy characteristics of the number formats investigated in this work. 
This plot shows only the local minima (measured graphs see figure \ref{fig:formats-measured}). 
FP16C reduces number range, but increases accuracy in the normalized regime (horizontal part). 
For floating-point formats, the downward slope indicates the denormalized part, where accuracy behaves like fixed-point. 
We also show 16-bit fixed-point scaled by $\times2^{-14}$ (INT16S). 
Posits (P16) have slopes left and right, with highest accuracy in the middle, which here is shifted from $1$ to $\frac{1}{128}$, hence the "S". P16$_0$S/P16$_1$S have a 0/1-bit exponents, making the slopes more/less steep and decreasing/increasing number range. P16$_2$C is a custom format with 2-bit exponent, but asymmetric slope.
} \label{fig:formats}
\end{figure}

A 16-bit number can represent only 65536 different values. The task is to spread these along the number axis in a way that the most commonly used numbers are represented with the best possible accuracy. 
There is a variety of number representations that come to mind as a 16-bit storage format: fixed-point, floating-point as well as the recently developed Posit format \cite{gustafson2017beating}, and each of them can be adjusted specifically for the LBM. 
Figure \ref{fig:formats-overview} illustrates the number formats investigated in this work and figure \ref{fig:formats} shows their accuracy characteristics.
\begin{figure}[!htb] \begin{center}
\begin{tabular}{|c|c|p{4.9cm}|}
\hline
0 & 01111111111 & 00000000000000000000000000...\newline ...00000000000000000000000000\\
\hline
\end{tabular}\\IEEE-754 FP64 ( 1 $|$ 11 $|$ 52 )\\\hfill\\
\begin{tabular}{|c|c|c|}
\hline
0 & 01111111 & 00000000000000000000000\\
\hline
\end{tabular}\\IEEE-754 FP32 ( 1 $|$ 8 $|$ 23 )\\\hfill\\
\begin{tabular}{|c|c|c|}
\hline
0 & 01111 & 0000000000\\
\hline
\end{tabular}\\IEEE-754 FP16 ( 1 $|$ 5 $|$ 10 )\\\hfill\\
\begin{tabular}{|c|c|c|}
\hline
0 & 11110 & 0000000000\\
\hline
\end{tabular}\\FP16S ( 1 $|$ 5 $|$ 10 )\\\hfill\\
\begin{tabular}{|c|c|c|}
\hline
0 & 1111 & 00000000000\\
\hline
\end{tabular}\\custom FP16C ( 1 $|$ 4 $|$ 11 )\\\hfill\\
\begin{tabular}{|c|c c|}
\hline
0 & 111111110 & 000000\\
\hline
\end{tabular}\\Posit P16$_0$S ( 1 $|$ $n_r$+1 $|$ 14-$n_r$ )\\\hfill\\
\begin{tabular}{|c|c c c|}
\hline
0 & 11110 & 1 & 000000000\\
\hline
\end{tabular}\\Posit P16$_1$S ( 1 $|$ $n_r$+1 $|$ 1 $|$ 13-$n_r$ )\\\hfill\\
\begin{tabular}{|c|c c c|}
\hline
0 & 1 & 11 & 000000000000\\
\hline
\end{tabular}\\custom asymmetric Posit P16$_2$C ( 1 $|$ $n_r$ $|$ 2 $|$ 13-$n_r$ )\\\hfill\\
\begin{tabular}{|c|c|}
\hline
0 & 100000000000000\\
\hline
\end{tabular}\\INT16S ( 1 $|$ 15 )
\end{center} \caption{The number $1.0$ represented by the different formats we investigate here. The leftmost single bit is the sign $s$ and the rightmost segment is the mantissa $m$. For floating-point (FP), the center segment is the exponent $e$. FP16S and FP16C are new formats specifically designed to store the DDFs. Fixed-point (INT16S) does not have an exponent. Posits have dynamic partitioning of the segments, with an extra regime segment and an optional exponent segment next to the mantissa.} \label{fig:formats-overview}
\vspace*{-0.4cm}
\end{figure}

\subsection{Floating-point}
\subsubsection{Overview}
In the normalized number range, a floating-point number \cite{ieee1985ieee, goldberg1991every, is2008754, kahan1996ieee} is represented as
\begin{equation}
x=\underbrace{(-1)^s}_\text{sign}\cdot\underbrace{\ 2^{e-b}\ \mathstrut}_\text{exponent}\cdot\underbrace{(1+2^{-n_m}\,m)}_\text{mantissa}
\end{equation}
with $s$ being the sign bit, $e$ being an integer representing the exponent and $m$ being an integer representing the mantissa. $b$ is a constant called exponent bias and $n_m$ is the number of bits in the mantissa (values in table \ref{tab:formats}). The precision is $\log_{10}(2^{n_m+1})$ decimal digits\footnote{The '$+1$' refers to the implicit leading bit of the mantissa.} and the truncation error is $\epsilon=2^{-n_m}$.\\
When the exponent $e$ is zero, the mantissa is shifted to the right as a way to represent even smaller numbers close to zero, although at less precision. This is called the denormalized number range and making use of it during the conversioans that will be described below is not straight-forward, but essential alongside correct rounding to keep decent accuracy with the 16-bit formats.

\begin{table*}[!htb] \begin{center}
\begin{tabular}{l|p{0.35cm} p{0.65cm} r r p{1.65cm} p{2.6cm} p{2.5cm} p{2.6cm}}
\hline
 & bits & $n_m$ & $b\ \ \ \ $ & digits & $\epsilon$ & range & smallest norm.\newline  number & smallest denorm.\newline number\\
\hline
IEEE FP64 & $64$ & $\ 52$ & $1023$ & $16.0$ & $\ 2.2\cdot10^{-16}$ & $\pm1.797693\cdot10^{308}$ & $2.225074\cdot10^{-308}$ & $4.940656\cdot10^{-324}$\\
IEEE FP32 & $32$ & $\ 23$ & $127$ &  $7.2$ & $\ \ 1.2\cdot10^{-7}$ & $\pm3.402823\cdot10^{38}$  & $1.175494\cdot10^{-38}$  & $1.401298\cdot10^{-45}$ \\
IEEE FP16 & $16$ & $\ 10$ & $15$ &  $3.3$ & $\ \ 9.8\cdot10^{-4}$ & $\pm6.550400\cdot10^{4}$   & $6.103516\cdot10^{-5}$   & $5.960464\cdot10^{-8}$  \\
FP16S         & $16$ & $\ 10$ & $30$ &  $3.3$ & $\ \ 9.8\cdot10^{-4}$ & $\pm1.999023\cdot10^{0}$   & $1.864464\cdot10^{-9}$   & $1.818989\cdot10^{-12}$ \\
FP16C         & $16$ & $\ 11$ & $15$ &  $3.6$ & $\ \ 4.9\cdot10^{-4}$ & $\pm1.999512\cdot10^{0}$   & $6.103516\cdot10^{-5}$   & $2.980232\cdot10^{-8}$  \\
Posit P16$_0$S         & $16$ & $0$-$13$ & $-$ &  $\leq4.2$ & $\geq1.2\cdot10^{-4}$ & $\pm1.280000\cdot10^{2}$   & $-$   & $4.768372\cdot10^{-7}$  \\
Posit P16$_1$S         & $16$ & $0$-$12$ & $0$ &  $\leq3.9$ & $\geq2.4\cdot10^{-4}$ & $\pm2.097152\cdot10^{6}$   & $-$   & $2.910383\cdot10^{-11}$ \\
Posit P16$_2$C         & $16$ & $0$-$12$ & $0$ &  $\leq3.9$ & $\geq2.4\cdot10^{-4}$ & $\pm1.999756\cdot10^{0}$   & $-$   & $1.734724\cdot10^{-18}$ \\
\hline
\end{tabular}
\end{center}
\vspace*{-0.2cm}
\caption{Comparing the properties of the number formats used here to store the LBM DDFs $f_i$.} \label{tab:formats}
\end{table*}

\subsubsection{Customized FP16 formats for the LBM}
In our lattice Boltzmann simulations, we implement and test three different 16-bit floating-point formats:
\begin{itemize}\itemsep0em
\item FP16: Standard IEEE-754 FP16, with FP32$\leftrightarrow$FP16 conversion supported on all CPUs and GPUs from within the last 12 years.
\item FP16S: We down-scale the number range of IEEE-754 FP16 by $\times2^{-15}$ to $\pm2$ and use the convenience that all modern CPUs and GPUs can do IEEE-754 floating-point conversion in hardware.
\item FP16C: We allocate one bit less for the exponent (to decrease number range towards small numbers) and one bit more for the mantissa (to gain accuracy). The number range is also limited to $\pm2$. This custom format requires manual conversion from and to FP32.
\end{itemize}
When looking at table \ref{tab:formats} and figure \ref{fig:formats}, FP16S and FP16C differ in extended range towards small numbers versus halved truncation error $\epsilon$. The question arises which of these two traits is more important for LBM. FP16 is inferior to both FP16S and FP16C as it combines lower mantissa accuracy and less range towards small numbers. Since FP16S comes at no additional computational cost and complexity compared to FP16, FP16S should always be preferred over FP16 for storing the DDFs.

\subsubsection{Floating-point conversion: FP32$\leftrightarrow$FP16S}
The IEEE-754 FP32$\leftrightarrow$FP16/FP16S conversion is supported in hardware and therefore only briefly described below.

\noindent\textbf{FP32$\rightarrow$FP16S:} For the FP32$\rightarrow$FP16 conversion, OpenCL provides the function \texttt{vstore\_half\_rte} that is executed in hardware. To convert to the FP16S format instead, we shift the number range up by $2^{15}$ via regular FP32 multiplication right before conversion. This is equivalent to increasing the exponent bias $b$ by $15$.\\\\
\noindent\textbf{FP16S$\rightarrow$FP32:} For the FP16$\rightarrow$FP32 conversion, OpenCL provides the function \texttt{vload\_half} that is executed in hardware. To convert from the FP16S format instead, we shift the number range down by $2^{-15}$ via regular FP32 multiplication right after conversion.

\subsubsection{Floating-point conversion: FP32$\leftrightarrow$FP16C}
For FP32$\leftrightarrow$FP16C conversion, we developed a set of fast conversion algorithms that work in any programming language and on any hardware which we describe in some more detail further below.
An OpenCL C version is presented in listing \ref{list:conversion}.

We ditch \texttt{NaN} and \texttt{Inf} definitions for an extended number range by a factor $2$ and less complicated and faster conversion. In PTX assembly \cite{nvidia2021ptx}, the FP32$\rightarrow$FP16C conversion takes $25$ instructions and FP16C$\rightarrow$FP32 takes $26$ instructions.\\\\
\noindent\textbf{FP32$\rightarrow$FP16C:} The first step is to interpret the bits of the FP32 input number as \texttt{uint}, for which there is the \texttt{as\_uint(float x)} function provided by OpenCL. The sign bit remains identical as the leftmost bit via bit-masking and bit-shifting. To assure correct rounding, we add a \texttt{1} to the $12^\text{th}$ bit from left (\texttt{0x00000800}), because mantissa bits at positions $12$ to $0$ later are truncated. Next, we extract the exponent $e$ by bit-masking and bit shift by $23$ places to the left.\\
For normalized numbers, the exponent is decreased by the difference in bias $127-15=112$ and bit-shifted to the right by $11$ places. A final bit-mask ensures that there is no overflow into the sign bit. The mantissa is bit-shifted in place and or-ed to sign and exponent.\\
For denormalized numbers, we first add a \texttt{1} to place $24$ (\texttt{0x00800000}) of the mantissa (to later figure out how many places the mantissa was shifted) and then bit-shift it to the right by as many places as the new exponent is below zero. Correct rounding however makes this a bit more difficult: We need to add \texttt{1} for rounding to the leftmost place of the mantissa that is cut off. To undo the initial rounding we did earlier, instead of \texttt{0x00800000}, we add \texttt{0x00800000-0x00000800=0x007FF800}, then shift by one place less than the new exponent is below zero, add \texttt{1} to the rightmost bit and finally shift right the one remaining place.\\
The exponent itself is the switch deciding whether the normalized or denormalized conversion is used. As an optional safety measure, we add saturation: If the number is larger than the maximum value, we override all exponent and mantissa bits to \texttt{1} (bitwise or with \texttt{0x7FFF}).\\\\
\noindent\textbf{FP16C$\rightarrow$FP32:} To convert back to FP32, we first extract the exponent $e$ and the mantissa $m$ by bit-masking and bit-shifting. Additionally, since we intend to avoid branching, we already count the number of leading zeros\footnote{The OpenCL function \texttt{clz(m)} also counts the number of leading zeros. While translated into a single \texttt{clz.b32} PTX instruction (instead of \texttt{cvt.rn.f32.u32} \texttt{mov.b32} \texttt{shr.u32}), \texttt{clz.b32} executes much slower, leading to noticeably worse performance.
} $v$ in the mantissa for decoding the denormalized format: We cast $m$ to \texttt{float}\footnote{Casting an \texttt{int} to \texttt{float} implicitly does a \texttt{log2} operation to determine the exponent.}, reinterpret the result as \texttt{uint}, bit-shift the exponent right by $23$ bits and subtract the exponent bias, giving us the base-2 logarithm of $m$, equivalent to $31$ minus the number of leading zeroes.\\
The sign bit is bit-masked and bit-shifted in place. The exponent $e$ again decides for normalized or denormalized numbers: For normalized numbers ($e\neq0$), the exponent is increased by the difference in bias $127-15=112$ and or-ed together with the bit-shifted mantissa. For denormalized numbers ($e=0$ and $m\neq0$), the mantissa is bit-shifted to the right by the number of leading zeroes and the shift-indicator \texttt{1} is removed by bit-masking. The mantissa is or-ed with the exponent which is set by the number of leading zeroes and bit-shifted in place.\\
Finally, the \texttt{uint} result is reinterpret as \texttt{float} via the OpenCL function \texttt{as\_float(uint x)}.

\subsection{Posit}
\subsubsection{Overview}
The novel Posit format (Type III Unum) \cite{gustafson2017beating, gustafson2017posit, de2019posits, klower2020number} is different from floating-point in that the bit segment for the mantissa (and also exponent) is variable in size and there is another bit segment, the regime, with variable size as well. The Posit number representation is
\begin{equation}
x=\underbrace{(-1)^s}_\text{sign}\cdot\underbrace{(2^{n_e+1})^r}_\text{regime}\cdot\underbrace{\ 2^e\ \mathstrut}_\text{exponent}\cdot\underbrace{(1+2^{-n_m}\,m)}_\text{mantissa}
\end{equation}
with sign $s$, regime $r$, exponent $e$ and mantissa $m$. $n=1+(n_r+1)+n_e+n_m$ is the total number of bits, whereby $n_r$, $n_e$ and $n_m$ are the variable numbers of bits in regime, exponent and mantissa, respectively.\\
For very small numbers, the regime bit pattern looks like \texttt{000..01} (negative $r$), then gets shorter towards \texttt{01} ($r=-1$), flips to \texttt{10} ($r=0$) and then gets longer again, looking like \texttt{111..10} (positive $r$). The last bit is the regime terminator bit that unambiguously tells the length of the regime. This bit is not included in the regime size $n_r$, so the size of the regime bit pattern is $n_r+1$. $n_r$ determines the value of the regime: $r=-n_r$ if the regime terminator bit is \texttt{1} or $r=n_r-1$ if the regime terminator bit is \texttt{0}.\\
For increasing regime size, the remaining bits for exponent and mantissa are shifted to the right, so the mantissa (and if no mantissa bits are left also the exponent) become shorter and precision is reduced.\\
Posits can be designed with different (fixed) exponent size or no exponent at all. Just like for floating-point, larger exponent increases the range but decreases accuracy.\\
This way, Posit is designed to deliver variable accuracy based on where the number is in the regime: best accuracy is around $\pm1.0$ where the regime is shortest (superior to floating-point) but for both tiny and large numbers, much precision is lost \cite{de2019posits}. 

\subsubsection{Customized Posit formats for the LBM}
As storage format for LBM DDFs, where numbers close to $0$ need to be resolved best and numbers outside the $\pm2$ range are not required at all, the standard 16-bit Posit formats seems an unfavorable choice. However, by multiplying a constant before and after conversion, similar to FP16S, we shift the most accurate part down to smaller numbers. We take a closer look at three different Posit formats:
\begin{itemize}\itemsep0em
\item P16$_0$S: 16-bit Posit without exponent, shifted down by $\times2^{7}$. In the interval $[2^{-11},2^{-3}]$, accuracy is equal to or better than FP16S and in the interval $[2^{-10},2^{-4}]$, accuracy is equal to or better than FP16C. The range towards small numbers is very poor and for numbers $>2^{-3}$, accuracy is vastly degraded.
\item P16$_1$S: 16-bit Posit with 1-bit exponent, shifted down by $\times2^{7}$. In the interval $[2^{-13},2^{-1}]$, accuracy is equal to or better than FP16S and in the interval $[2^{-11},2^{-3}]$, accuracy is equal to or better than FP16C. For numbers $<2^{-13}$ or $>2^{-1}$ accuracy is reduced. The range towards small numbers is between FP16S and FP16C. This format poses no limitations on the density $\rho$ because its number range is $\pm2^{21}$.
\item P16$_2$C: Custom asymmetric 16-bit Posit with 2-bit exponent, not shifted. By only covering the lower flank, we can get rid of the bit reserved for the regime sign, thus making the regime shorter by 1 bit and increasing the mantissa size by 1 bit in turn. The conversion algorithms are vastly simplified with the asymmetric regime. Accuracy is better or equal to FP16C in the interval $[2^{-7},2]$ and equal or better than FP16S in the interval $[2^{-11},2]$. For smaller numbers, accuracy is slowly reduced, but the range towards small numbers is excellent.
\end{itemize}
Both P16$_0$S and P16$_1$S provide numbers $>2$ that are unused in the LBM. Shifting the number range further down would degrade accuracy for larger numbers too much. Since the LBM with DDF-shifting uses numbers around $0$ and it is not entirely clear in which order of magnitude accuracy is most important, it is also unclear if the increased accuracy in the center interval will benefit more than the decreased accuracy further away from the center will adversely affect.

\subsubsection{FP32$\leftrightarrow$Posit conversion}
Conversion between FP32 and Posit is not supported in hardware (yet). Since the reference conversion algorithm in the SoftPosit library \cite{leong2019softposit} is not written for speed primarily, we provide self-written, ultrafast conversion algorithms in listing \ref{list:conversion} in OpenCL C. 
These work on any hardware. 
A detailed description of how the algorithms work is omitted here, but can be inferred by studying the provided listings.
Note that the Posit specification \cite{gustafson2017beating} does two's complement for negative numbers in order to have no duplicate zero and an infinity definition instead.
To simplify the conversion algorithms and since infinity is not required in our applications, we just use the sign bit to reduce operations, so that there is positive and negative zero.

\subsection{Fixed-point}
16-bit fixed-point format with a range scaling of $\pm2$ has discrete additive steps of $2^{-14}\approx6.1\cdot10^{-5}$, so this is also the smallest possible value. Compared to floating-point, precision is worse for small numbers and better for large numbers. For the LBM, this is insufficient and does not work.

\subsection{Required code interventions}
At all places where the DDFs are used as kernel parameters, their data type is made switchable with a macro (\texttt{fpXX}). At any location where the DDFs are loaded or stored in memory, the load/store operation is replaced with another macro as provided in listing \ref{list:conversion} for FP32, FP16S, FP16C, P16$_0$S, P16$_1$S and P16$_2$C. In the appendix in listing \ref{list:lbm-core} we provide the core of our LBM implementation, exemplary for D3Q19 SRT.

\section{Accuracy comparison}
\subsection{3D Poiseuille flow}
A standard setup for LBM validation is a Poiseuille flow through a cylindrical channel \cite{kruger2011unit}. For the channel walls, we use standard non-moving mid-grid bounce-back boundaries \cite{lehmann2019high, kruger2017lattice} and we drive the flow with a body force as proposed by Guo et al. \cite{guo2002discrete}. Simulations are done with the D3Q19 velocity set and a single-relaxation-time (SRT) collision operator. We compare the simulated flow profile $u_\text{sim}(r)$ with the analytic solution \cite{batchelor2000introduction}
\begin{equation} \label{eq:poiseuille-u}
u_\text{theo}(r)\,=\frac{f}{4\,\rho\,\nu}\,(R^2-r^2)
\end{equation}
to compute the error. 
Here, $\rho=1$ is the average fluid density,
\begin{equation}
r=\sqrt{\left(y-\frac{L_y}{2}\right)^2+\left(z-\frac{L_z}{2}\right)^2}
\end{equation}
is the radial distance from the channel center, $R$ is the channel radius,
\begin{equation}
\nu=\frac{2\,R\,u_\text{max}}{\textit{Re}}=\frac{\tau}{3}-\frac{1}{6}
\end{equation}
is the kinematic shear viscosity and $\tau$ is the relaxation time. 
The dimensions of the simulation box are
\begin{equation}
L_x=1,\ \ \ \ \ \ \ \ \ L_y=L_z:=2\,(R+1).
\end{equation}
The flow is driven by a force per volume $f$ that is calculated by rearranging equation (\ref{eq:poiseuille-u}) with $r=0$:
\begin{equation} \label{eq:poiseuille-f}
f=\frac{4\,\rho\,\nu\,u_\text{max}}{R^2}
\end{equation}
In accordance to \cite{lehmann2019high}, we define the error as the $L_2$-norm \cite[p.138]{kruger2017lattice}:
\begin{equation} \label{eq:poiseuille-E}
E=\sqrt{\frac{\sum_{r=0}^R\left|u_\text{sim}(r)-u_\text{theo}(r)\right|^2}{\sum_{r=0}^R\left|u_\text{theo}(r)\right|^2}}
\end{equation}
\begin{figure}[!htb]
\centering \includegraphics[width=8cm]{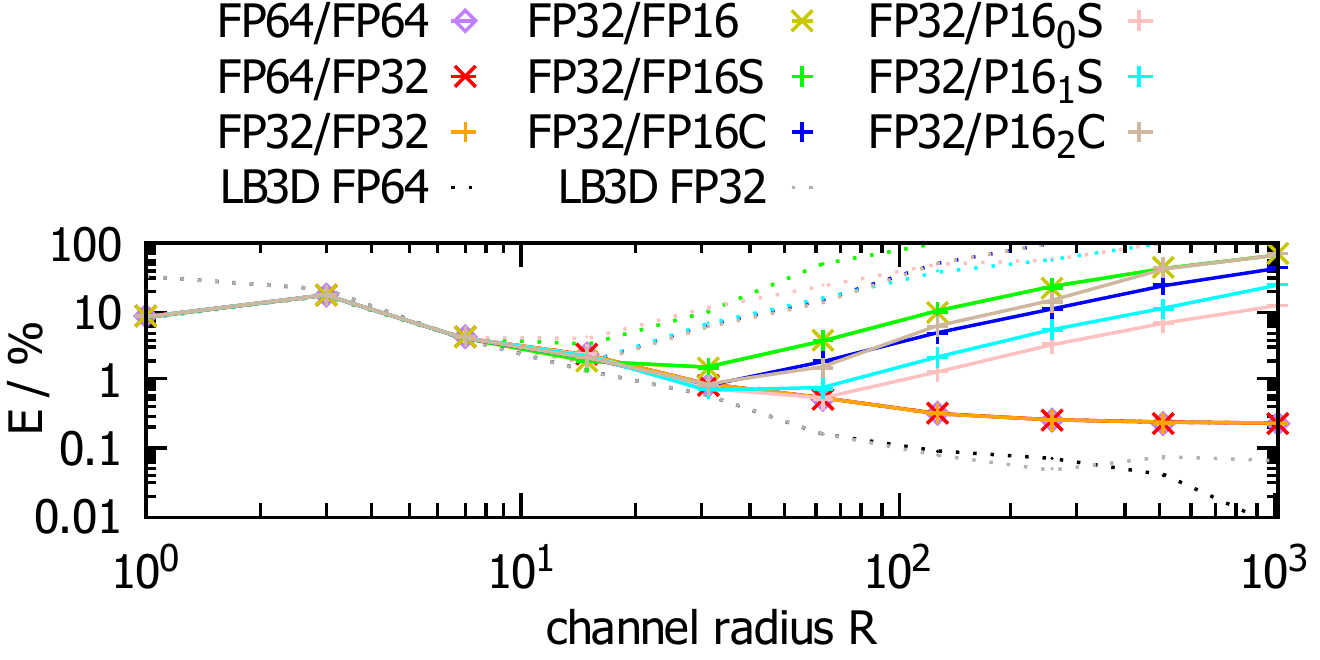}
\vspace*{0.1cm}
\caption{Error of D3Q19 SRT Poiseuille flow for varying channel radius $R$ (lattice resolution) at constant Reynolds number $\textit{Re}=10$ and constant center flow velocity $u_\text{max}=0.1$. The dashed lines represent corresponding simulations without DDF-shifting.} \label{fig:Re=const-R}
\end{figure}
\begin{figure*}[!t]
\vspace*{-0.4cm}
\begin{center}
  \begin{subfigure}{8cm}
\centering \includegraphics[width=8cm]{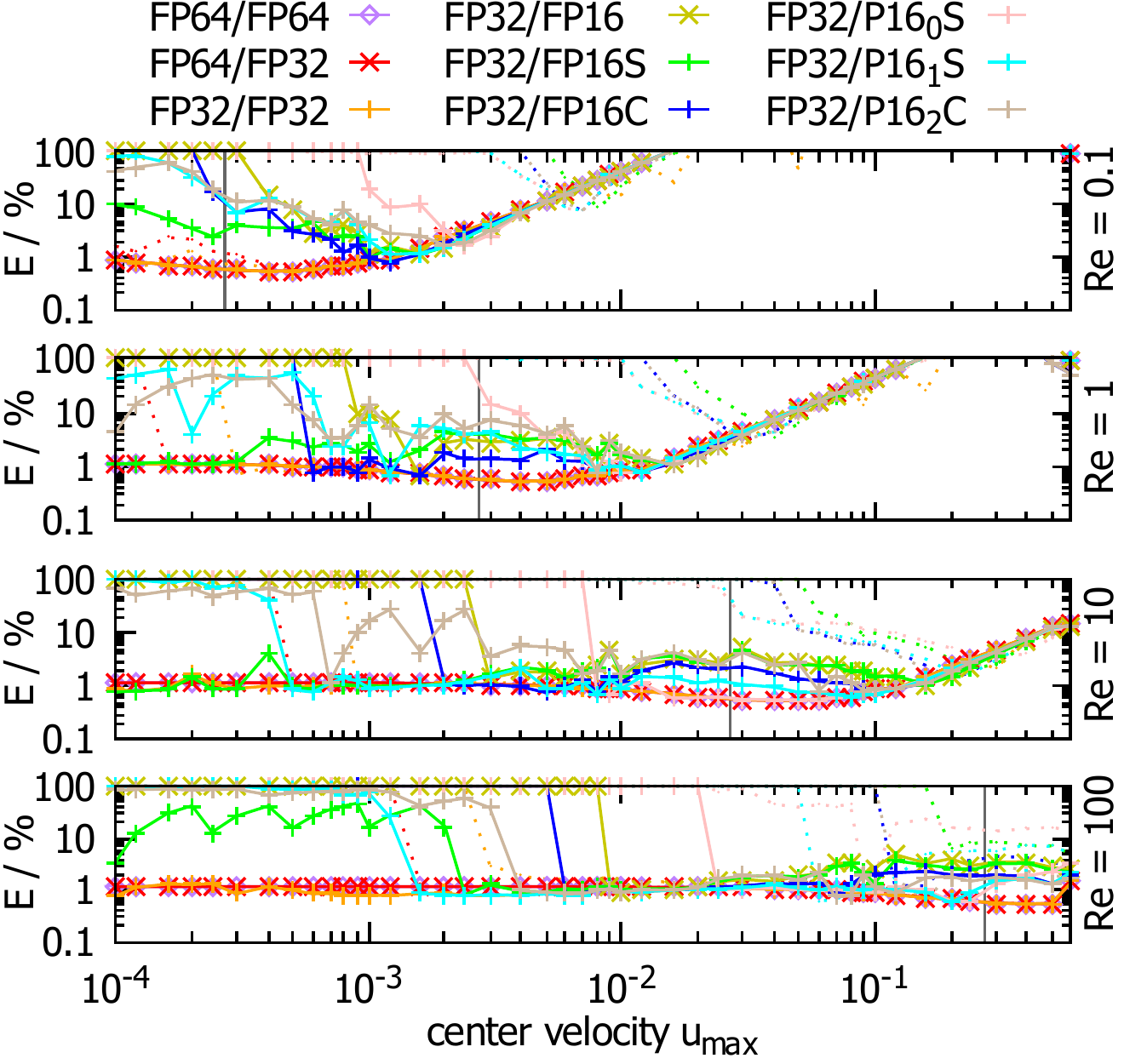}
    \vspace*{-0.6cm}\caption{$R=31$}
  \end{subfigure}\hfill
  \begin{subfigure}{8cm}
\centering \includegraphics[width=8cm]{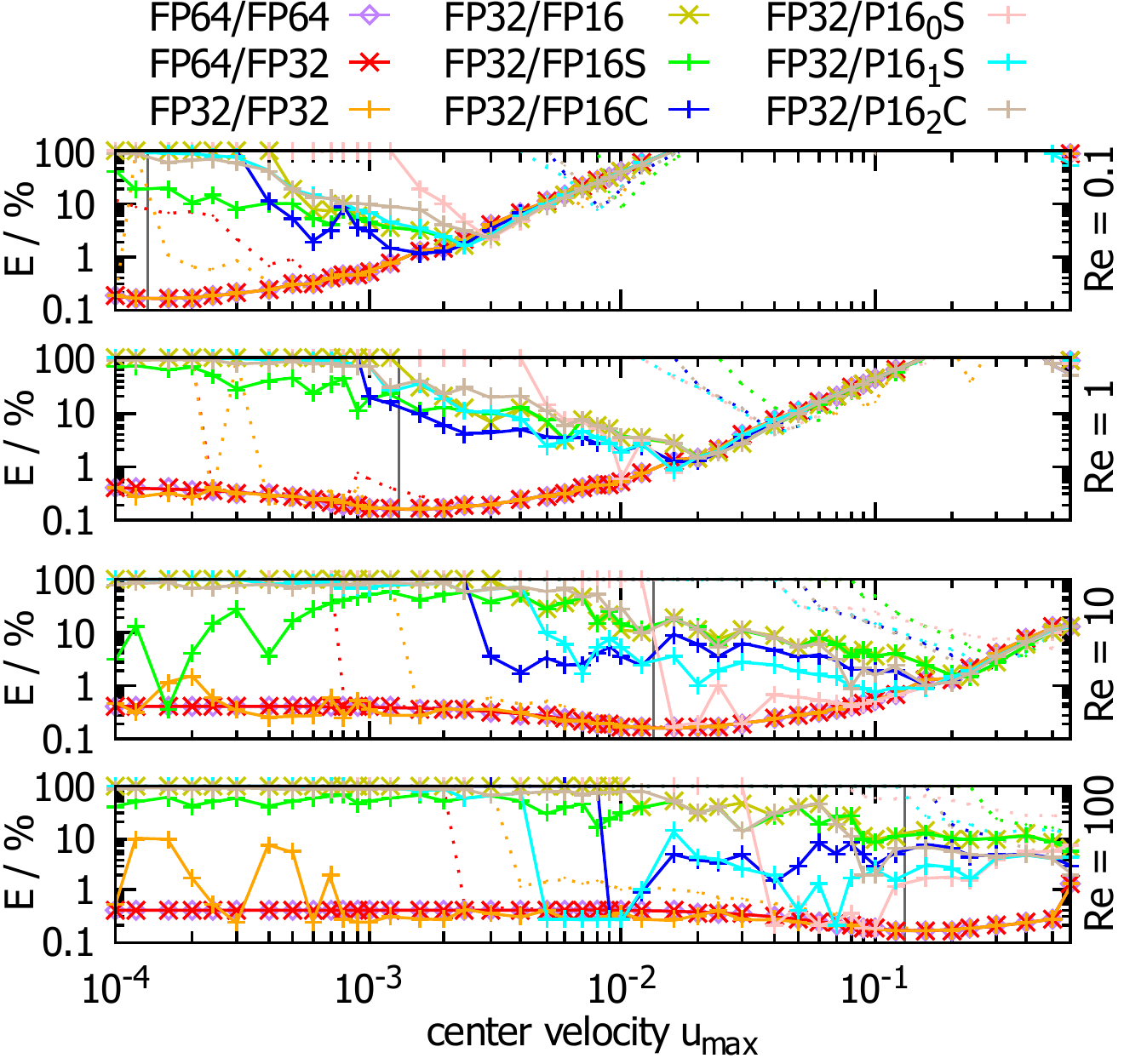}
    \vspace*{-0.6cm}\caption{$R=63$}
  \end{subfigure}
\end{center}
\vspace*{-0.3cm}
\caption{Error of D3Q19 SRT Poiseuille flow for varying center velocity $u_\text{max}$ at constant Reynolds number $\textit{Re}\in\{0.1,1,10,100\}$ and constant channel radius (a) $R=31$ and (b) $R=63$. The dashed curves represent corresponding simulations without DDF-shifting. The vertical lines represent the LBM relaxation time $\tau=1$.} \label{fig:Re=const-umax}
\end{figure*}\\
In figure \ref{fig:Re=const-R} we keep the Reynolds number and center velocity constant at $\textit{Re}=10$ and $u_\text{max}=0.1$, and vary channel radius $R$ and kinematic shear viscosity $\nu$ accordingly. For $R\leq15$, we see almost no difference between any of the floating-point variants. Here the staircase effect of the channel walls dominates the error. Moving towards larger radii, the error increases at first for FP32/FP16 and FP32/FP16S and later for FP32/FP16C as well, while FP64/xx and FP32/FP32 show no difference in this regime either. 16-bit Posit formats hold up even better here with their increased peak accuracy. P16$_2$C for small $R$ behaves like FP16C and then migrates over to FP16S as $R$ becomes larger and the DDFs become smaller. 
We also simulate the same system without DDF-shifting (dashed lines) to quantify the difference. Already here we see that the 16-bit formats become unfeasible without DDF-shifting.

To confirm that the observed agreement between FP32/FP32 and FP64/FP64 is not a coincidence of our implementation, we include in figure~\ref{fig:Re=const-R} data from a simulation of the very same system with the LB3D code \cite{schmieschek17} that is further described in the appendix. 

We now investigate the error in more detail for a constant channel radius $R\in\{31,63\}$ in figure \ref{fig:Re=const-umax}. We simulate the flow in the channel for different Reynolds numbers $\textit{Re}\in\{0.1,1,10,100\}$ and vary the center velocity $u_\text{max}$ and kinematic shear viscosity $\nu$ accordingly.\\
We find that the higher the Reynolds number, the further the minimal error is shifted towards larger $u_\text{max}$, always staying close to where $\tau=1$ (vertical lines). The better small numbers can be resolved, the lower $u_\text{max}$ can be chosen before the error suddenly becomes large. The better the accuracy of the mantissa, the lower is the overall error, up to a certain point where discretization errors dominate at large $u_\text{max}$.\\
It is important to consider that compute time is proportional to $u_\text{max}$ and that $u_\text{max}<u_{\text{max},\tau=1}=\frac{\textit{Re}}{12\,R}$ smaller than at the error minimum is thus less practically relevant. In the domain $u_\text{max}\geq u_{\text{max},\tau=1}$ (in figure \ref{fig:Re=const-umax} right of the vertical lines), FP16C is almost always superior to FP16S, especially at higher $\textit{Re}$. Posits show their superior precision most of the time, if the DDFs are just in the right interval.\\
We find that without DDF-shifting, the 16-bit formats become very inaccurate. For FP32/FP32, there is some benefit at higher $\textit{Re}$ and especially low velocities $u_\text{max}$. For FP64, the DDF-shifting does not make any noticeable difference in this setup as discretization errors dominate.
\begin{figure}[!htb]
\centering \includegraphics[width=8cm]{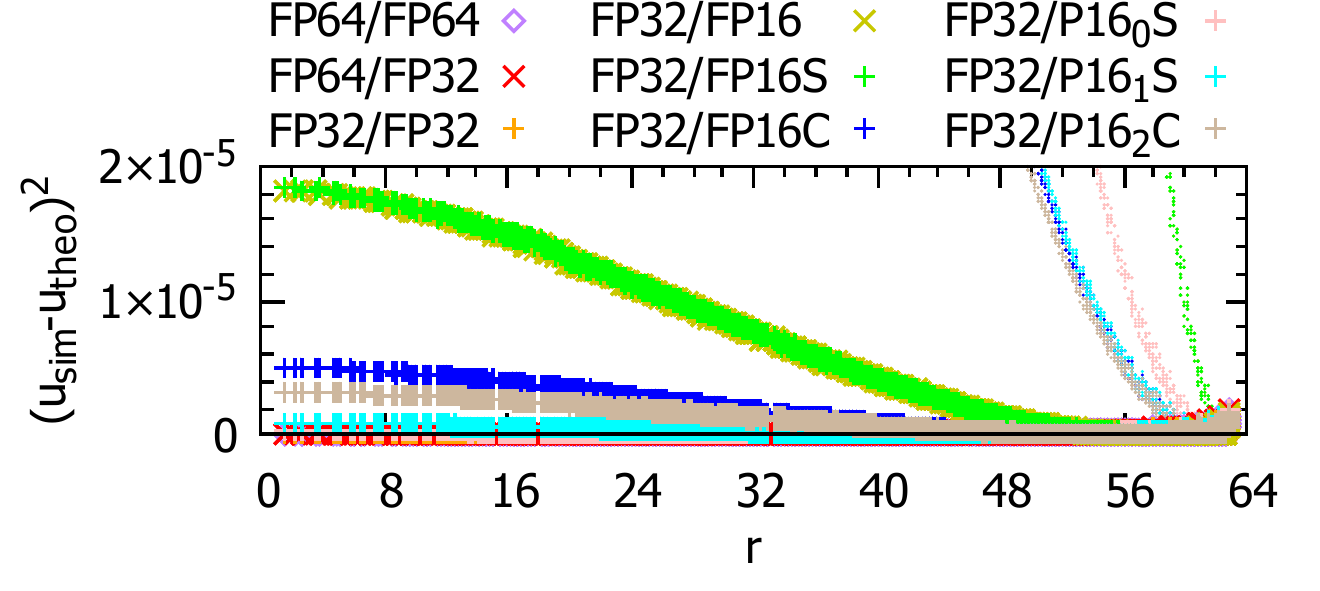}
\caption{Radial error profile of D3Q19 SRT Poiseuille flow for a channel radius $R=63$ and center flow velocity $u_\text{max}=0.1$ at constant Reynolds number $\textit{Re}=10$. The small dots on the right represent corresponding simulations without DDF-shifting. Note that the error contribution is on a linear scale here.} \label{fig:radial-error}
\end{figure}\\
To better understand where the error comes from in the Poiseuille channel radially, we exemplary plot the error contribution as a function of the radial coordinate $r$ for the parameters $R=63$, $u_\text{max}=0.1$, $\textit{Re}=10$ in figure \ref{fig:radial-error}. We find that for FP64 to FP32, the largest error contribution is near the channel wall (staircase effect along the no-slip bounce-back boundaries). For FP32/16-bit, there is equal error contribution near the wall, but the majority of the error comes from close to the channel center. The wall poses a boundary condition not only for velocity ($u(R)=0$), but also for the velocity error. Going radially inward from the channel wall, at first the staircase effect smooths out, lowering the error, but then each concentric ring of lattice points accumulates systematic floating-point errors, so at the channel center the error is largest. For FP32/FP32 this error behavior is barely noticeable, but visible upon close inspection. For FP64, the floating-point errors are so tiny that the staircase smoothing continues all the way through the radial profile, making the error smallest in the center. Without DDF-shifting, there is no noticeable difference for FP64 and FP32 compared to when DDF-shifting is done, but the 16-bit formats become unfeasible.

\subsection{Taylor-Green vortices}
\begin{figure}[!htb]
\centering \includegraphics[width=7cm]{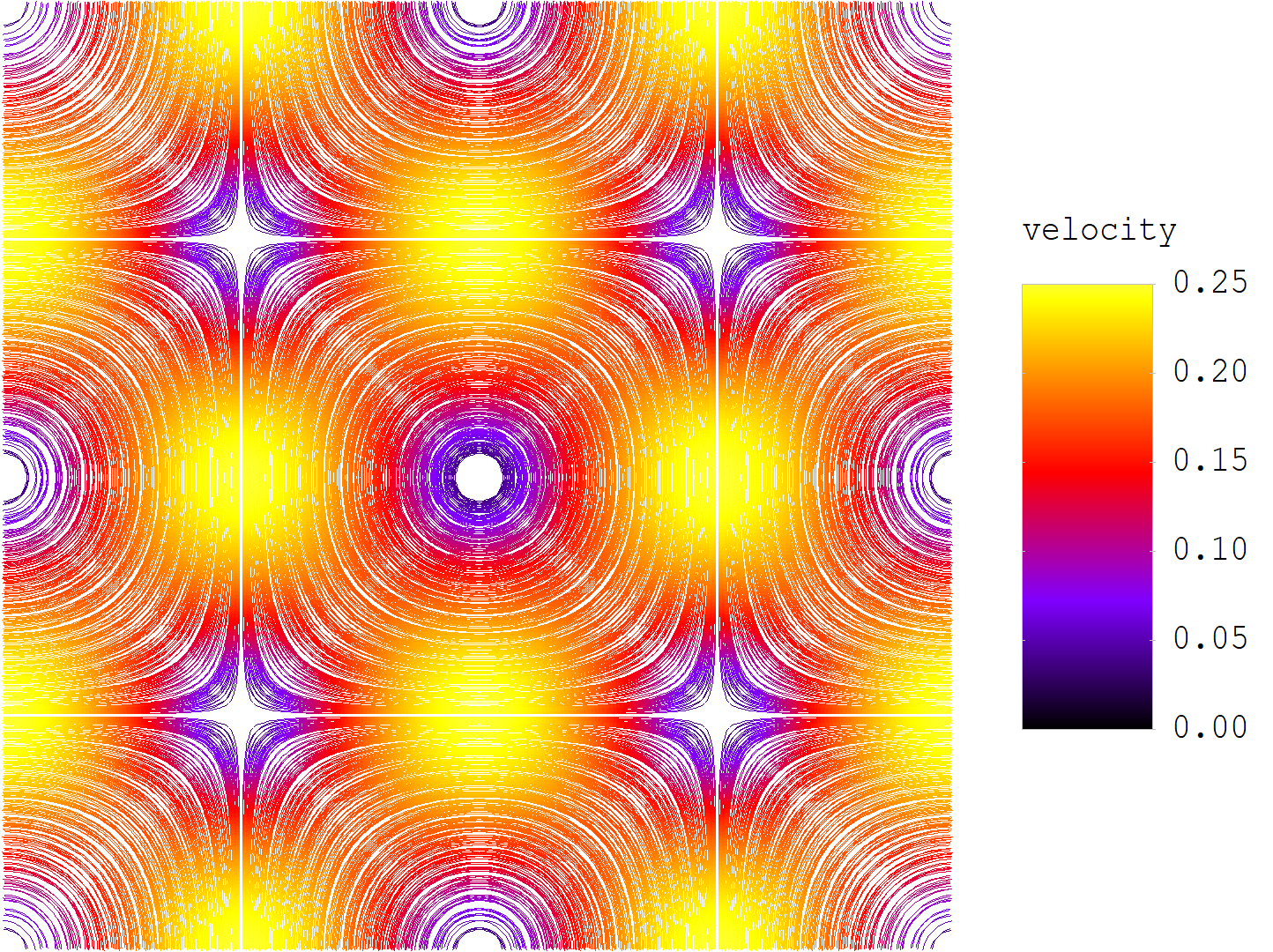}
\vspace*{0.1cm} \caption{Illustration of the velocity field at $t=0$ with colored streamlines.} \label{fig:taylor-green:setup}
\end{figure}

An especially well suited setup for testing the behavior at low velocities is Taylor-Green vortices. A periodic grid of vortices is initialized with velocity magnitude $u_0$ (illustrated in figure \ref{fig:taylor-green:setup}) and then over time viscous friction slows down the vortices while they remain in place on the grid. 
In 2D, the analytic solution \cite{taylor1937mechanism, skordos1993initial} reads
\begin{align}
u_x(t)&=+u_0\,\cos(k\,x)\,\sin(k\,y)\ e^{-2\,\nu\,k^2\,t}\\
u_y(t)&=-u_0\,\sin(k\,x)\,\cos(k\,y)\ e^{-2\,\nu\,k^2\,t}\\
\rho(t)&=1-\frac{3\,u_0^2}{4}\,(\cos(2\,k\,x)+\cos(2\,k\,y))\ e^{-4\,\nu\,k^2\,t}
\end{align}
and at $t=0$ is used to initialize the simulation with $u_0=0.25$. Here $\nu=\frac{\tau}{3}-\frac{1}{6}=\frac{1}{6}$ is the kinematic shear viscosity at $\tau=1$ and $k=\frac{2\,\pi\,N}{L}$. $L=256$ is the side length of the square lattice and $N=1$ is the number of periodic tiles in one direction. The kinetic energy
\begin{align}
E(t)&=\int_0^{L}\int_0^{L}\frac{\rho}{2}\,(u_x^2+u_y^2)\,dx\,dy=\nonumber\\
&=u_0^2\,\pi^2\,e^{-4\,\nu\,k^2\,t} \label{eq:taylor-green:energy}
\end{align}
drops exponentially with time $t$. $E_0=E(t=0)$ denotes the initial kinetic energy. 
\begin{figure}[!htb]
\centering \includegraphics[width=8cm]{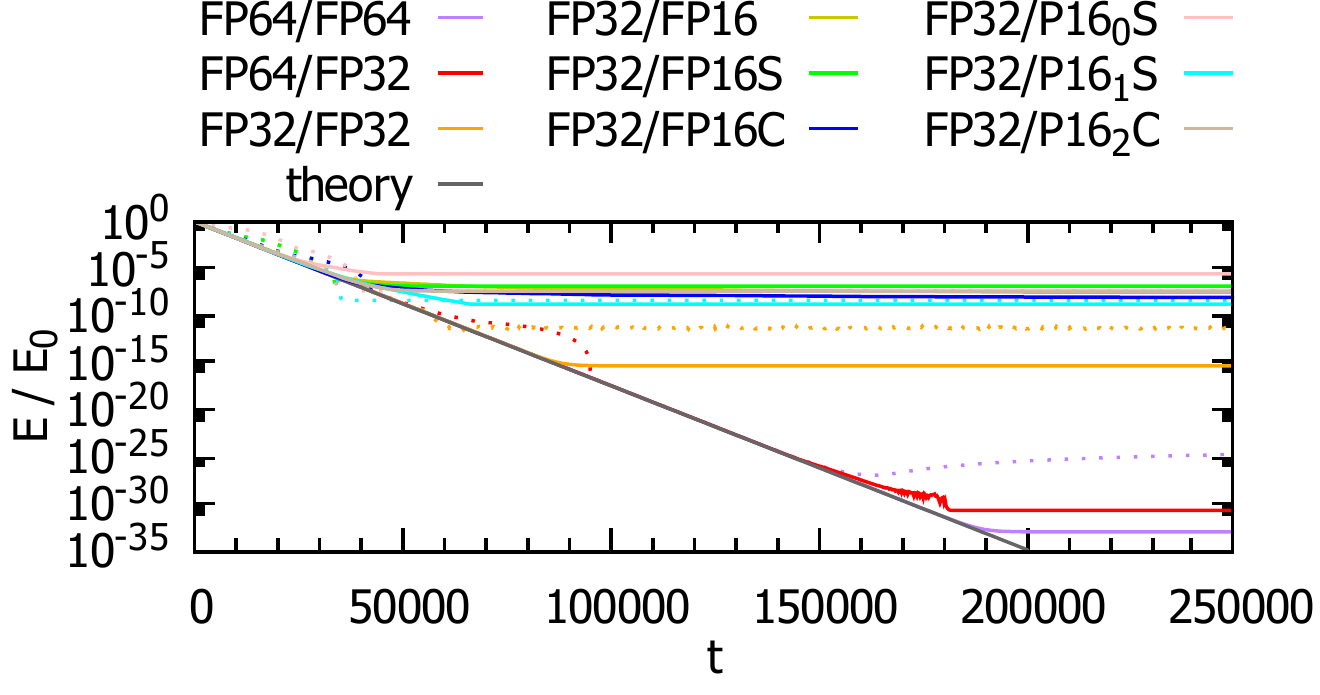}
\caption{Relative kinetic energy $E(t)/E_0$ of a D2Q9 SRT simulation of Taylor-Green vortices compared to the analytic solution in eq. \eqref{eq:taylor-green:energy}. Dashed lines represent corresponding simulations without DDF-shifting.} \label{fig:taylor-green}
\end{figure}
We compute the kinetic energy from the simulation as the discrete sum across all lattice points and compare it to the analytic solution in figure \ref{fig:taylor-green}. The simulated kinetic energy drops exponentially as well, but at some point it does not drop further and remains constant as a result of floating-point errors. The relative energies of the plateaus are no coincidence: The plateaus are located at approximately the truncation error $\epsilon$ squared (table \ref{tab:formats}) for the respective number format in use. 
Particularly interesting is that for FP64/FP32 the plateau is much lower than for FP32 $\epsilon^2$, being closer to FP64 $\epsilon^2$. 
With P16$_0$S, the DDFs are outside of the most accurate interval, so accuracy is poor overall.

Finally, we note that the plateaus only reach down to $\epsilon^2$ if DDF-shifting is properly implemented as presented in equation \eqref{eq:lbm-equilibrium-shifted}. 
Without DDF-shifting, there is significant loss in accuracy across all number formats.

\subsection{Karman vortex street}
Our next setup is a Karman vortex street in two dimensions \cite{karman1911ueber}: a cylinder with radius $R=32$ is placed into a simulation box with dimensions $512\times1024$. 
At the upper and lower box perimeter, a velocity of $\vec{u}=(0,\,0.15)$ is enforced using non-reflecting equilibrium boundaries \cite{lehmann2019high, izquierdo2009analysis}. 
The Reynolds number is set to $\textit{Re}=\frac{2\,R\,|u|}{\nu}=250$, defining the kinematic shear viscosity $\nu=\frac{\tau}{3}-\frac{1}{6}$ and relaxation time $\tau$. 

If starting the simulation with perfectly symmetric initial conditions, only floating-point errors can eventually trigger the Karman vortex instability. 
We notice that in some cases, the instability would not start at all even after several hundred thousand time steps.
To avoid this non-physical behavior, we initialize the velocity $\vec{u}=(0,\,0.15)$ not only at the simulation box perimeter, but also on the left half $x<256$. 
This immediately triggers the Karman vortex instability regardless of floating-point setting.

\begin{figure}[!htb]
\centering \includegraphics[width=8cm]{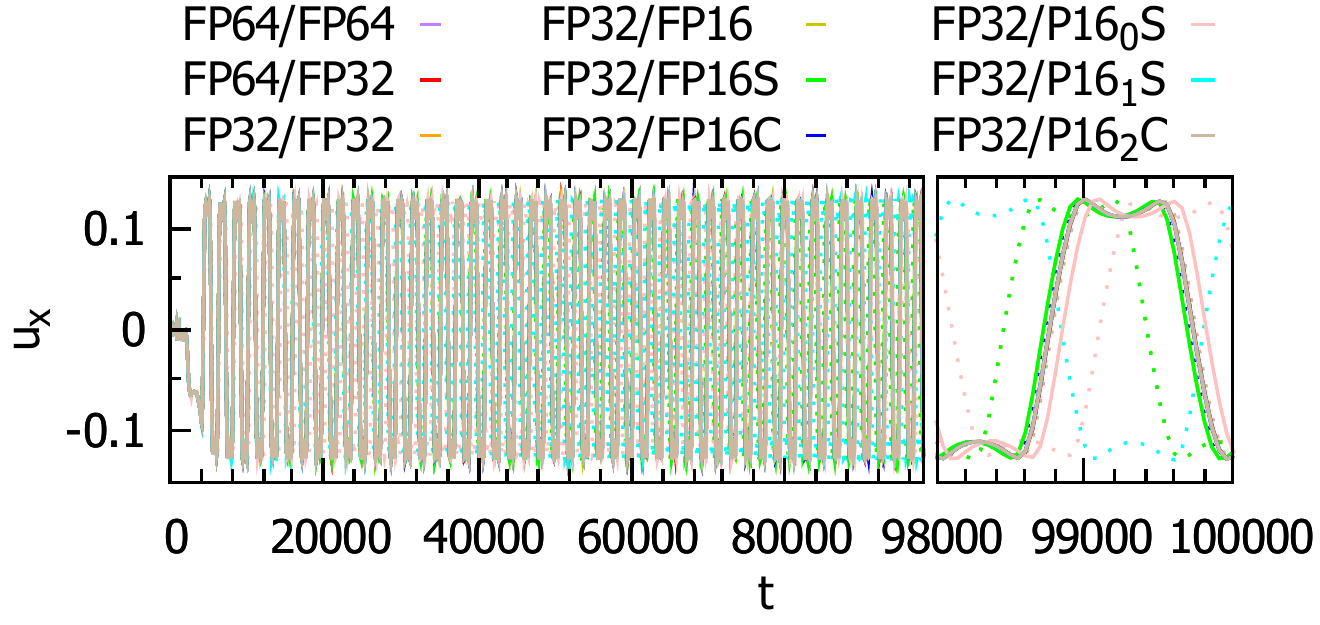}
\caption{Velocity $x$-component of the Karman vortex street at the simulation box center $(256,\,512)$ over time for various floating-point precision. 
Dashed lines represent corresponding simulations without DDF-shifting. 
Even after $100000$ LBM time steps ($50$ vortex periods), the 16-bit graphs still cover the FP64 ground truth as amplitude, frequency and even phase appear indistinguishable. Only zooming in at the last oscillation period reveals minuscule differences in phase for FP16, FP16S and P16$_0$S. The phase-shift in the 16-bit graphs is large without the DDF-shifting optimization. FP32/FP16C, FP32/FP32 and FP64/FP32 are indistinguishable from FP64 ground truth even when zooming in.} \label{fig:karman-ux}
\end{figure}

\begin{figure}[!htb]
\begin{center}

\includegraphics[width=0.76cm]{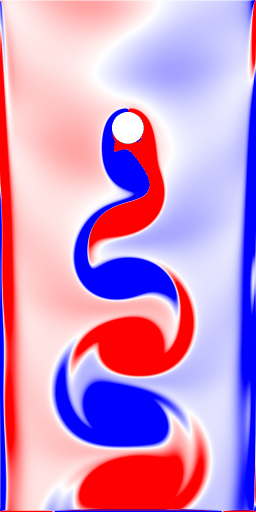}
\includegraphics[width=0.76cm]{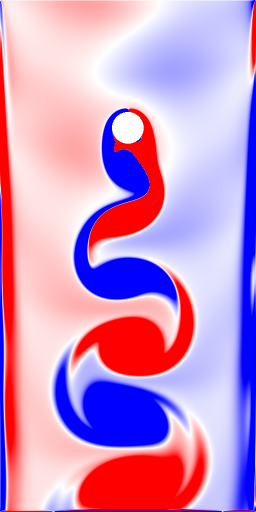}
\includegraphics[width=0.76cm]{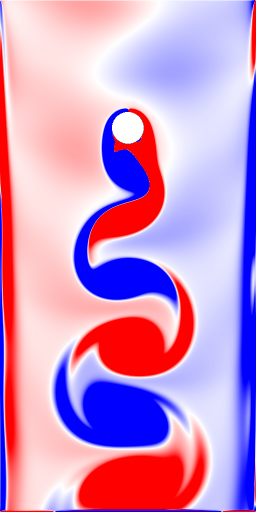}
\includegraphics[width=0.76cm]{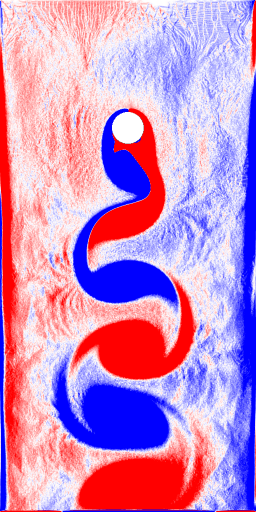}
\includegraphics[width=0.76cm]{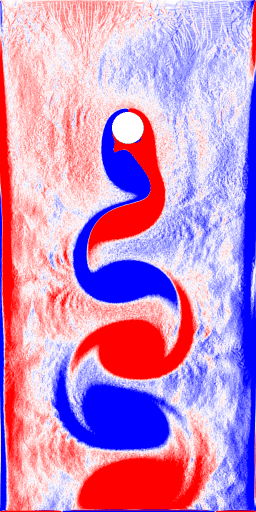}
\includegraphics[width=0.76cm]{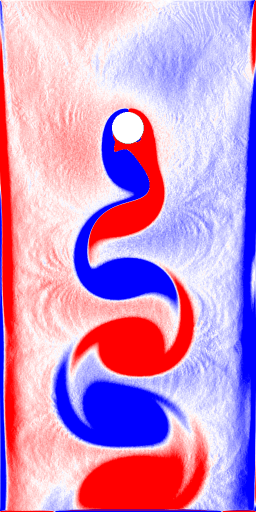}
\includegraphics[width=0.76cm]{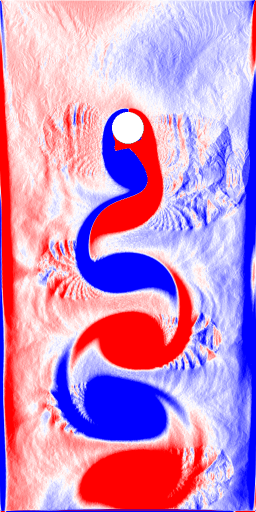}
\includegraphics[width=0.76cm]{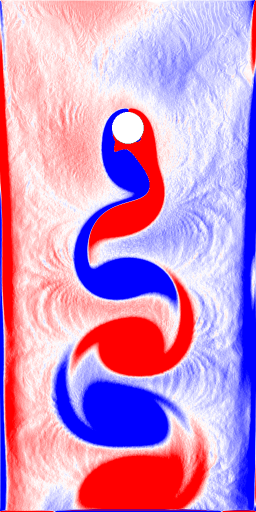}
\includegraphics[width=0.76cm]{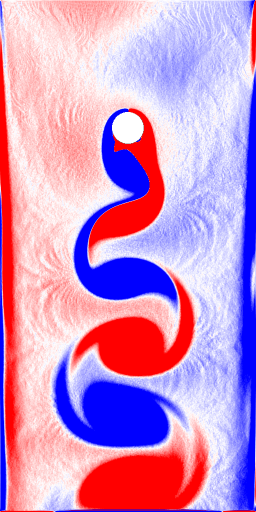}
\includegraphics[width=0.76cm]{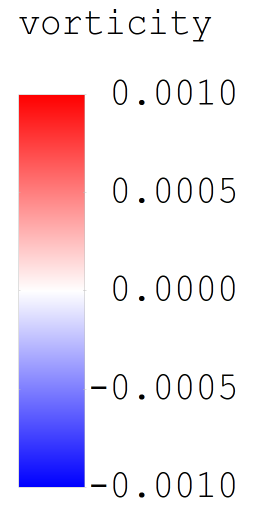}\\
\vspace*{-0.2cm}\raggedright\scalebox{0.4}{\ FP64/FP64\ \ \ FP64/FP32\ \ \ FP32/FP32\ \ \ FP32/FP16\ \ FP32/FP16S\ \ FP32/FP16C\ FP32/P16$_0$S\ \ FP32/P16$_1$S\ \ FP32/P16$_2$C\hfill}\\
\centering\normalsize(a) simulations with DDF-shifting
\end{center}
\begin{center}
\includegraphics[width=0.76cm]{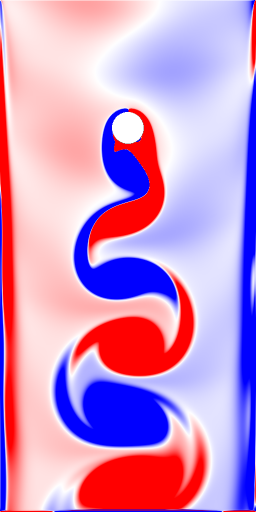}
\includegraphics[width=0.76cm]{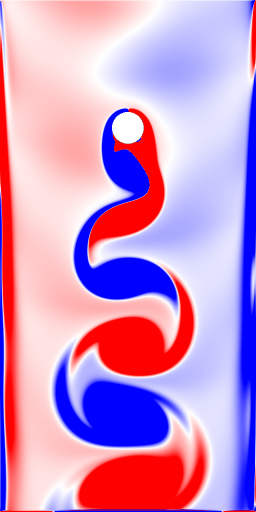}
\includegraphics[width=0.76cm]{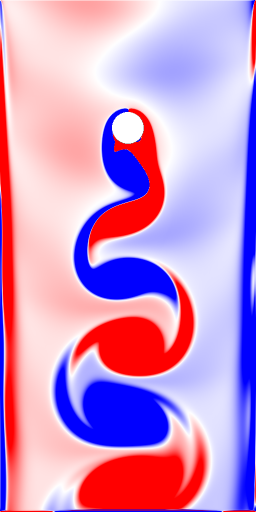}
\includegraphics[width=0.76cm]{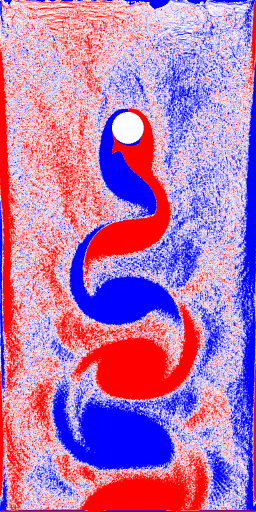}
\includegraphics[width=0.76cm]{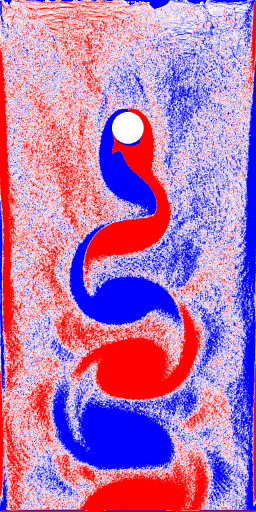}
\includegraphics[width=0.76cm]{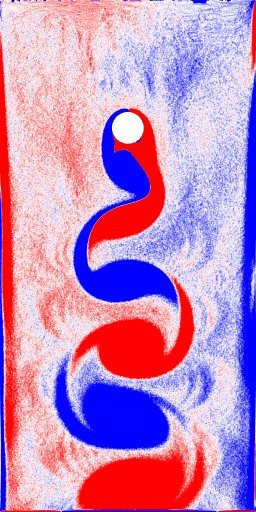}
\includegraphics[width=0.76cm]{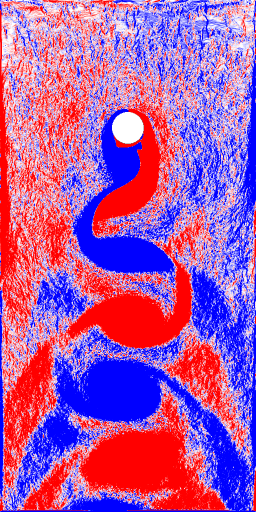}
\includegraphics[width=0.76cm]{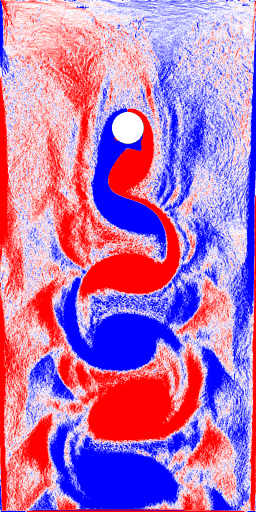}
\includegraphics[width=0.76cm]{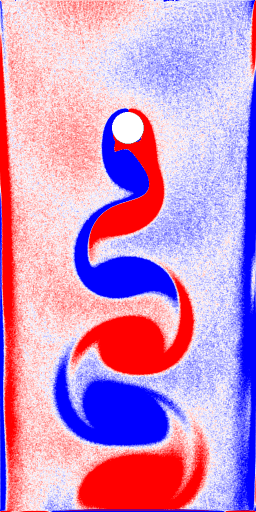}
\includegraphics[width=0.76cm]{images/karman/vorticity-1000/scale.png}\\
\vspace*{-0.2cm}\raggedright\scalebox{0.4}{\ FP64/FP64\ \ \ FP64/FP32\ \ \ FP32/FP32\ \ \ FP32/FP16\ \ FP32/FP16S\ \ FP32/FP16C\ FP32/P16$_0$S\ \ FP32/P16$_1$S\ \ FP32/P16$_2$C\hfill}\\
\centering\normalsize(b) simulations without DDF-shifting
\end{center}
\caption{Vorticity in the vastly overexposed range $\pm0.001$ for simulations (a) with and (b) without DDF-shifting, after $100000$ LBM time steps. All simulations very accurately predict the vortex street, with frequency and amplitude of the vortices being identical and only insignificant differences in phase-shift even after $50$ vortex periods. FP32 is indistinguishable from FP64 ground truth. For 16-bit, in the low vorticity range there is noise present, equally for FP16 and FP16S, but vastly reduced for FP16C and P16$_2$C. Omitting DDF-shifting vastly increases this noise and also adds significant phase-shift as can be seen by comparing the position of the last red vortex at the bottom.} \label{fig:karman-w}
\end{figure}

We probe the velocity at the simulation box center $(256,\,512)$ over time in figure \ref{fig:karman-ux}. 
This demonstrates that, when DDF-shifting is done, the 16-bit formats are almost indistinguishable from FP64 ground truth both qualitatively and quantitatively, with only minimal phase-shift for FP16, FP16S and P16$_0$S.\\
To assess in detail where eventual differences may be present beyond a single velocity point-probe, we look at the vorticity throughout the simulation box. In figure \ref{fig:karman-w} we show the vorticity in the very much zoomed in range of $\pm0.001$. For the 16-bit formats, in low vorticity areas there is noise present. Comparing FP16 and FP16S, the extended number range towards small numbers has no benefit here. FP16C with DDF-shifting mostly mitigates this noise, showing that the noise purely originates in smaller mantissa accuracy and numeric loss off significance. Our custom Posit P16$_2$C has similarly low noise. P16$_0$S shows artifacts.

\subsection{Lid-driven cavity} \label{sec:ldc}
\begin{figure}[!htb]
\centering \includegraphics[width=8cm]{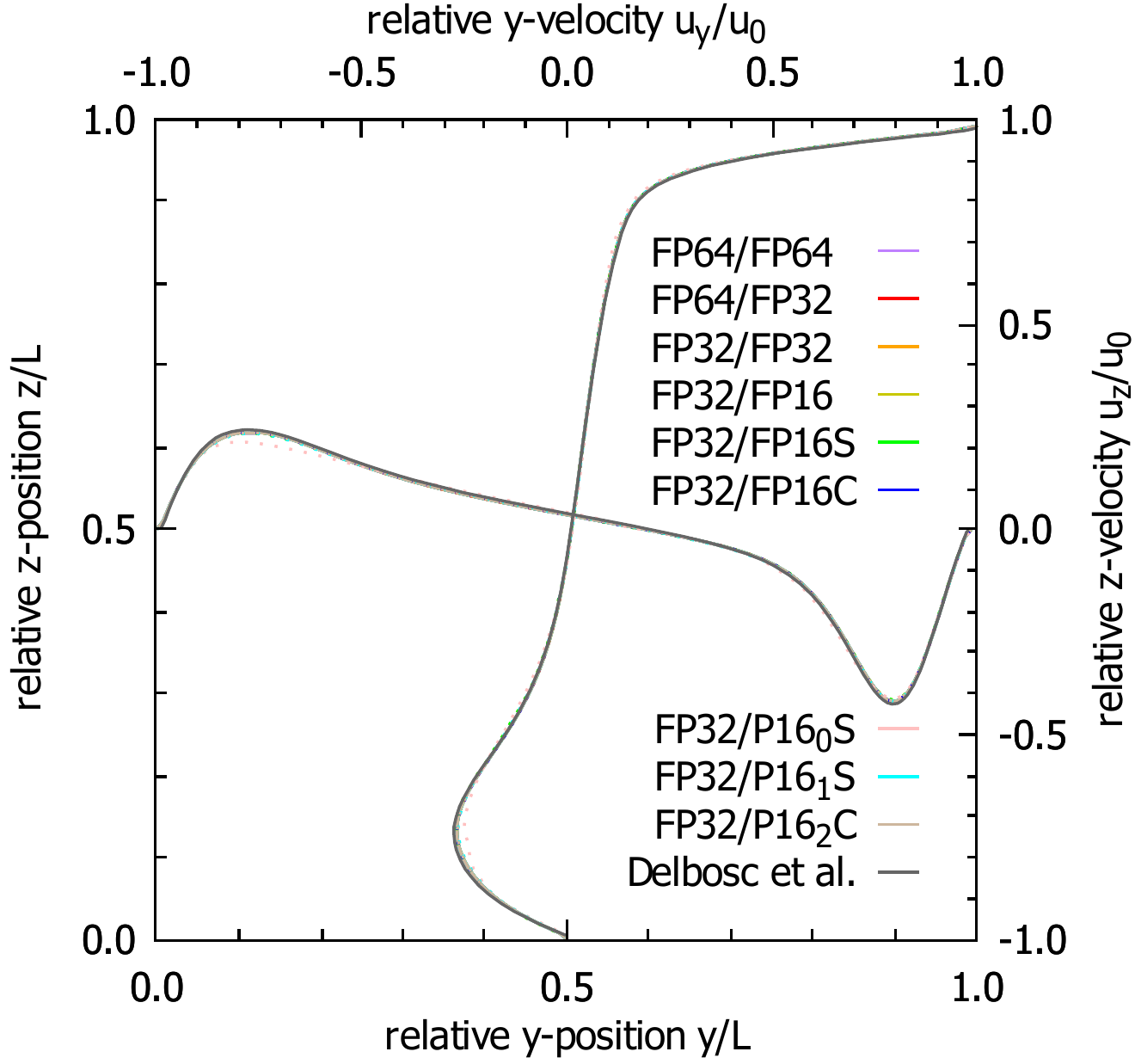}
\caption{The $y$-velocity along a vertical probe line through the simulation box center as well as the $z$-velocity along a horizontal line through the simulation box center. At the top lid, the velocity is fixed. As the flow goes one rotation clockwise, the width of the high-velocity peak increases and the height decreases. Dashed lines represent corresponding simulations without DDF-shifting. As a reference, we show the data from Delbosc et al. \cite{delbosc2014optimized}.} \label{fig:ldc-velocity}
\end{figure}

The lid-driven cavity is a common test setup for the LBM \cite{rinaldi2012lattice, obrecht2011new, kuznik2010lbm, delbosc2014optimized, mawson2014memory, hong2015scalable, cattenone2020basis, alim2009lattice, neumann2013dynamic} and other Navier-Stokes solvers \cite{ghia1982high, jiang1994large, yang1998implicit}.
We here implement it in a cubic box at Reynolds number $\textit{Re}=1000$.
On the lid, velocity parallel to the $y$-axis is enforced through moving bounce-back boundaries \cite{lehmann2019high, kruger2017lattice}. 
The box edge length is $L=128$, the velocity at the top lid is $u_0=0.1$ in lattice units and the kinematic shear viscosity is set by the Reynolds number $\textit{Re}=\frac{L\,u_0}{\nu}=1000$. We simulate $100000$ LBM time steps with the D3Q19 SRT scheme.\\
Figure \ref{fig:ldc-velocity} shows the $y$-/$z$-velocity along horizontal/vertical probe lines through the simulation box center. All number formats look indistinguishable, even without DDF-shifting. Only when zooming in (not shown), for the simulations without DDF-shifting, deviations in relative velocity in the 2nd digit become visible. With DDF-shifting, deviations are present only in the 4th digit, being smallest for FP16C, P16$_1$S and P16$_2$C.

\subsection{Capsule in shear flow}
\begin{figure}[!htb]
\begin{center}
\includegraphics[width=1.33cm]{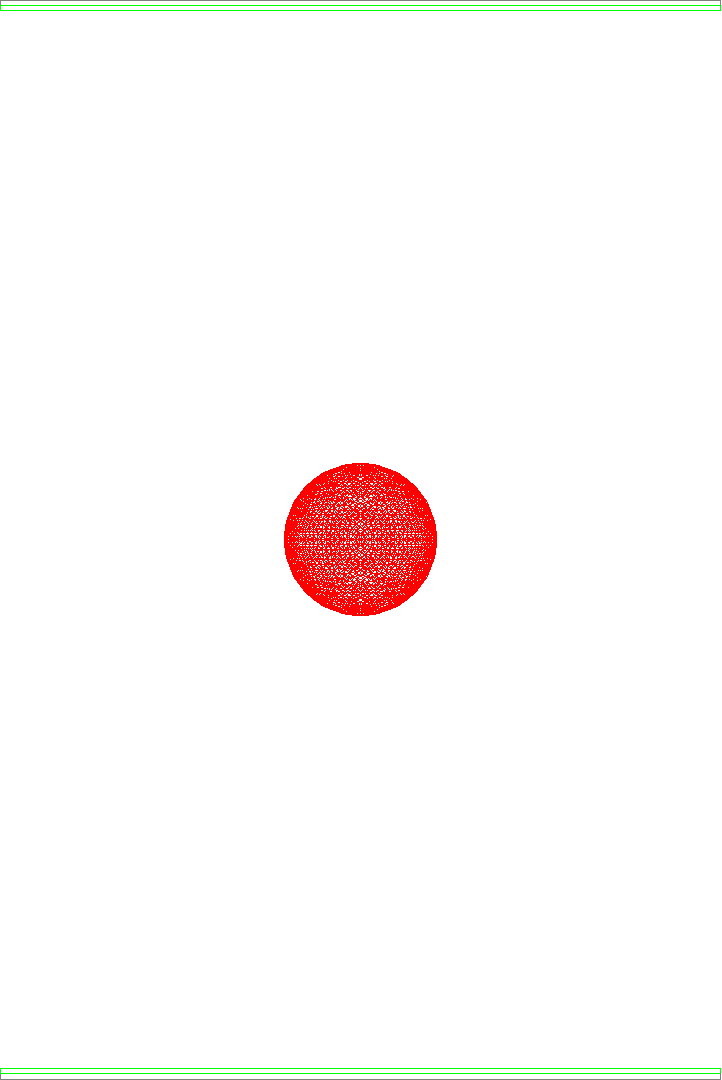}
\includegraphics[width=1.33cm]{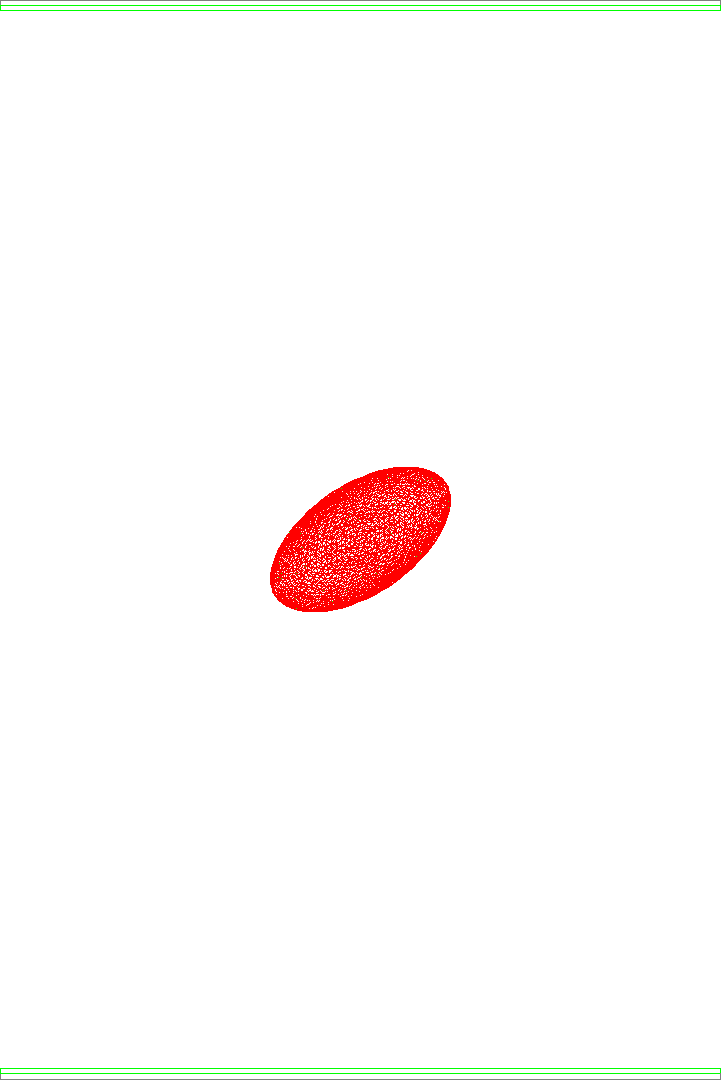}
\includegraphics[width=1.33cm]{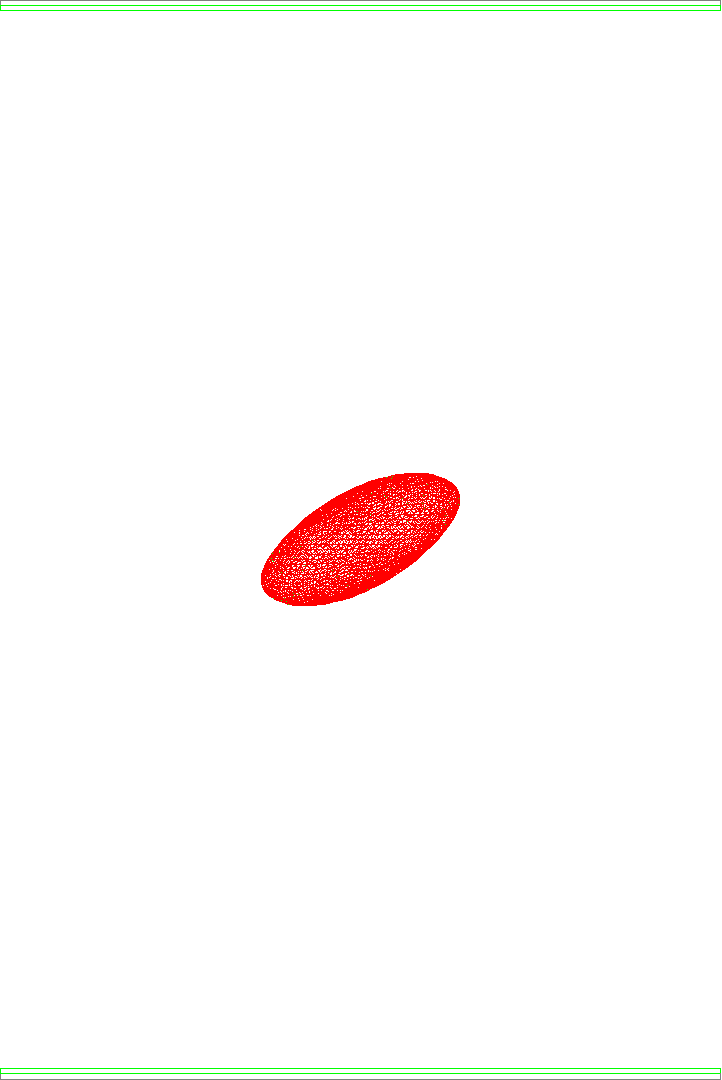}
\includegraphics[width=1.33cm]{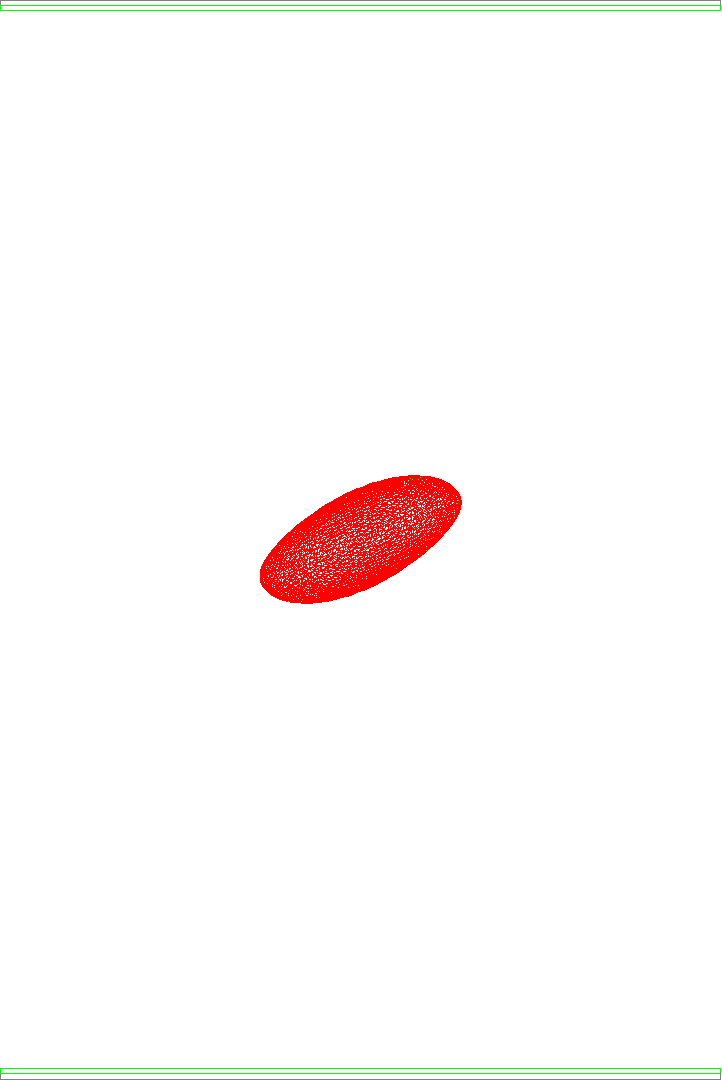}
\includegraphics[width=1.33cm]{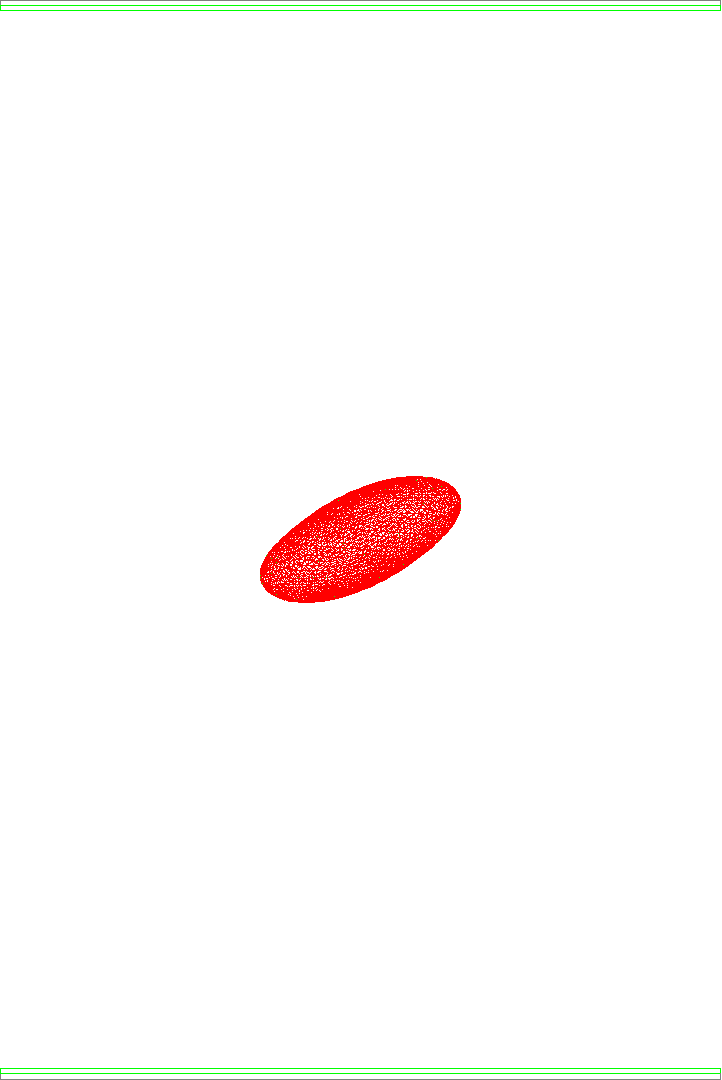}
\includegraphics[width=1.33cm]{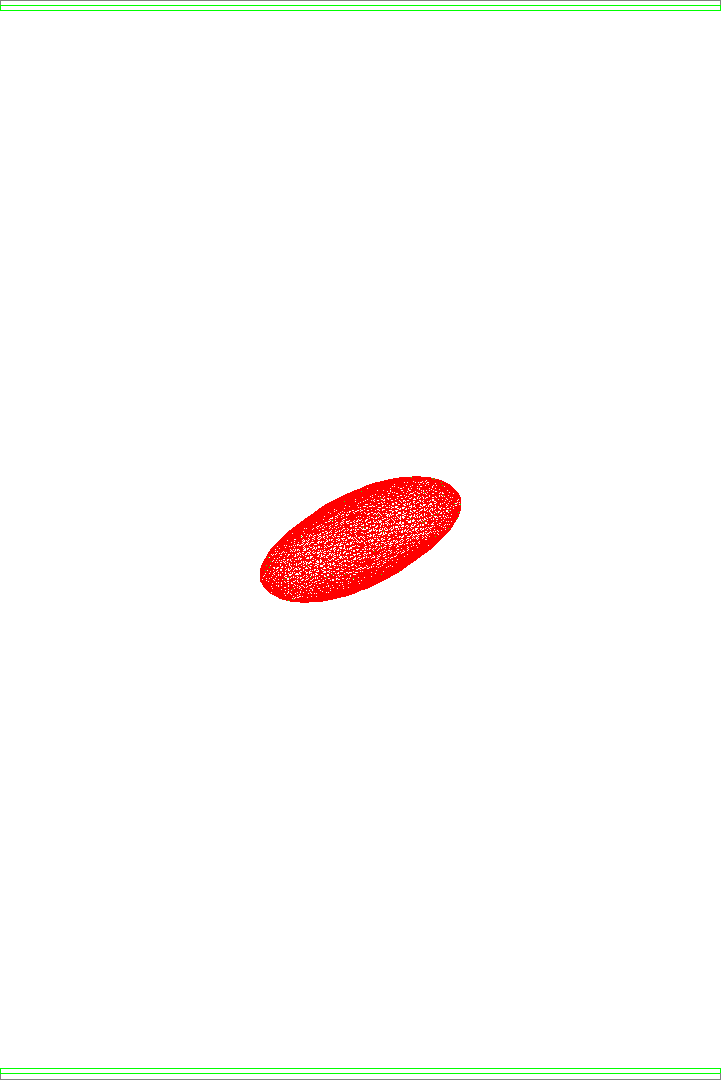}
\end{center}
\caption{Illustration of the capsule in shear flow (FP32/FP32, $\textit{Ca}=0.1$) simulation at dimensionless times $\dot{\gamma}\,t\in\{1,\,2,\,3,\,4,\,5\}$. Each image shows the simulation box from the side, with the top and bottom moving bounce-back boundaries marked in green. The capsule initially deforms to an elongated shape and then performs tank-treading, i.e. rotating the membrane while keeping its deformed shape.} \label{fig:capsule-illustration}
\end{figure}

Here we test the number formats on a microcapsule in shear flow, one of the standard tests for microfluidics simulations in medical applications \cite{guckenberger2016bending, guckenberger2017theory}. The D3Q19 multi-relaxation-time (MRT) \cite{lehmann2019high, kruger2017lattice} LBM is extended with the immersed-boundary method (IBM) \cite{kruger2011introduction} to simulate the deformable microcapsule in flow. 
For the IBM we use the same level of precision as for the LBM arithmetic, so either FP64 or FP32. 
As illustrated in figure \ref{fig:capsule-illustration}, we place an initially spherical capsule of radius $R=13.5$ in the center of a simulation box with the dimensions $128\times64\times192$, and we compute $385000$ time steps. 
At the top and bottom of the simulation box, a shear flow is enforced via moving bounce-back boundaries \cite{ladd1994numerical}. 
The membrane of the capsule is discretized into $5120$ triangles and membrane forces, consisting of shear forces (neo-Hookean) \cite{barthes2010flow, barthes2016motion, guckenberger2016bending} as well as volume forces (volume has to be conserved), are computed as in \cite{guckenberger2016bending}.

\begin{figure}[!htb]
\centering \includegraphics[width=8cm]{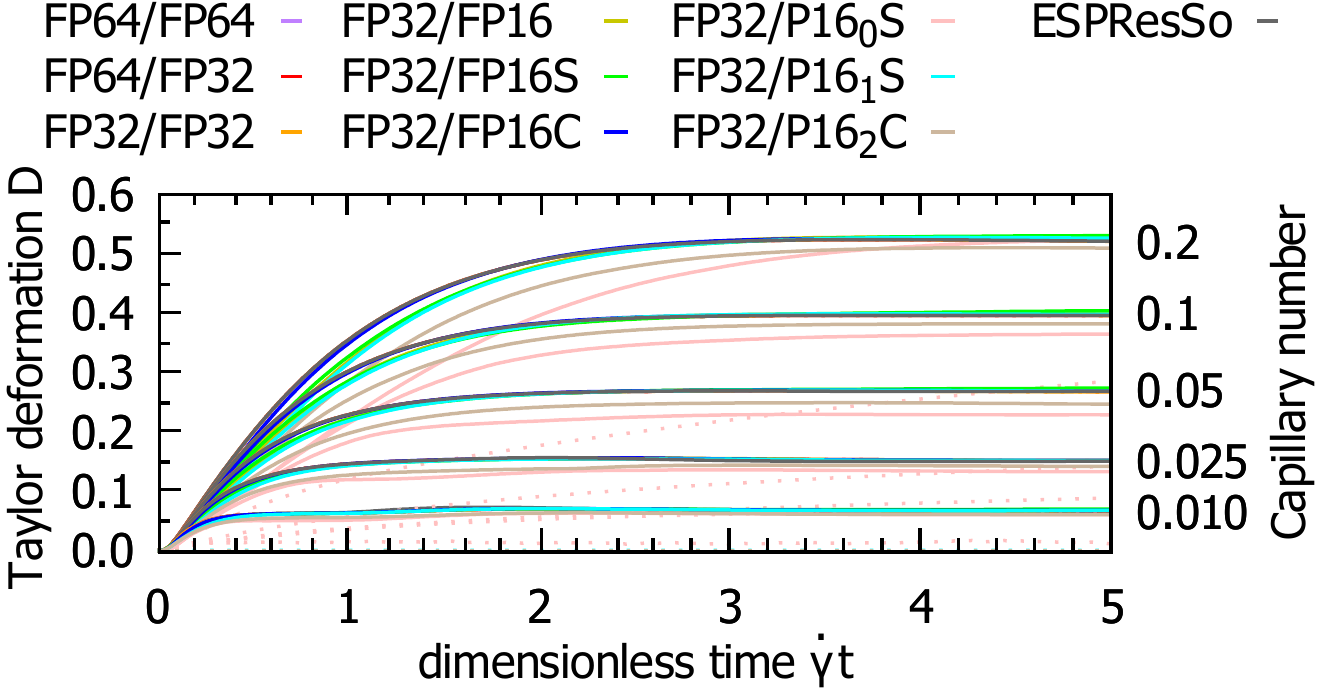}
\caption{Taylor deformation of the capsule first increases and then plateaus as the capsule starts tank-treading. This plateau depends on the Capillary number. Dashed lines represent corresponding simulations without DDF-shifting.} \label{fig:capsule-deformation}
\end{figure}

The Reynolds number is $\textit{Re}=0.05$, the kinematic shear viscosity is $\nu=\frac{1}{3}$ and we simulate various capillary numbers $\textit{Ca}=\frac{\dot{\gamma}\,\mu\,R}{k_1}\in\{0.010,\,0.025,\,0.05,\,0.1,\,0.2\}$ by varying the membrane shear modulus $k_1$. 
The shear rate is $\dot{\gamma}=1.3\cdot10^{-5}$ in simulation units. 
To cross-validate our results, we perform the same simulations with ESPResSo (FP32 for LBM, FP64 for IBM) \cite{griebel2005meshfree}, which has been cross-validated with boundary-integral simulations and many others in \cite{guckenberger2016bending}. 
In figure \ref{fig:capsule-deformation} we plot the Taylor deformation $D=\frac{a-c}{a+c}$ over time, with the largest and smallest semi axes $a$ and $c$ of the deformed capsule \cite{guckenberger2016bending}. We see that even in this complex scenario, the FP16C simulations produce physically accurate results with only insignificant deviations from FP64. The other 16-bit formats, especially Posits, perform noticeably worse here. Without DDF-shifting, while FP32 still appears identical to ground truth, the 16-bit simulations all do not produce the correct outcome (deformation remains close to zero). This emphasizes that DDF-shifting is essential for the lower precision formats.

\subsection{Raindrop impact}
\begin{figure}[!b]
\begin{center}
\vspace*{-0.1cm}
\includegraphics[width=1.99cm]{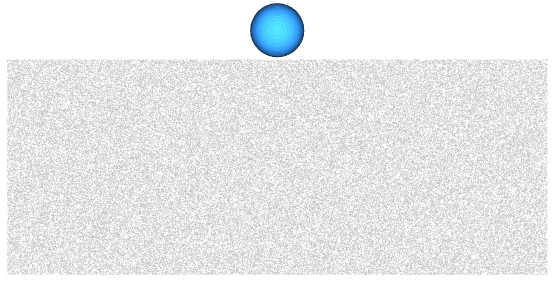}
\includegraphics[width=1.99cm]{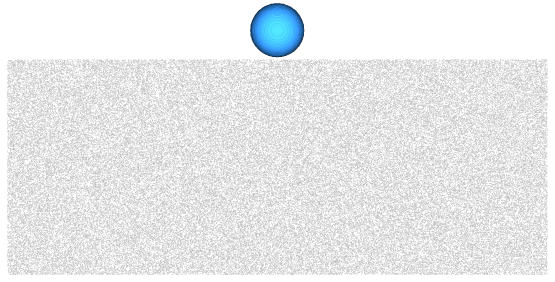}
\includegraphics[width=1.99cm]{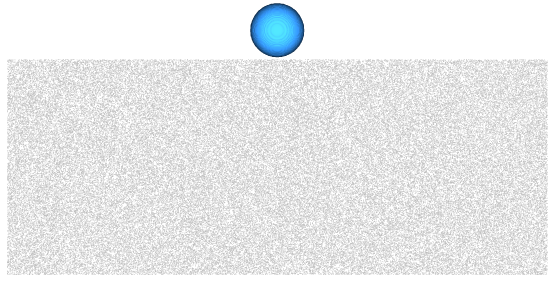}
\includegraphics[width=1.99cm]{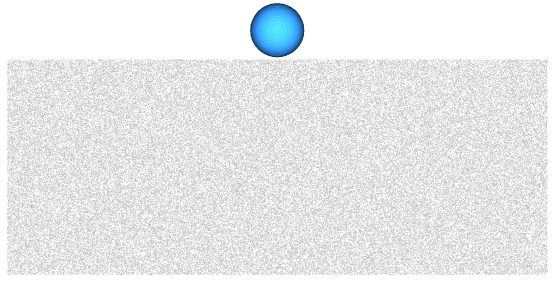}\\
\includegraphics[width=1.99cm]{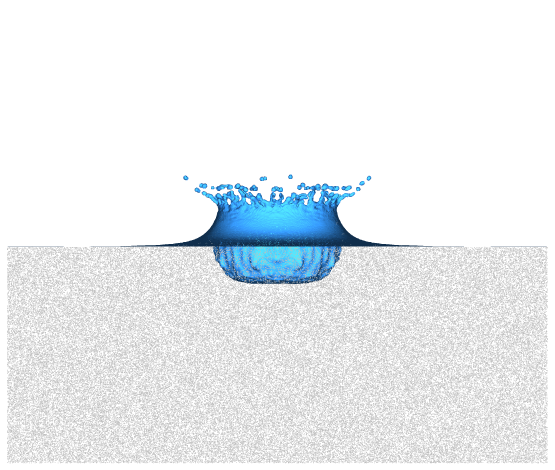}
\includegraphics[width=1.99cm]{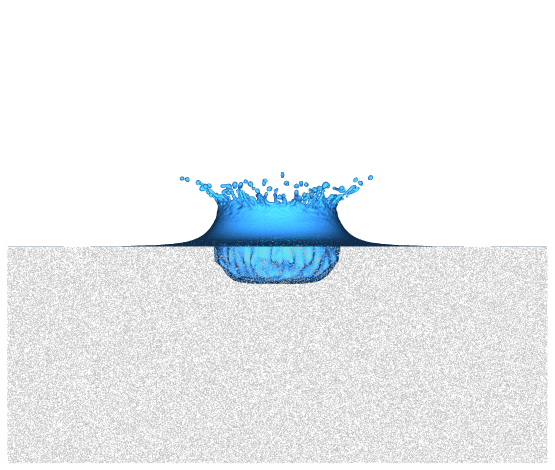}
\includegraphics[width=1.99cm]{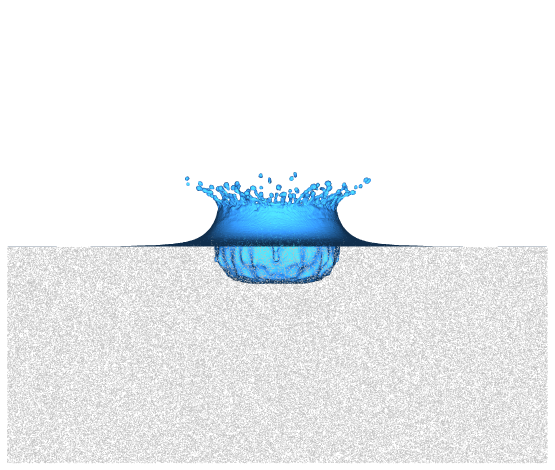}
\includegraphics[width=1.99cm]{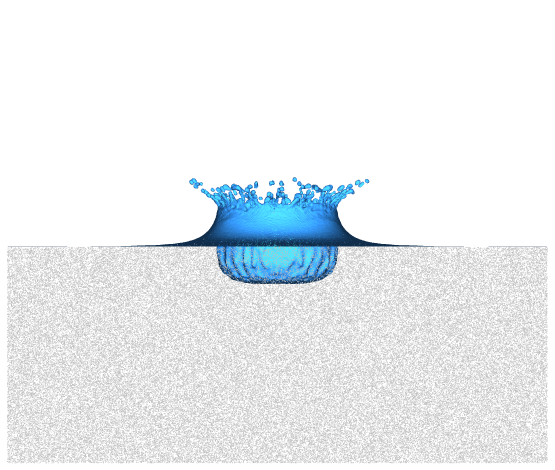}\\
\includegraphics[width=1.99cm]{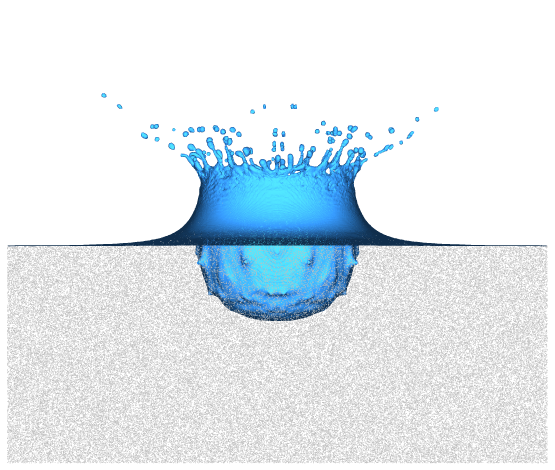}
\includegraphics[width=1.99cm]{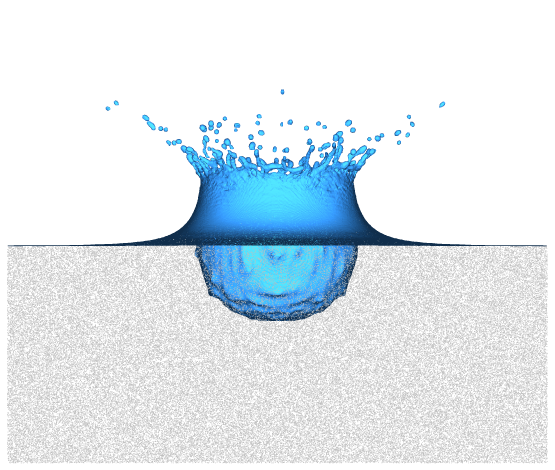}
\includegraphics[width=1.99cm]{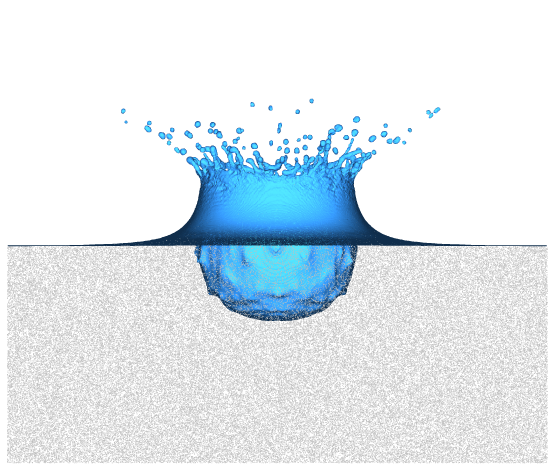}
\includegraphics[width=1.99cm]{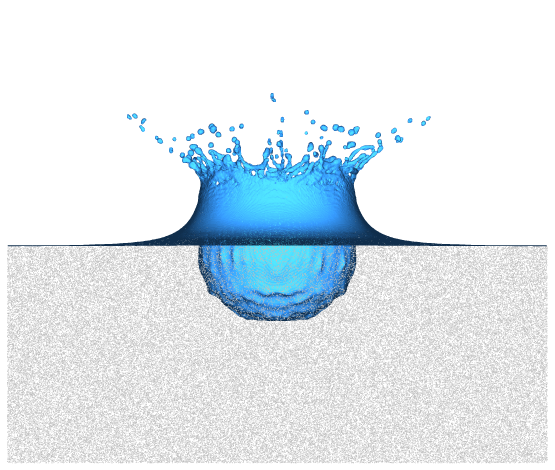}\\
\includegraphics[width=1.99cm]{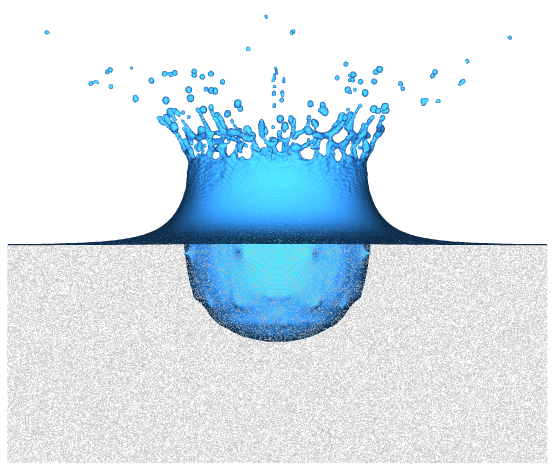}
\includegraphics[width=1.99cm]{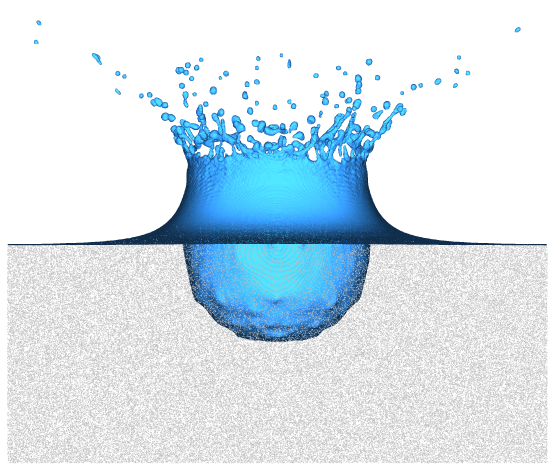}
\includegraphics[width=1.99cm]{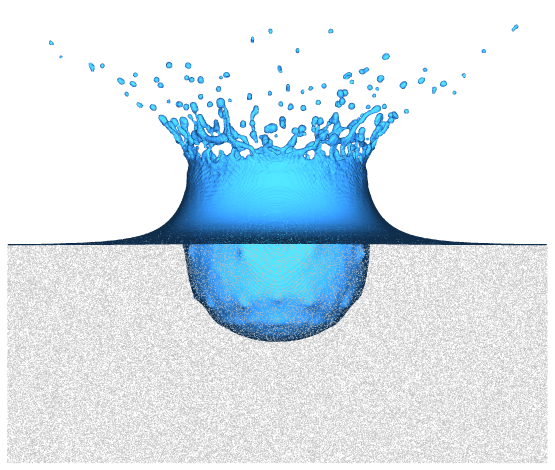}
\includegraphics[width=1.99cm]{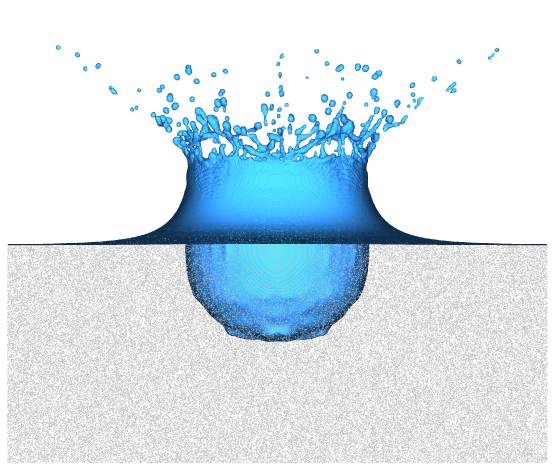}\\
\includegraphics[width=1.99cm]{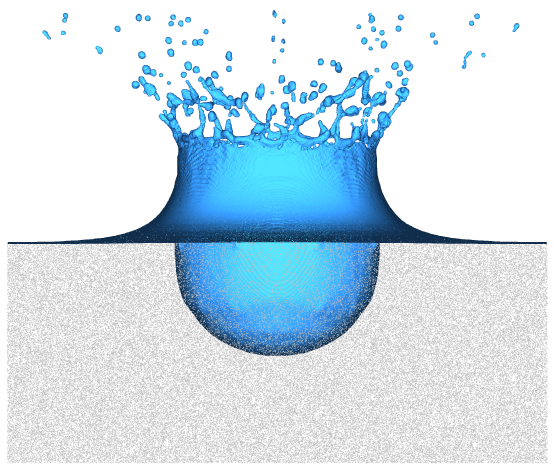}
\includegraphics[width=1.99cm]{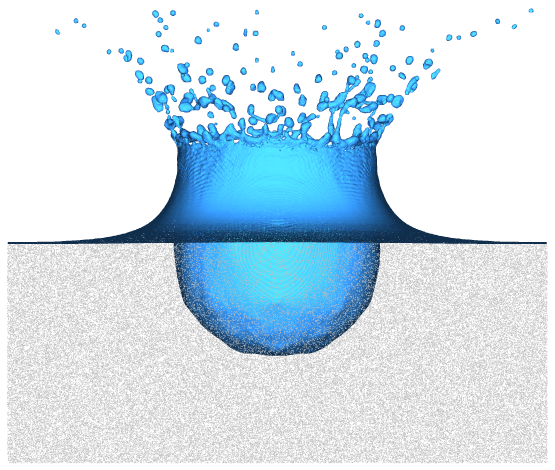}
\includegraphics[width=1.99cm]{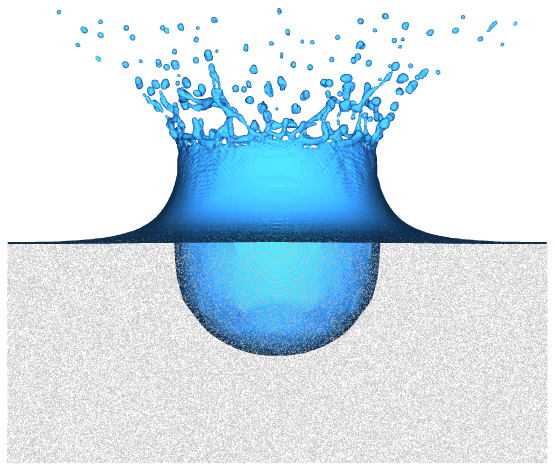}
\includegraphics[width=1.99cm]{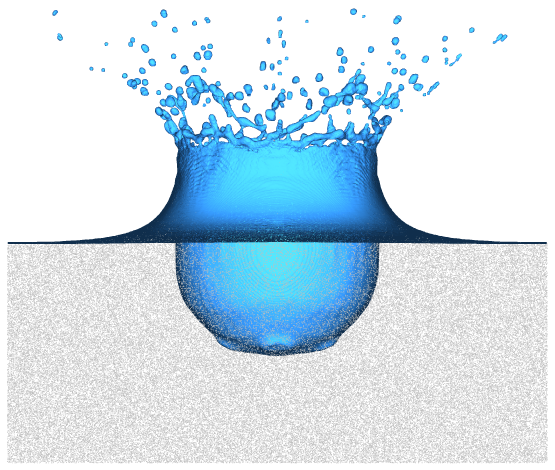}\\
\includegraphics[width=1.99cm]{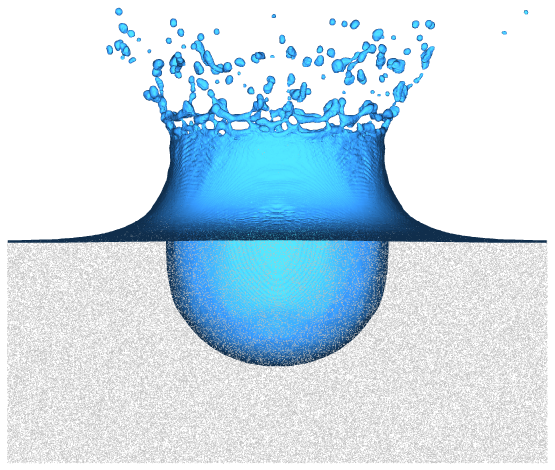}
\includegraphics[width=1.99cm]{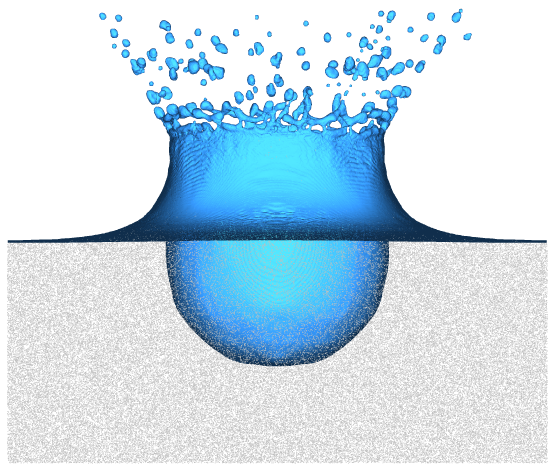}
\includegraphics[width=1.99cm]{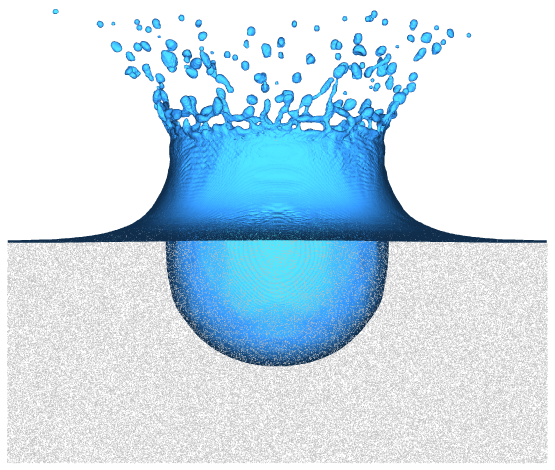}
\includegraphics[width=1.99cm]{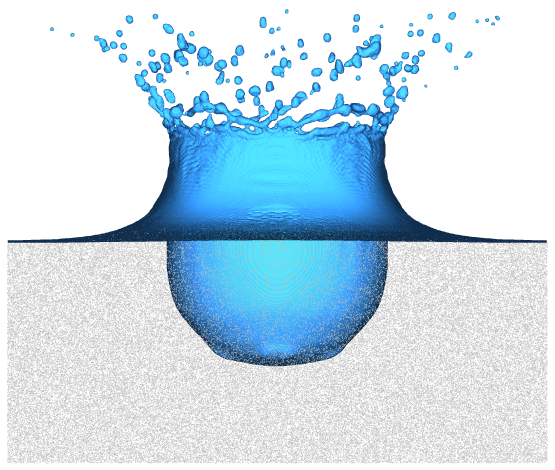}\\
\vspace*{-0.4cm}
\end{center}
\raggedright\scalebox{0.9}{\ \ \ FP32/FP32\ \ \ FP32/FP16S\ \ \ FP32/FP16C\ \ \ FP32/P16$_1$S\hfill}\\
\vspace*{0.1cm}
\caption{A $4\,\text{mm}$ diameter raindrop impacting a deep pool at $8.8\,\frac{\text{m}}{\text{s}}$ terminal velocity, illustrated at times $t\in\{0,1,2,3,4,5\}\,\text{ms}$ after impact as used in \cite{lehmann2021ejection}.
} \label{fig:raindrop}
\end{figure}
\begin{figure}[!t]
\begin{center}
  \begin{subfigure}{8cm}
    \includegraphics[width=8cm]{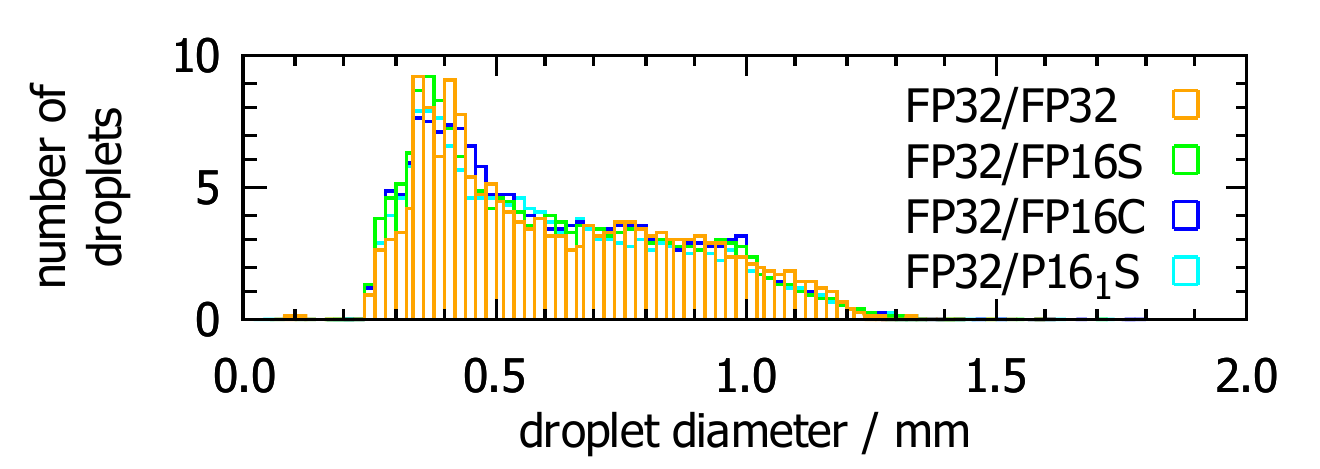}
    \vspace*{-0.6cm}\caption{}
  \end{subfigure}\\
  \begin{subfigure}{8cm}
    \includegraphics[width=8cm]{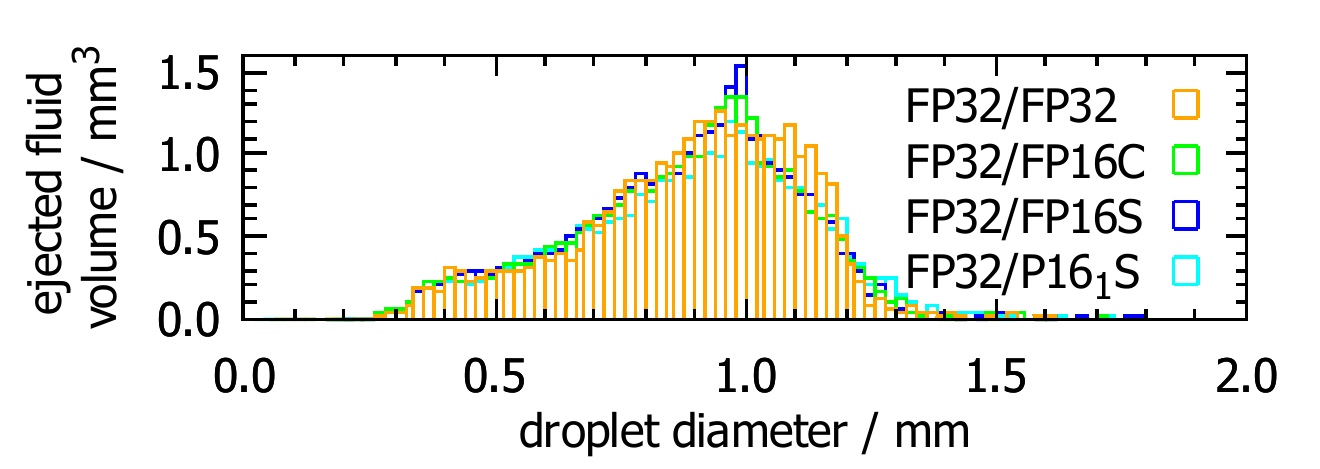}
    \vspace*{-0.6cm}\caption{}
  \end{subfigure}\\
  \begin{subfigure}{8cm}
    \includegraphics[width=8cm]{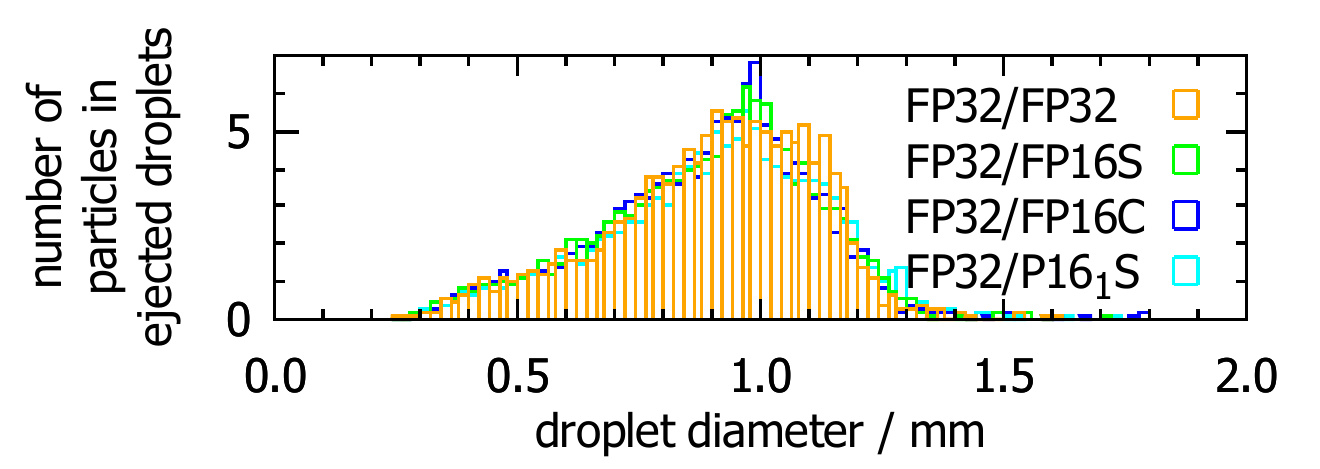}
    \vspace*{-0.6cm}\caption{}
  \end{subfigure}
\end{center}
\vspace*{-0.3cm}
\caption{(a) The size distribution of droplets, (b) the distribution of ejected fluid volume by droplet diameter and (c) the distribution of microplastic particles in droplets for 100 simulations each conducted with FP32/FP32, FP32/FP16S, FP32/FP16C and FP32/P16$_1$S.} \label{fig:drop-statistics:histograms}
\end{figure}
Finally, we examine how number formats affect a Volume-of-Fluid LBM simulation of a $4\,\text{mm}$ diameter raindrop impacting a deep pool at $8.8\,\frac{\text{m}}{\text{s}}$ terminal velocity.
This system is described and extensively validated in \cite{lehmann2021ejection} to study microplastic particle transport from the ocean into the atmosphere. The particles are simulated with the immersed-boundary method. There, simulations are performed in FP32/FP32 with the maximum lattice size that fits into memory, so FP64 is not used here as it does not fit into the memory of a single GPU. 
The dimensionless numbers for this setup are Reynolds number $\textit{Re}=\frac{d\,u}{\nu}=33498$, Weber number $\textit{We}=\frac{d\,u^2\,\rho}{\sigma}=4301$, Froude number $\textit{Fr}=\frac{u}{\sqrt{d\,g}}=44.4$, Capillary number $\textit{Ca}=\frac{u\,\rho\,\nu}{\sigma}=0.1284$ and Bond number $\textit{Bo}=\frac{d^2\,\rho\,g}{\sigma}=2.179$. 
The simulated domain is $464\times464\times394$ lattice points and runs on a single AMD Radeon VII GPU. The impact is simulated for $10\,\text{ms}$ time, equivalent to $20416$ time steps in LBM units. 

The raindrop impact is illustrated in figure \ref{fig:raindrop}. Note that the fully parallelized GPU implementation of the IBM with floating-point \texttt{atomic\_add\_f} makes the simulation non-deterministic \cite{lehmann2021ejection, lehmann2019high} and that the exact breakup of the crown into droplets is expected to be randomly different every time. We see minor artifacts at the bottom of the cavity for FP32/P16$_1$S, but otherwise no qualitative differences in random crown breakup. 

To be able to obtain statistics of ejected droplets and particles, we run the simulation $100$ times each with FP32/FP32, FP32/FP16S, FP32/FP16C and FP32/P16$_1$S. 
The microplastic particles each time are initialized at different random positions, resulting in slightly different random crown breakup. 
Ejected droplets that touch the top of the simulation box are measured and then deleted as detailed in \cite{lehmann2021ejection}.\\
In histograms of the size, volume and particle count depending on droplet diameter (figure \ref{fig:drop-statistics:histograms}), we see no significant differences across the data sets.

To conclude this section, we find that all FP32/FP16S, FP32/FP16C and FP32/P16$_1$S are able to recreate the results of FP32/FP32 in raindrop impact simulations without negative impact on the accuracy of the results, while significantly reducing the memory footprint of these simulations. This in turn enables simulations higher lattice resolution, potentially increasing accuracy by resolving smaller droplets.

\begin{figure*}[!htb]
\centering \includegraphics[width=17cm]{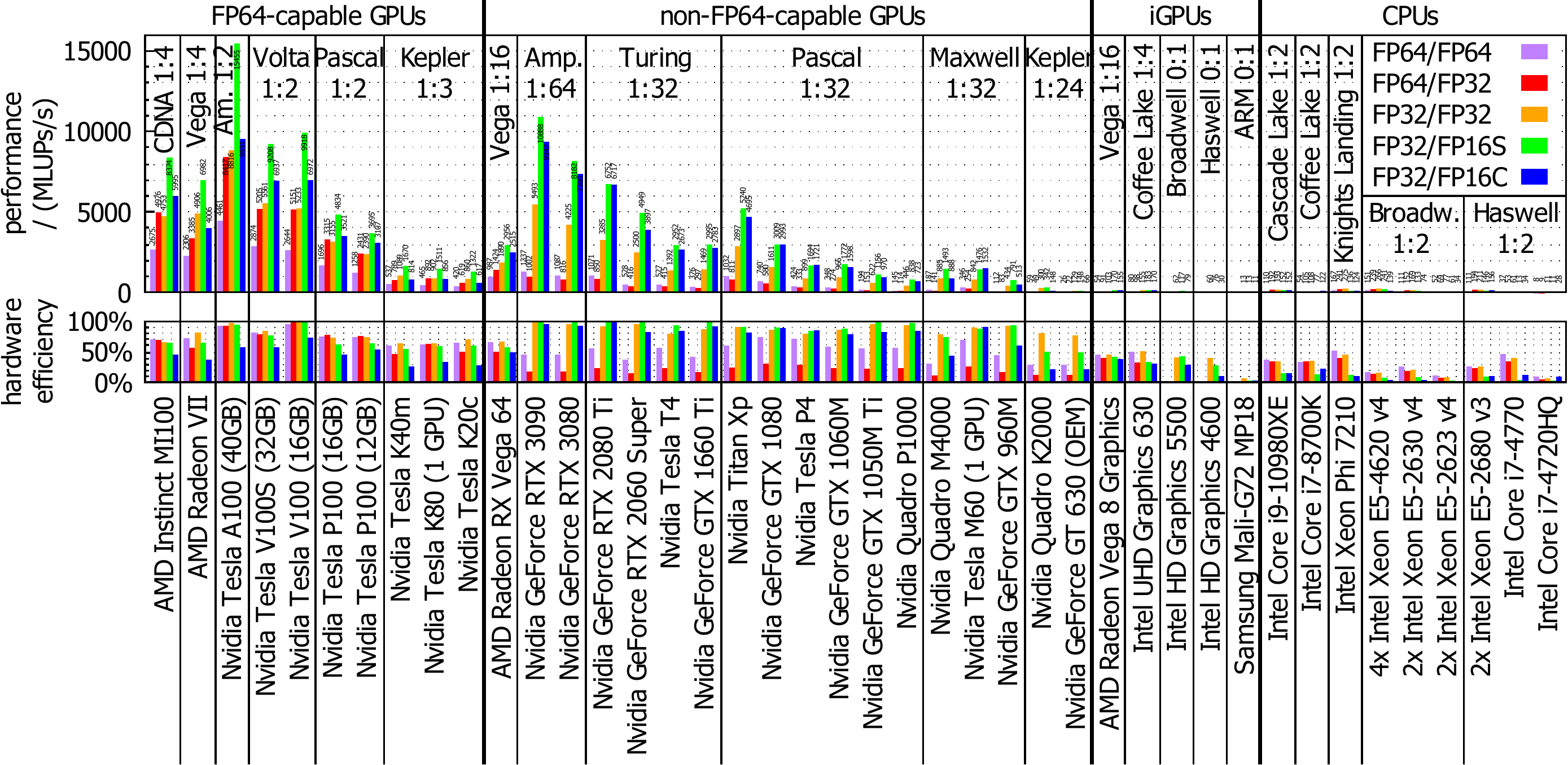}
\vspace*{0.1cm}
\caption{Performance of \textsl{FluidX3D} with D3Q19 SRT on different hardware (code as in listing \ref{list:lbm-core}). The unit MLUPs/s is an acronym for Mega lattice updates per second, meaning how many times $10^6$ LBM lattice points are computed every second. To obtain the hardware efficiency, we divide the measured MLUPs/s by the data sheet memory bandwidth times the number of bytes transferred per lattice point and time step (table \ref{tab:memory-transfers}). Performance characteristics differ depending on the FP64 arithmetic capabilities as indicated by the FP64:FP32 compute ratios of the microarchitectures. CPU benchmarks are on all cores. Values in table \ref{tab:hardware}.
} \label{fig:benchmarks}
\end{figure*}

\section{Memory and performance comparison}
For GPUs, the most efficient streaming step implementation \cite{wittmann2013comparison} is the one-step-pull scheme (A-B pattern) with two copies of the DDFs in memory, because the non-coalesced memory read penalty is lower than non-coalesced write penalty on GPUs \cite{lehmann2019high, mawson2014memory, delbosc2014optimized, tran2017performance, obrecht2013multi, obrecht2010global, obrecht2011new}, see figure \ref{fig:memory} in the appendix. One-step-pull further greatly facilitates implementing LBM extensions like Volume-of-Fluid, so it is a popular choice. 
Our \textsl{FluidX3D} base implementation (no-slip bounce-back boundaries, no extensions, as in listing \ref{list:lbm-core}) with DdQq velocity set has memory requirements per lattice point as shown in table \ref{tab:memory-requirements}. For D3Q19, going from FP32/FP32 to FP32-16x reduces the memory footprint by $\approx45\%$.
\begin{table}[!htb] \begin{center}
\begin{tabular}{l|r r r r|r}
\hline
 & $\vec{u}\ \ $ & $\rho$ & flags & $f_i\ \ \,$ & $\sum\ \ \ \ \ \ \ \ \ \ \ \ \ \ \,$\\
\hline
FP64/FP64 & $8\,d$ & $8$ & $1$ & $16\,q$ & $8\,d+9+16\,q$\\
FP64/FP32 & $8\,d$ & $8$ & $1$ &  $8\,q$ &  $8\,d+9+\ \,8\,q$\\
FP32/FP32 & $4\,d$ & $4$ & $1$ &  $8\,q$ &  $4\,d+5+\ \,8\,q$\\
FP32/16-bit & $4\,d$ & $4$ & $1$ &  $4\,q$ &  $4\,d+5+\ \,4\,q$\\
\hline
\end{tabular}
\end{center}
\vspace*{-0.2cm}
\caption{Memory requirements in byte per lattice point of LBM floating-point variants for the one-step-pull swap algorithm with two copies of the DDFs for the DdQq velocity set.} \label{tab:memory-requirements}
\end{table}
\begin{table}[!htb] \begin{center}
\begin{tabular}{l|r r|r}
\hline
 & flags & $f_i\ \ \,$ & $\sum\ \ $\\
\hline
FP64/FP64 & $q$ & $16\,q$ & $17\,q$\\
FP64/FP32 & $q$ &  $8\,q$ &  $9\,q$\\
FP32/FP32 & $q$ &  $8\,q$ &  $9\,q$\\
FP32/16-bit & $q$ &  $4\,q$ &  $5\,q$\\
\hline
\end{tabular}
\end{center}
\vspace*{-0.2cm}
\caption{Memory transfer in byte per lattice point per time step of LBM floating-point variants for the DdQq velocity set.} \label{tab:memory-transfers}
\end{table}\\
Although our main goal with FP16 is to reduce memory footprint and allow for larger simulation domains, as a side effect, performance is vastly increased as a result of less memory transfer in every LBM time step. 
For our base implementation with the DdQq velocity set, the amount of memory transfers per lattice point per time step is shown in table \ref{tab:memory-transfers}. Writing velocity and density to memory in each time step is not required for LBM without extensions. Theoretical speedup from FP32/FP32 to FP32/16-bit is $80\%$ for all velocity sets and swap algorithms.\\\\
While most LBM implementations are limited to one particular hardware platform -- either CPUs \cite{wellein2006towards, krause2021openlb, olb03, krause:08, olb14, latt2021palabos, min2013performance, mountrakis2015parallel, kotsalos2019bridging, kotsalos2020palabos, wellein2006single, lintermann2020lattice, SSFKSHC17}, Nvidia GPUs \cite{mawson2014memory, tolke2008teraflop, herschlag2018gpu, delbosc2014optimized, bailey2009accelerating, obrecht2013multi, de2019performance, obrecht2011new, tran2017performance, rinaldi2012lattice, rinaldifluid, beny2019efficient, ames2020multi, xiong2012efficient, zhu2020efficient, duchateau2015accelerating, janssen2015validation, habich2011performance, calore2015optimizing, hong2015scalable, xian2011multi, obrecht2010global, kuznik2010lbm, feichtinger2011flexible, calore2016massively, horgalattice, onodera2020locally}, CPUs and Nvidia GPUs \cite{riesinger2017holistic, aksnes2010porous, kummerlander2021implicit, geveler2010lattice, beny2019toward, boroni2017full, griebel2005meshfree, limbach2006espresso, institute2016espresso, walsh2009accelerating, ZSH21} or mobile SoCs \cite{harwood2017parallelisation, neumann2016lattice} -- only few use OpenCL \cite{tekic2012implementation, haeusl2019mpi, lehmann2020analytic, lehmann2021ejection, laermanns2021tracing, lehmann2019high, schreiber2011free, holzer2020highly, takavc2021cross, ho2017improving, habich2013performance}.
With \textsl{FluidX3D} also being implemented in OpenCL, we are able to benchmark our code across a large variety of hardware, from the worlds fastest data-center GPUs over gaming GPUs and CPUs to even the GPUs of mobile phone ARM SoCs. This enables us to determine LBM performance characteristics on various hardware microarchitectures. In figure \ref{fig:benchmarks} we show performance and efficiency on various hardware for D3Q19 SRT without extensions (only no-slip bounce-back boundaries are enabled in the code). The benchmark setup consists of a cubic box without any boundary nodes and with periodic boundary conditions in all directions. The standard domain size for the benchmark is $256^3$, except where device memory is not enough; there we use the largest cubic box that fits into memory.\\
We group the tested devices into four classes with different performance characteristics:
\begin{itemize}\itemsep0em
\item FP64-capable dedicated GPUs (high FP64:FP32 compute ratio) provide excellent efficiency for FP64/xx, FP32/FP32 and FP32/FP16S. They have such fast memory bandwidth that the FP32$\leftrightarrow$FP16C software conversion brings FP32/FP16C from the bandwidth limit into the compute limit, reducing its efficiency.
\item Non-FP64-capable dedicated GPUs (low FP64:FP32 compute ratio) have a particularly high FP32 arithmetic hardware limit, so even with the FP32$\leftrightarrow$FP16C software conversion the algorithm remains in the memory bandwidth limit. FP32/xx efficiency is excellent except for older Nvidia Kepler. However due to the poor FP64 arithmetic capabilities, FP64/xx efficiency is low as LBM here runs entirely in the compute limit rather than memory bandwidth limit. Surprisingly, FP64/FP32 runs even slower than FP64/FP64. This is because there is additional overhead for the FP64$\leftrightarrow$FP32 cast conversion in the compute limit, despite less memory bandwidth being used.
\item Integrated GPUs (iGPUs) overall show low performance and low efficiency. This is expected due to the slow system memory and cache hierarchy. Some older models do not support FP64 arithmetic at all.
\item CPUs also show low performance and low efficiency. The low efficiency on CPUs is less of a property of the implementation or a result of OpenCL, and more related to CPU microarchitectures in general \cite{wellein2006towards}. Other native CPU implementations of the LBM have equally low hardware efficiency \cite{olb14, krause:21, min2013performance, wellein2006towards} as a result of multi-level-caching, inter-CPU communication and other hardware properties unfavorable for LBM.
To illustrate this further, our implementation runs about as fast on the Mali-G72 MP18 mobile phone GPU as CPU codes on between $2$ and $16$ cores depending on the CPU model \cite{olb14, krause:21, krause:10b, min2013performance, wellein2006towards, aksnes2010porous, boroni2017full, walsh2009accelerating}.\\
It is of note that performance on CPUs with large cache greatly depends on the domain size: If a large fraction of the domain fits into L3 cache, efficiency (relative to memory bandwidth) is significantly better. Our CPU tests use a domain size of $256^3$, so only an insignificant $\approx1\%$ is covered by L3 cache -- a scenario representative of typical applications.
\end{itemize}
On the vast majority of hardware, we actually reach the theoretical $80\%$ speedup as indicated by the hardware efficiency remaining equal for FP32/FP32 and FP32/FP16S. Some hardware, namely the Nvidia Turing and Volta microarchitectures, do actually reach $100\%$ efficiency with FP32/FP32 and FP32/FP16S. The Nvidia RTX 2080 Ti is at $100\%$ efficiency even with FP32/FP16C, since the Nvidia Turing microarchitecture can do concurrent floating-point and integer computation and the 2080 Ti has high enough compute power per memory bandwidth to entirely remain in the memory bandwidth limit. Some efficiency values are even above $100\%$ as Nvidia Turing and Ampere A100 are capable of memory compression to increase effective bandwidth beyond the memory specifications \cite{nvidia2018turing, nvidia2020ampere}. Nvidia Pascal GeForce and Titan GPUs (that lack ECC memory) lock into P2 power state with reduced memory clock for compute applications in order to prevent memory errors \cite{gonzales2021nvidia}, lowering maximum bandwidth and making perfect (data sheet) efficiency impossible.\\
FP32/P16$_0$S and FP32/P16$_2$C performance is very similar to FP32/FP16C (data not shown), since the conversion needs to be emulated in software as well. FP32/P16$_1$S performance is a bit lower because the conversion algorithm is slightly more complex.
\begin{figure}[!htb]
\vspace*{-0.1cm}
\centering \includegraphics[width=8cm]{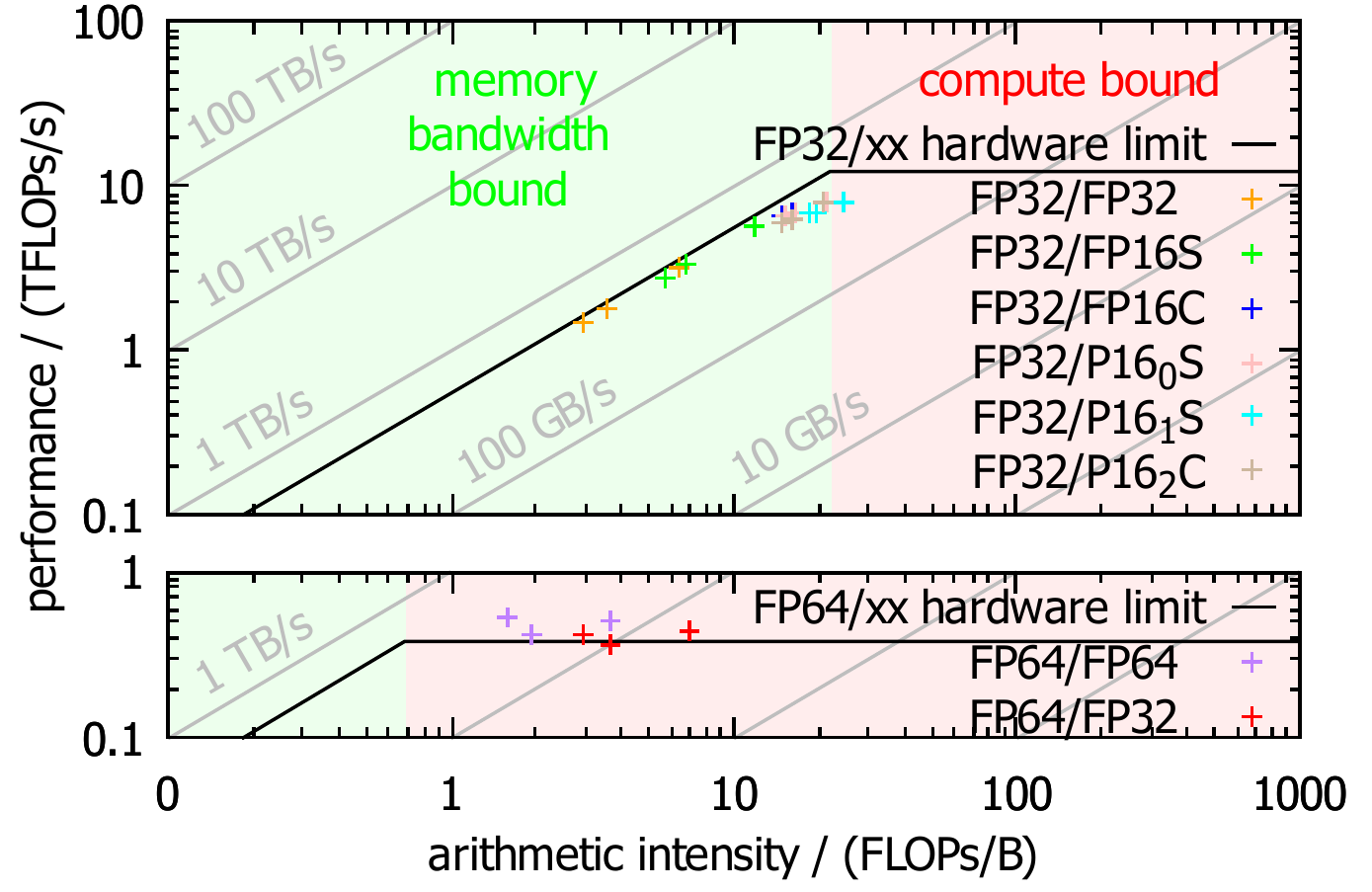}
\vspace*{0.1cm}
\caption{Roofline model analysis of \textsl{FluidX3D} with the D3Q19 velocity set, running on an Nvidia Titan Xp GPU. For each floating-point type, the three data points (left to right) correspond to the SRT, TRT and MRT collision operators. The arithmetic hardware limit is different for FP64/xx and FP32/xx, so we use two plots.} \label{fig:roofline}
\end{figure}\\
To better understand why performance is excellent with FP32/xx but not with FP64/xx on non-FP64-capable GPUs, we perform a roofline analysis \cite{wittmann2016hardware, williams2009roofline} for the Nvidia Titan Xp in figure \ref{fig:roofline}. The number of arithmetic operations and memory transfers is determined by automated counting of the corresponding PTX assembly instructions \cite{nvidia2021ptx} of the stream-collide kernel. We note that we count the arithmetic intensity as the sum of floating-point and integer operations, because the Pascal microarchitecture computes floating-point and integer on the same CUDA cores. For D3Q19 SRT FP32/FP32 for example, we count $255$ floating-point operations and $248$ integer operations. LBM performance scales proportionally to memory bandwidth, which is indicated by diagonal lines. The factor of proportionality is different for FPxx/64 ($323\,\text{byte}$ memory transfer per LBM time step), FPxx/32 ($171\,\text{byte}$) and FPxx/16 ($95\,\text{byte}$) as the amount of memory transfer is different (table \ref{tab:memory-transfers}). FP16 reduces the number of memory transfers, so the arithmetic intensity (number of arithmetic operations divided by memory transfers) is increased. The manual conversion from and to FP16C significantly increases the number of arithmetic operations, further raising arithmetic intensity. Nevertheless, even with the arithmetic-heavy matrix multiplication of the MRT collision operator, all data points are still within the memory bandwidth limit and thus almost equally efficient compared to FP32. Actual memory clocks during the benchmark are $3.5\%$ lower than the data sheet value (hardware limit) due to the Titan Xp locking into P2 power state \cite{gonzales2021nvidia}, inhibiting perfect efficiency for FP32/xx. In contrast, FP64/xx is in the compute limit, greatly reducing performance. The data points in the compute limit can be a bit above the hardware limit if core clocks are boosted beyond official data sheet values.

\section{Conclusions}
In this work, we studied the consequences of the employed floating-point number format on accuracy and performance of lattice Boltzmann simulations. We used six different test systems ranging from simple, pure fluid cases (Poiseuille flow, Taylor-Green vortices, Karman vortex streets, lid-driven cavity) to more complex situations such as immersed-boundary simulations for a microcapsule in shear flow or a Volume-of-Fluid simulation of an impacting raindrop. 
For all of these, we thoroughly compared how FP64, FP32, FP16 and Posit16 (mixed) precision affect accuracy and performance of the lattice Boltzmann method. 
In the mixed variants, a higher precision floating-point format is used for arithmetics and a lower precision format is used for storing the density distribution functions (DDFs). 
Based on the observation that a number range of $\pm2$ is sufficient for storing DDFs, we designed two novel 16-bit number formats specifically tailored to the needs of LBM simulations: 
a custom 16-bit floating-point format (FP16C) with halved truncation error compared to the standard IEEE-754 FP16 format by taking one bit from the exponent to increase the mantissa size, and a specifically designed asymmetric Posit variant (P16$_2$C). 
Conversion to these formats can be implemented highly efficiently and code interventions are only a few lines.

In all setups that we have tested and for the majority of parameters, FP32 turned out to be as accurate as FP64, provided that proper DDF-shifting \cite{skordos1993initial} is used.
Our custom FP16C format considerably diminishes errors and noise and turned out to be a viable option for FP32/16-bit mixed precision in many cases. 
16-bit Posits with their variable precision have shown to be very compelling options, too. 
Especially P16$_1$S in some cases could beat our FP16C. 
In other cases however, where the DDFs are outside the most favorable number range, the simulation error is increased significantly for the FP32/Posit16 simulations.

Regarding performance, we find that pure FP64 runs very poorly on the vast majority of GPUs, with the exception of very few data-center GPUs with extended FP64 arithmetic capabilities such as the MI100/A100/V100(S)/P100. 
FP64/FP32 mixed precision can be almost as fast as pure FP32 on these special data-center GPUs. 
However, somewhat counter-intuitively, on all GPUs with poor FP64 capabilities, FP64/FP32 is even slower than pure FP64 due to the conversion overhead. 
In general, pure FP32 then is a better choice since it enables excellent computational efficiency across all GPUs, especially considering that it is equally accurate to FP64 in all but edge cases. 
Computational efficiency is also excellent for FP32/FP16S mixed precision across all GPUs, reaching a maximum performance of $15455\,\text{MLUPs}$ (D3Q19) on a single $40\,\text{GB}$ Nvidia A100. 
On almost all GPUs that we have tested, we see the theoretical speedup of $80\%$ that FP32/16-bit mixed precision offers for D3Q19, alongside $45\%$ reduced memory footprint. 
Our custom format FP32/FP16C requires manual floating-point conversion which is heavy on integer computation. 
Nevertheless, FP32/FP16C runs efficiently on most GPUs with good FP32 arithmetic capabilities compared to their respective memory bandwidth and the theoretically expected $80\%$ speedup can be achieved.

In conclusion, we show that pure FP32 precision is sufficient for most application scenarios of the LBM and that with our specifically tailored FP16C number format in many cases even mixed FP32/FP16C precision can be used without significant loss of accuracy.

\section*{Acknowledgements}
We acknowledge support through the computational resources provided by BZHPC, LRZ and CINECA. 
We acknowledge the NVIDIA Corporation for donating a Titan Xp GPU for our research. 
We thank Maximilian Lehmann, Marcel Meinhart and Richard Kellnberger for running the benchmarks on their PCs.

\section*{Funding}
This study was funded by the Deutsche Forschungsgemeinschaft (DFG, German Research Foundation) - Project Number 391977956 - SFB 1357. 
We further acknowledge funding from Deutsche Forschungsgemeinschaft in the framework of FOR 2688 ''Instabilities, Bifurcations and Migration in Pulsating Flow'', projects B3 (417989940) and B2 (417989464). 
Open Access funding enabled and organized by Projekt DEAL.

\section*{Author's contributions}
ML and MK contributed the original concept. ML conducted the simulations and wrote the manuscript. ML, MK, GA and SG contributed essential ideas, manuscript review and literature research. ML, GA, MS and JH contributed test setups and benchmarks.

\section{References}
\printbibliography[heading=none]

\cleardoublepage
\onecolumn
\section{Appendix}

\subsection{The LB3D code}

While in the most part of this manuscript, we use the \textsl{FluidX3D} code \cite{lehmann2021ejection, laermanns2021tracing, lehmann2019high, haeusl2019mpi, lehmann2020analytic}, we also confirmed selected results with the LB3D lattice Boltzmann simulation package \cite{schmieschek17}. 
For this, we ported the FP64/FP64 routines to FP32/FP32 also in LB3D. 
LB3D is an MPI-based, general-purpose simulation package that includes various multicomponent and multiphase lattice Boltzmann methods, coupled to point particle molecular dynamics, discrete elements \cite{harting2014recent} and immersed boundary \cite{kruger2013numerical, KKH14} methods, as well as finite element solvers for advection-diffusion problems, including the Nernst-Planck equation \cite{rivas2018mesoscopic}. 
For the Poiseuille test, we used second-order accurate, mid-plane boundary conditions.

\subsection{LBM equations in a nutshell} \label{sec:appendix:lbm}
The coloring indicates the level of precision for the equations below:\\\begingroup\color{blue}lower precision storage\endgroup, \begingroup\color{magenta}conversion\endgroup, \begingroup\color{orange}higher precision arithmetic\endgroup

\subsubsection{Without DDF-shifting} \label{sec:appendix:lbm:without}
\begin{enumerate}
\item Streaming
\begin{equation} \label{eq:lbm-streaming}
\begingroup\color{orange}f_i^\text{temp}(\vec{x},t)\endgroup\begingroup\color{magenta}=\endgroup\begingroup\color{blue}f_i^\text{A}(\vec{x}-\vec{e}_i,\,t)\endgroup
\end{equation}
\item Collision (SRT)
\begin{align} \label{eq:lbm-fields}
\color{orange}\rho(\vec{x},t)&\color{orange}=\sum_i f_i^\text{temp}(\vec{x},t)\\
\color{orange}\vec{u}(\vec{x},t)&\color{orange}=\frac{1}{\rho(\vec{x},t)}\sum_i\vec{c}_i\,f_i^\text{temp}(\vec{x},t)
\end{align}
\begin{equation} \label{eq:lbm-equilibrium}
\begingroup\color{orange}f_i^\text{eq}(\vec{x},t)=f_i^\text{eq}(\rho(\vec{x},t),\,\vec{u}(\vec{x},t))=w_i\,\rho\,\cdot\left(\frac{(\vec{u}\scalar\vec{c}_i)^2}{2\,c^4}+\frac{\vec{u}\scalar\vec{c}_i}{c^2}+1-\frac{\vec{u}\scalar\vec{u}}{2\,c^2}\right)\endgroup
\end{equation}
\begin{equation} \label{eq:lbm-collision}
\begingroup\color{blue}f_i^\text{B}(\vec{x},\,t+\Delta t)\endgroup\begingroup\color{magenta}=\endgroup\begingroup\color{orange}\left(1-\frac{\Delta t}{\tau}\right)f_i^\text{temp}(\vec{x},t)+\frac{\Delta t}{\tau}\,f_i^\text{eq}(\vec{x},t)\endgroup
\end{equation}
\end{enumerate}

\subsubsection{With DDF-shifting} \label{sec:appendix:lbm:with}
\begin{enumerate}
\item Streaming
\begin{equation}
\begingroup\color{orange}f_i^\text{temp}(\vec{x},t)\endgroup\begingroup\color{magenta}=\endgroup\begingroup\color{blue}f_i^\text{A}(\vec{x}-\vec{e}_i,\,t)\endgroup
\end{equation}
\item Collision (SRT)
\begin{align}
\color{orange}\rho(\vec{x},t)&\color{orange}=\left(\sum_i f_i^\text{temp}(\vec{x},t)\right)+1\\
\color{orange}\vec{u}(\vec{x},t)&\color{orange}=\frac{1}{\rho(\vec{x},t)}\sum_i\vec{c}_i\,f_i^\text{temp}(\vec{x},t)
\end{align}
\begin{equation}
\begingroup\color{orange}f_i^\text{eq-shifted}(\vec{x},t)=f_i^\text{eq-shifted}(\rho(\vec{x},t),\,\vec{u}(\vec{x},t))=w_i\,\rho\,\cdot\left(\frac{(\vec{u}\scalar\vec{c}_i)^2}{2\,c^4}-\frac{\vec{u}\scalar\vec{u}}{2\,c^2}+\frac{\vec{u}\scalar\vec{c}_i}{c^2}\right)+w_i\,(\rho-1)\endgroup
\end{equation}
\begin{equation}
\begingroup\color{blue}f_i^\text{B}(\vec{x},\,t+\Delta t)\endgroup\begingroup\color{magenta}=\endgroup\begingroup\color{orange}\left(1-\frac{\Delta t}{\tau}\right)f_i^\text{temp}(\vec{x},t)+\frac{\Delta t}{\tau}\,f_i^\text{eq}(\vec{x},t)\endgroup
\end{equation}
\end{enumerate}

\newpage
\subsection{List of physical quantities and nomenclature} \label{sec-list-of-physical-quantities}
\vspace*{-0.5cm} 
\begin{table}[!htb] \begin{center} \begin{tabular}{|c|c|c|l|}
\hline
quantity & SI-units & defining equation(s) & description \\
\hline
$\vec{x}$ & $m$ & $\vec{x}=(x,y,z)$ & 3D position in Cartesian coordinates \\
\hline
$t$ & $s$ & $-$ & time \\
\hline
$\Delta x$ & $m$ & $\Delta x:=1$ & lattice constant (in lattice units) \\
\hline
$\Delta t$ & $s$ & $\Delta t:=1$ & simulation time step (in lattice units) \\
\hline
$c$ & $\frac{m}{s}$ & $c:=\frac{1}{\sqrt{3}}\frac{\Delta x}{\Delta t}$ & lattice speed of sound (in lattice units) \\
\hline
$\rho$ & $\frac{kg}{m^3}$ & $\rho=\sum_i f_i$ & mass density \\
\hline
$\vec{u}$ & $\frac{m}{s}$ & $\vec{u}=\sum_i \vec{c}_i f_i$ & velocity \\
\hline
$f_i$ & $\frac{kg}{m^3}$ & (\ref{eq:lbm-streaming}) & density distribution functions (DDFs) \\
\hline
$f_i^\text{eq}$ & $\frac{kg}{m^3}$ & (\ref{eq:lbm-equilibrium}) & equilibrium DDFs \\
\hline
$i$ & $1$ & $0\leq i<q$ & LBM streaming direction index \\
\hline
$q$ & $1$ & $q\in\{7,9,13,15,19,27\}$ & number of LBM streaming directions \\
\hline
$\vec{c}_i$ & $\frac{m}{s}$ & \cite[eq.(11)]{lehmann2019high} & streaming velocities \\
\hline
$\vec{e}_i$ & $m$ & $\vec{e}_i=\vec{c}_i\,\Delta t$ & streaming directions \\
\hline
$w_i$ & $1$ & \cite[eq.(10)]{lehmann2019high}, $\sum_i w_i=1$ & velocity set weights \\
\hline
$\tau$ & $s$ & $\tau=\frac{\nu}{c^2}+\frac{\Delta t}{2}$ & LBM relaxation time \\
\hline
$\nu$ & $\frac{m^2}{s}$ & $\nu=\frac{\mu}{\rho}$ & kinematic shear viscosity \\
\hline
$\vec{f}$ & $\frac{kg}{m^2\,s^2}$ & $\vec{f}=\frac{\vec{F}}{V}$ & force per volume \\
\hline
$(L_x,L_y,L_z)$ & $(m,m,m)$ & $L_x\,L_y\,L_z=V$ & simulation box dimensions \\
\hline
$g$ & $\frac{m}{s^2}$ & $g:=9.81\frac{m}{s^2}$ & gravitational acceleration \\
\hline
$\sigma$ & $\frac{kg}{s^2}$ & $-$ & surface tension coefficient \\
\hline
\end{tabular} \end{center} \end{table}

\subsection{Measured number format characteristics}
\begin{figure}[!htb]
\centering \includegraphics[width=8cm]{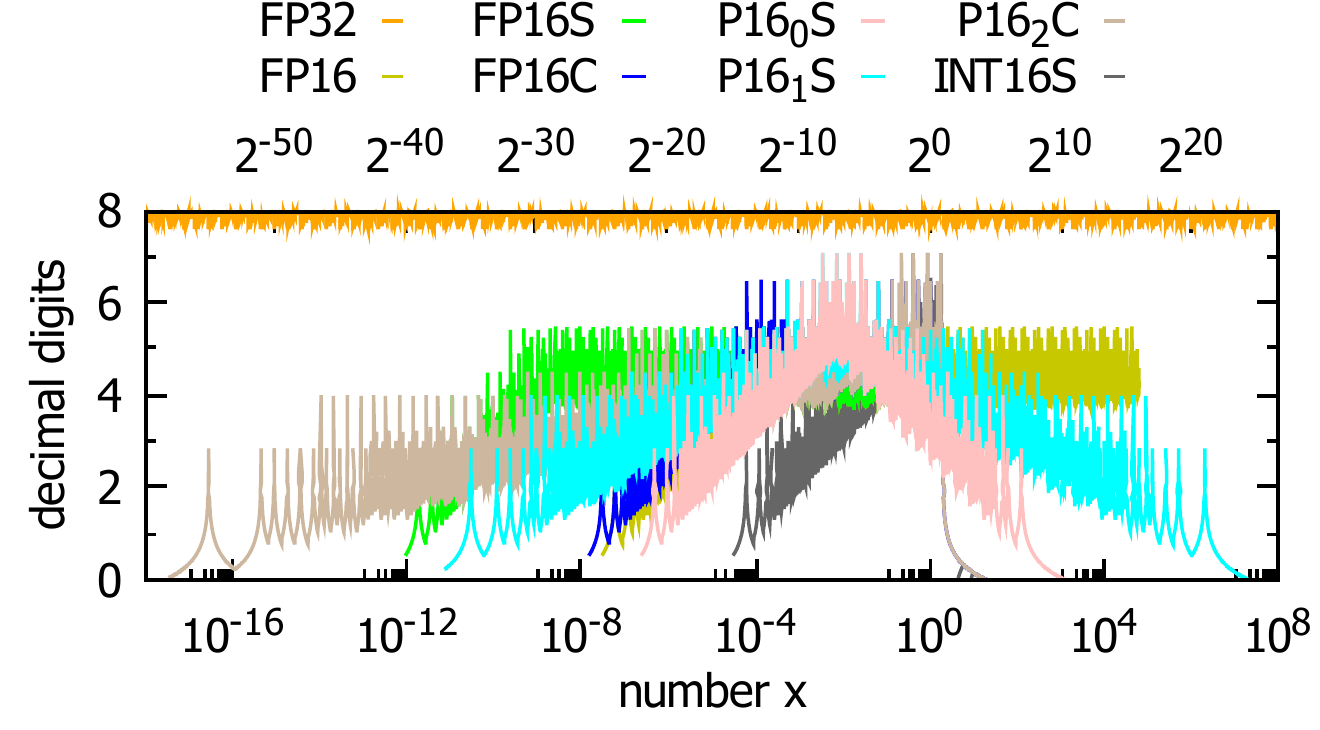}\vspace*{0.1cm}
\caption{Measured accuracy characteristics of the number formats investigated in this work. 
The number of decimal digits for a given number $x$ is computed via $-\log_{10}(|\log_{10}(\frac{x_\text{represented}}{x})|)$ \cite{gustafson2017beating, gustafson2017posit, klower2020number}. 
Only the local minima are the relevant criterion for the error. Note that this definition for the number of decimal digits is off from the $\log_{10}(2^{n_m+1})$ definition by about $0.4$.
} \label{fig:formats-measured}
\end{figure}

\newpage
\subsection{Numerical values of $f_i$ and $f_i^\text{shifted}$ for all setups (FP32/FP32)}
\begin{figure}[!htb]
\begin{center}
  \begin{subfigure}{16cm}
  	\includegraphics[width=8cm]{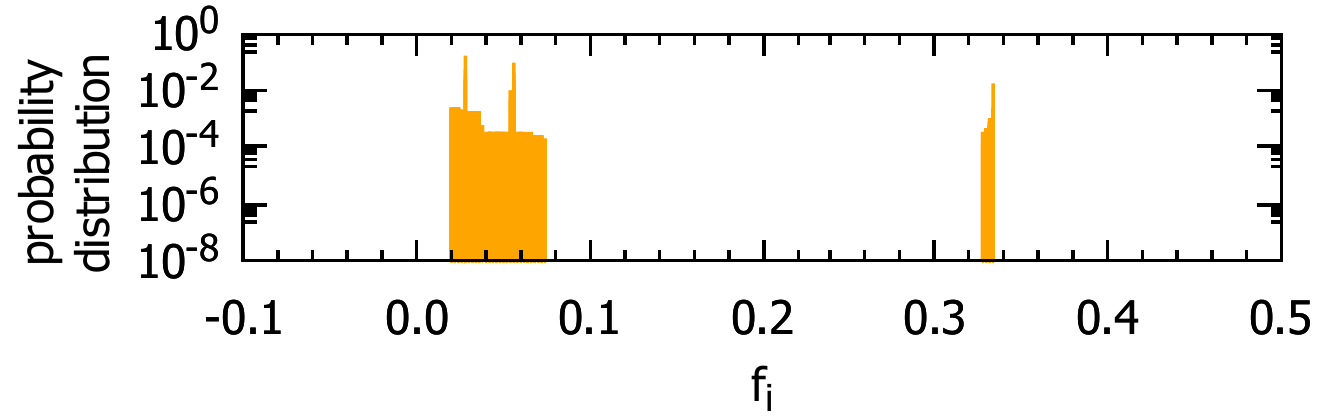}\hfill
  	\includegraphics[width=8cm]{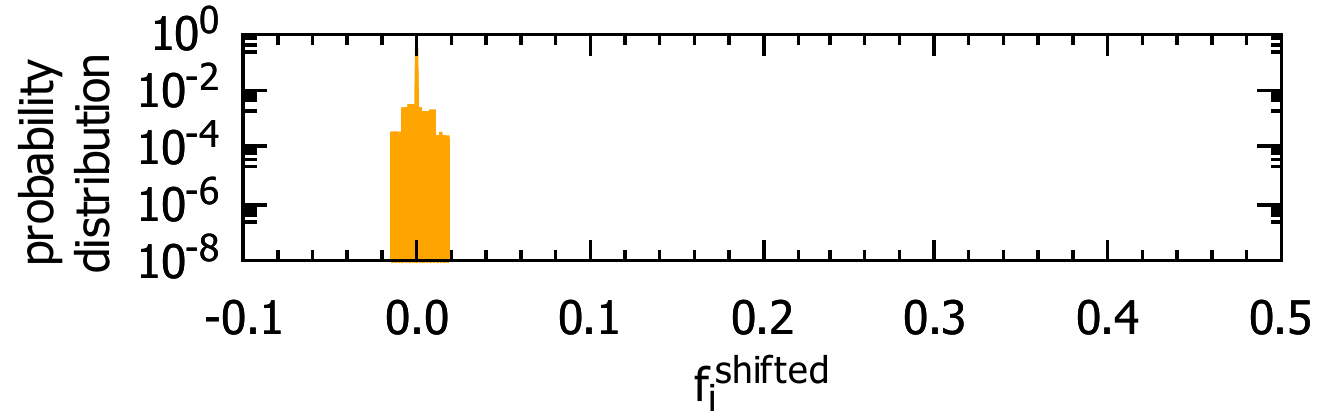}
    \vspace*{-0.6cm}\caption{Poiseuille flow, $R=63$, $u_\text{max}=0.1$, $\textit{Re}=10$, at the time of convergence, like in figure \ref{fig:radial-error}.}\vspace*{0.2cm}
  \end{subfigure}\\
  \begin{subfigure}{16cm}
  	\includegraphics[width=8cm]{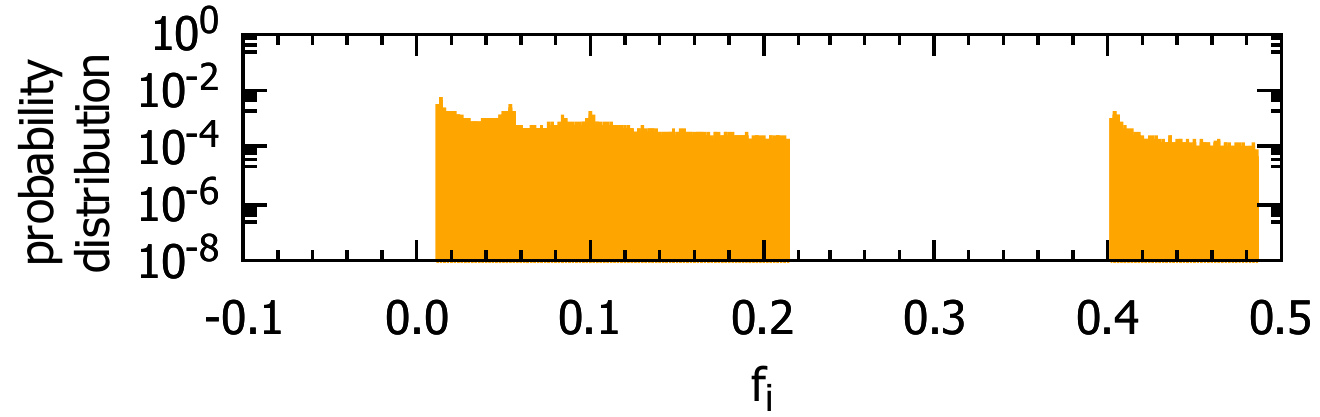}\hfill
  	\includegraphics[width=8cm]{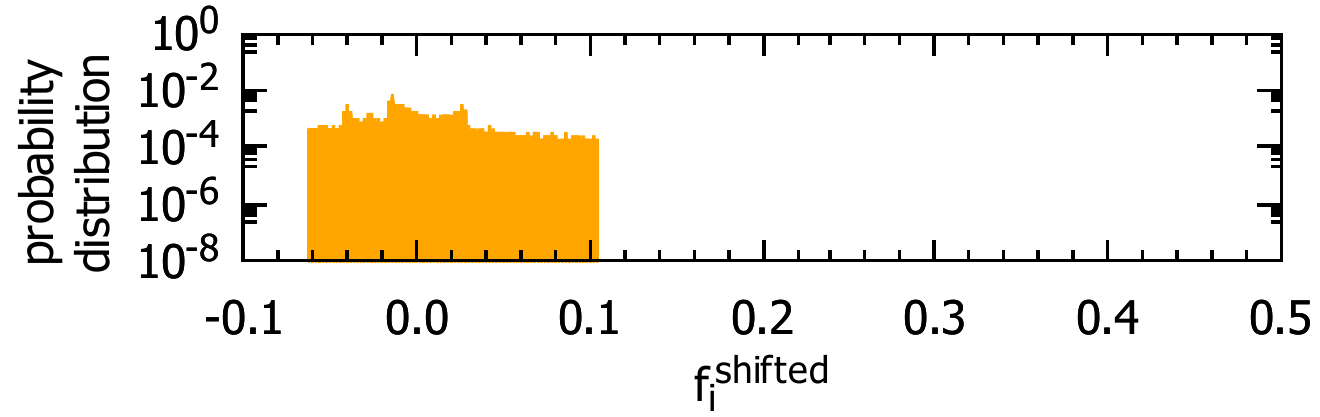}
    \vspace*{-0.6cm}\caption{Taylor-green vortices, at time of initialization $t=0$.}\vspace*{0.2cm}
  \end{subfigure}\\
  \begin{subfigure}{16cm}
  	\includegraphics[width=8cm]{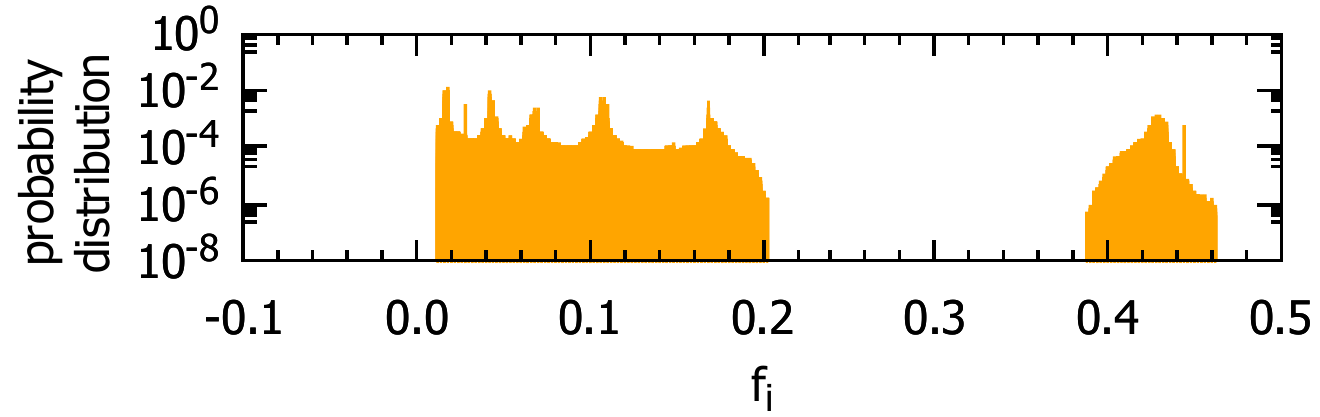}\hfill
  	\includegraphics[width=8cm]{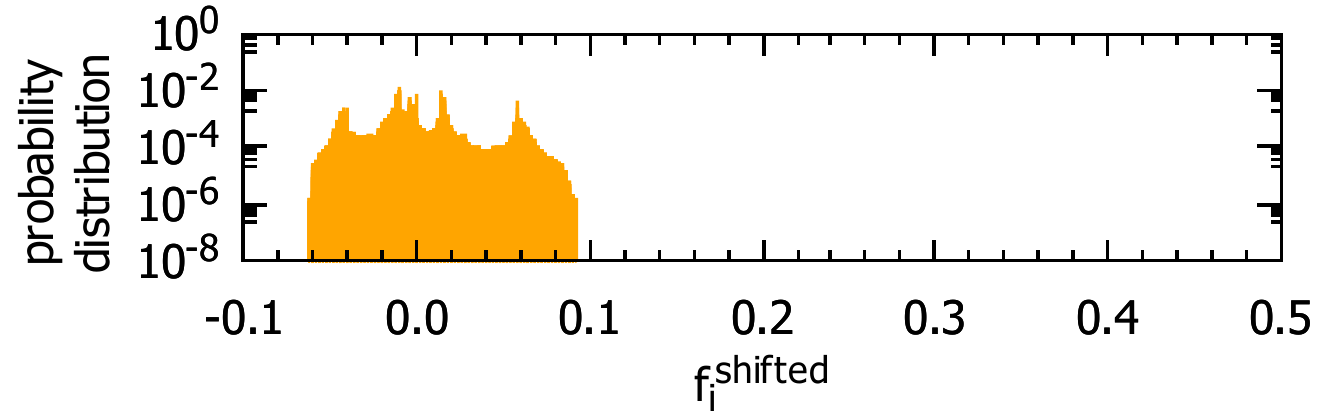}
    \vspace*{-0.6cm}\caption{Karman vortex street, after $t=100000$ time steps.}\vspace*{0.2cm}
  \end{subfigure}\\
  \begin{subfigure}{16cm}
  	\includegraphics[width=8cm]{images/lid-driven-cavity/fi.pdf}\hfill
  	\includegraphics[width=8cm]{images/lid-driven-cavity/fi-shifted.pdf}
    \vspace*{-0.6cm}\caption{Lid-driven cavity, after $t=100000$ time steps, like in figures \ref{fig:fi} and \ref{fig:ldc-velocity}.}\vspace*{0.2cm}
  \end{subfigure}\\
  \begin{subfigure}{16cm}
  	\includegraphics[width=8cm]{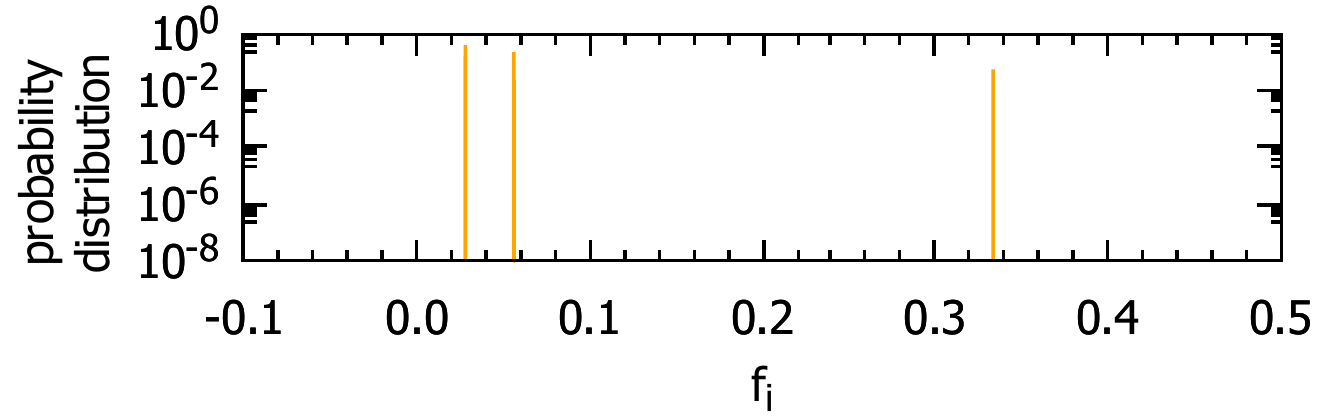}\hfill
  	\includegraphics[width=8cm]{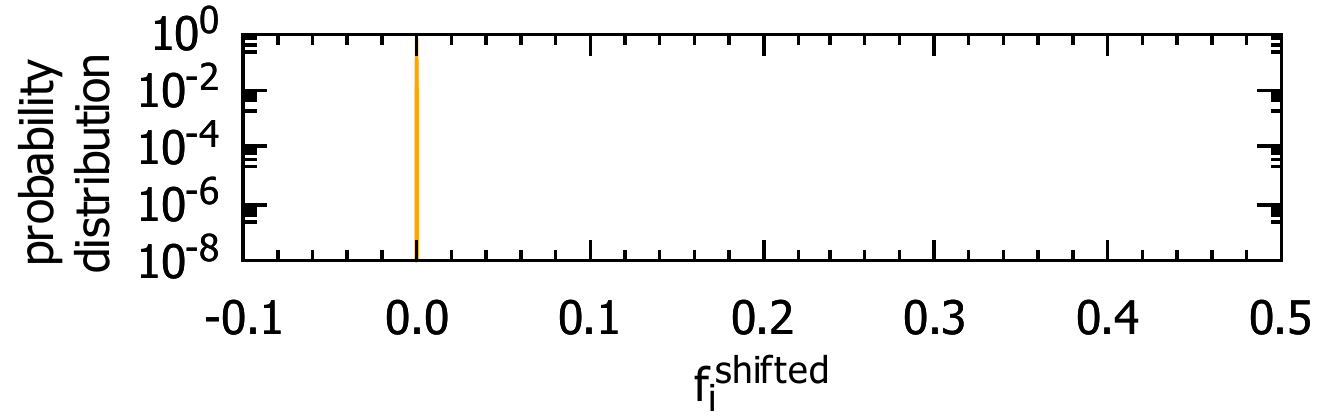}
    \vspace*{-0.6cm}\caption{Microcapsule in shear flow at $\textit{Ca}=0.1$, after dimensionless time $\dot{\gamma}\,t=5$.}\vspace*{0.2cm}
  \end{subfigure}\\
  \begin{subfigure}{16cm}
  	\hfill
  	\includegraphics[width=8cm]{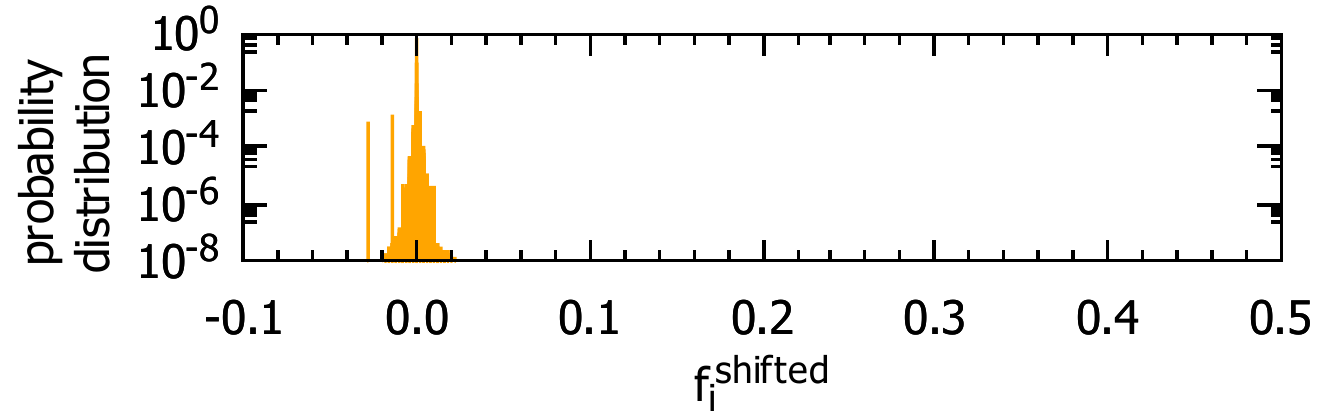}
    \vspace*{-0.1cm}\caption{Raindrop impact at resolution $464\times464\times394$, with IBM particles, after $t=10\,\text{ms}$ time.}\vspace*{0.2cm}
  \end{subfigure}
\end{center}
\caption{Numerical values of $f_i$ and $f_i^\text{shifted}$ for all setups (FP32/FP32).} \label{fig:setups:fi}
\end{figure}

\newpage
\subsection{Properties of benchmarked hardware}
\begin{table}[!htb] \begin{center}
\vspace*{-0.6cm}
\begin{tabular}{l|R{0.6cm} R{0.7cm} R{0.2cm} R{0.6cm}|R{0.5cm} R{0.5cm} R{0.5cm} R{0.5cm}|R{0.5cm} R{0.5cm} R{0.5cm} R{0.6cm} R{0.5cm}}
\hline
device & \multicolumn{3}{p{1.8cm}}{data sheet} & & \multicolumn{3}{p{2.4cm}}{measured memory bandwidth / GB/s} & & \multicolumn{5}{p{3.0cm}}{LBM performance (D3Q19 SRT) / MLUPs/s}\\
       &  \rotatebox[origin=c]{90}{ FP64 / TFLOPs/s} & \rotatebox[origin=c]{90}{ FP32 / TFLOPs/s} & \rotatebox[origin=c]{90}{memory / GB} & \rotatebox[origin=c]{90}{bandwidth / GB/s} & \rotatebox[origin=c]{90}{coalesced read} & \rotatebox[origin=c]{90}{misaligned read} & \rotatebox[origin=c]{90}{coalesced write} & \rotatebox[origin=c]{90}{misaligned write} & \rotatebox[origin=c]{90}{FP64/FP64} & \rotatebox[origin=c]{90}{FP64/FP32} & \rotatebox[origin=c]{90}{FP32/FP32} & \rotatebox[origin=c]{90}{FP32/FP16S} & \rotatebox[origin=c]{90}{FP32/FP16C}\\
\hline
AMD Instinct MI100			& 11.54	& 46.14	& 32	& 1229	&  812	&  936	&  776	& 567	& 2675	& 4976	& 4753	& 8374	& 5995\\
AMD Radeon VII				&  3.46	& 13.83	& 16	& 1024	&  647	&  825	&  698	& 397	& 2306	& 3385	& 4906	& 6982	& 4006\\
Nvidia Tesla A100 (40GB)	&  9.75	& 19.49	& 40	& 1555	& 1334	& 1235	& 1514	& 207	& 4461	& 8417	& 8816	& 15455	& 9531\\
Nvidia Tesla V100S (32GB)	&  8.18	& 16.35	& 32	& 1134	&  911	&  930	&  951	& 153	& 2874	& 5205	& 5561	& 9208	& 6937\\
Nvidia Tesla V100 (16GB)	&  7.07	& 14.13	& 16	&  900	&  811	&  806	&  888	& 133	& 2644	& 5151	& 5233	& 9918	& 6972\\
Nvidia Tesla P100 (16GB)	&  4.76	&  9.52	& 16	&  732	&  504	&  306	&  544	&  90	& 1696	& 3315	& 3155	& 4834	& 3521\\
Nvidia Tesla P100 (12GB)	&  4.76	&  9.52	& 12	&  549	&  426	&  308	&  403	&  47	& 1258	& 2431	& 2390	& 3695	& 3107\\
Nvidia Tesla K40m			&  1.43	&  4.29	& 12	&  288	&  158	&   42	&  182	&  42	& 537	& 789	& 1089	& 1670	& 814\\
Nvidia Tesla K80 (1 GPU)	&  1.37	&  4.11	& 12	&  240	&  135	&   42	&  153	&  43	& 465	& 892	& 902	& 1511	& 856\\
Nvidia Tesla K20c			&  1.17	&  3.52	&  5	&  208	&  131	&   34	&  147	&  35	& 420	& 619	& 860	& 1322	& 617\\\hline
AMD Radeon RX Vega 64		&  0.79	& 12.66	&  8	&  484	&  251	&  368	&  252	& 391	& 987	& 1424	& 1890	& 2956	& 2515\\
Nvidia GeForce RTX 3090		&  0.61	& 39.05	& 24	&  936	&  886	&  886	&  909	& 204	& 1337	& 1002	& 5493	& 10888	& 9347\\
Nvidia GeForce RTX 3080		&  0.47	& 29.77	& 10	&  760	&  692	&  693	&  709	& 171	& 1087	& 816	& 4225	& 8182	& 7380\\
Nvidia GeForce RTX 2080 Ti	&  0.42	& 13.45	& 11	&  616	&  533	&  531	&  579	& 159	& 1071	& 850	& 3285	& 6752	& 6717\\
Nvidia GeForce RTX 2060 S.	&  0.22	&  7.18	&  8	&  448	&  392	&  264	&  423	&  82	& 528	& 416	& 2500	& 4949	& 3897\\
Nvidia Tesla T4				&  0.25	&  8.14	& 16	&  300	&  259	&  264	&  226	&  85	& 527	& 415	& 1392	& 2952	& 2673\\
Nvidia GeForce GTX 1660 Ti	&  0.17	&  5.48	&  6	&  288	&  246	&  187	&  263	&  53	& 376	& 297	& 1469	& 2995	& 2783\\
Nvidia Titan Xp				&  0.38	& 12.15	& 12	&  548	&  445	&  163	&  514	& 112	& 1032	& 811	& 2897	& 5240	& 4695\\
Nvidia GeForce GTX 1080		&  0.31	&  9.78	&  8	&  320	&  257	&  201	&  275	&  80	& 740	& 580	& 1611	& 3009	& 2993\\
Nvidia Tesla P4				&  0.18	&  5.70	&  8	&  192	&  139	&  121	&  144	&  44	& 424	& 333	& 899	& 1694	& 1721\\
Nvidia GeForce GTX 1060M	&  0.14	&  4.44	&  6	&  192	&  156	&  104	&  165	&  44	& 348	& 274	& 966	& 1772	& 1598\\
Nvidia GeForce GTX 1050M Ti	&  0.08	&  2.49	&  4	&  112	&   98	&   65	&  102	&  27	& 194	& 153	& 622	& 1156	& 970\\
Nvidia Quadro P1000			&  0.06	&  1.89	&  4	&   82	&   70	&   47	&   73	&  21	& 145	& 114	& 446	& 838	& 723\\
Nvidia Quadro M4000			&  0.08	&  2.57	&  8	&  192	&  154	&   35	&  159	&  23	& 187	& 141	& 885	& 1493	& 888\\
Nvidia Tesla M60 (1 GPU)	&  0.15	&  4.82	&  8	&  160	&  143	&   56	&  145	&  39	& 346	& 254	& 842	& 1476	& 1532\\
Nvidia GeForce GTX 960M		&  0.05	&  1.51	&  4	&   80	&   73	&   31	&   70	&  14	& 112	& 83	& 434	& 791	& 513\\
Nvidia Quadro K2000			&  0.03	&  0.73	&  2	&   64	&   35	&    7	&   57	&   8	& 59	& 49	& 300	& 342	& 148\\
Nvidia GeForce GT 630 (OEM)	&  0.02	&  0.46	&  2	&   29	&   13	&    3	&   26	&   4	& 26	& 22	& 129	& 148	& 66\\\hline
AMD Radeon Vega 8 Graphics	&  0.08	&  1.23	&  7	&   38	&   30	&   33	&   26	&  33	& 54	& 91	& 103	& 170	& 155\\
Intel UHD Graphics 630		&  0.12	&  0.46	&  7	&   51	&   38	&   34	&   28	&  18	& 80	& 98	& 155	& 183	& 170\\
Intel HD Graphics 5500		&     -	&  0.35	&  3	&   26	&   16	&   19	&    8	&   9	&  -	&  -	& 62	& 117	& 79\\
Intel HD Graphics 4600		&     -	&  0.38	&  2	&   26	&   19	&   10	&   11	&   7	&  -	&  -	& 60	& 76	& 30\\
Samsung Mali-G72 MP18		&     -	&  0.24	&  4	&   29	&   22	&   21	&   21	&  21	&  -	&  -	& 13	& 13	& 11\\\hline
Intel Core i9-10980XE		&  1.81	&  3.23	& 128	&   94	&   60	&   65	&   26	&  30	& 110	& 192	& 193	& 152	& 152\\
Intel Core i7-8700K			&  0.36	&  0.71	& 16	&   51	&   38	&   51	&   26	&  18	& 54	& 105	& 108	& 77	& 122\\
Intel Xeon Phi 7210			&  2.66	&  5.32	& 192	&  102	&   35	&   65	&   62	&  61	& 167	& 241	& 275	& 136	& 124\\
4x Intel Xeon E5-4620 v4	&  1.34	&  2.69	& 512	&  273	&  104	&  126	&   52	&  54	& 151	& 239	& 266	& 241	& 139\\
2x Intel Xeon E5-2630 v4	&  0.70	&  1.41	& 64	&  137	&   65	&   88	&   66	&  86	& 111	& 152	& 169	& 133	& 74\\
2x Intel Xeon E5-2623 v4	&  0.33	&  0.67	& 64	&  137	&   25	&   43	&   24	&  26	& 52	& 69	& 77	& 61	& 34\\
2x Intel Xeon E5-2680 v3	&  0.96	&  1.92	& 64	&  137	&  110	&  102	&   43	&  50	& 111	& 194	& 211	& 146	& 156\\
Intel Core i7-4770			&  0.22	& 0.44	& 16	&   26	&   24	&   26	&   15	&  12	& 37	& 53	& 61	& 13	& 34\\
Intel Core i7-4720HQ		&  0.17	& 0.33	& 16	&   26	&   22	&   22	&    9	&  11	&  8	&  9	& 11	& 11	& 28\\
\hline
\end{tabular}
\end{center}
\vspace*{-0.2cm}
\caption{Properties of benchmarked hardware.} \label{tab:hardware}
\vspace*{-2.0cm}
\end{table}

\newpage
\subsection{Memory benchmarks}
\begin{figure*}[!htb]
\centering \includegraphics[width=17cm]{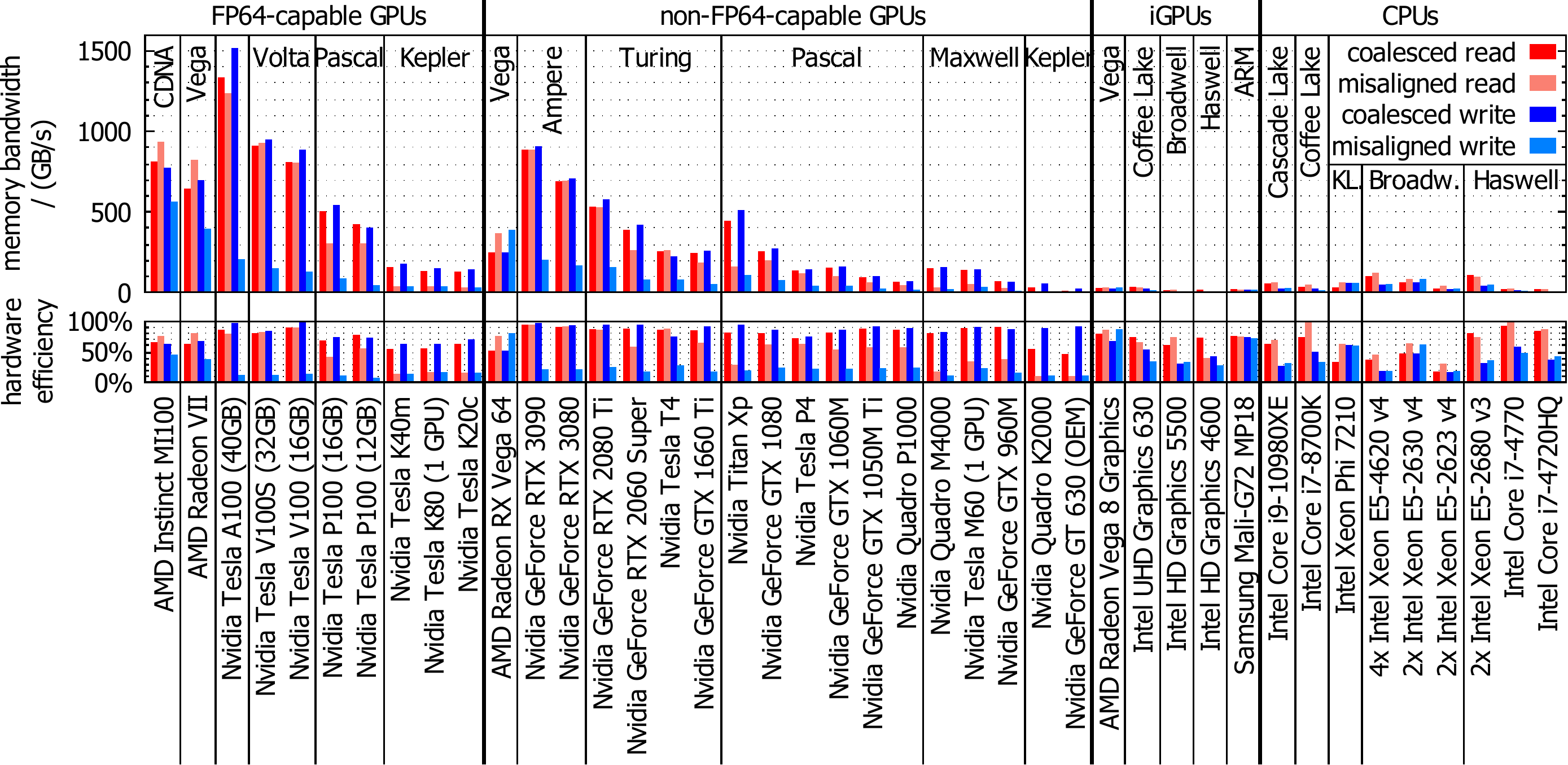}
\vspace*{0.1cm}
\caption{Synthetic OpenCL memory benchmarks to measure coalesced/misaligned read/write performance. The misaligned write penalty is much larger than the misaligned read penalty across almost all tested devices. Values in table \ref{tab:hardware}.} \label{fig:memory}
\end{figure*}

\newpage
\subsection{Ultrafast converseion algorithms}
\begin{lstlisting}[caption={OpenCL C macros for regular FP32, for FP16S using hardware-accelerated IEEE-754 FP16 floating-point conversion and for our FP16C format with calls to our manual floating-point conversion functions. Manual floating-point conversion functions for FP32$\leftrightarrow$FP16C (\texttt{float}$\leftrightarrow$\texttt{half}) in OpenCL C. We also provide macros and conversion algorithms for FP32$\leftrightarrow$P16$_0$S/P16$_1$S/P16$_2$C Posit formats. The saturation term in the algorithms can be omitted if it is made sure that larger than maximum numbers are never used, which is the case in this LBM application.},captionpos=b,label={list:conversion}, numbers=left]
// FP32/FP32 macros
#define fpXX float // switchable data type
#define load(p,o) p[o] // regular float read
#define store(p,o,x) p[o]=x // regular float write

// FP32/FP16S macros
#define fpXX half // switchable data type
#define load(p,o) vload_half(o,p)*3.0517578E-5f // use OpenCL function
#define store(p,o,x) vstore_half_rte((x)*32768.0f,o,p) // use OpenCL function

// FP32/FP16C macros
#define fpXX ushort // switchable data type
#define load(p,o) half_to_float_custom(p[o]) // call conversion function
#define store(p,o,x) p[o]=float_to_half_custom(x) // call conversion function

// FP32/P160S macros
#define fpXX ushort // switchable data type
#define load(p,o) p160_to_float(p[o])*0.0078125f // call conversion function
#define store(p,o,x) p[o]=float_to_p160((x)*128.0f) // call conversion function

// FP32/P161S macros
#define fpXX ushort // switchable data type
#define load(p,o) p161_to_float(p[o])*0.0078125f // call conversion function
#define store(p,o,x) p[o]=float_to_p161((x)*128.0f) // call conversion function

// FP32/P162C macros
#define fpXX ushort // switchable data type
#define load(p,o) p162C_to_float(p[o]) // call conversion function
#define store(p,o,x) p[o]=float_to_p162C(x) // call conversion function



ushort float_to_half_custom(const float x) {
	const uint b = as_uint(x)+0x00000800; // round-to-nearest-even: add last bit after truncated mantissa
	const uint e = (b&0x7F800000)>>23; // exponent
	const uint m = b&0x007FFFFF; // mantissa; in line below: 0x007FF800 = 0x00800000-0x00000800 = decimal indicator flag - initial rounding
	return (b&0x80000000)>>16 | (e>112)*((((e-112)<<11)&0x7800)|m>>12) | ((e<113)&(e>100))*((((0x007FF800+m)>>(124-e))+1)>>1) | (e>127)*0x7FFF; // sign | normalized | denormalized | saturate
}
float half_to_float_custom(const ushort x) {
	const uint e = (x&0x7800)>>11; // exponent
	const uint m = (x&0x07FF)<<12; // mantissa
	const uint v = as_uint((float)m)>>23; // evil log2 bit hack to count leading zeros in denormalized format
	return as_float((x&0x8000)<<16 | (e!=0)*((e+112)<<23|m) | ((e==0)&(m!=0))*((v-37)<<23|((m<<(150-v))&0x007FF000))); // sign | normalized | denormalized
}



ushort float_to_p160(const float x) {
	const uint b = as_uint(x);
	const int e = ((b&0x7F800000)>>23)-127; // exponent-bias
	int m = (b&0x007FFFFF)>>9; // mantissa
	const int v = abs(e); // shift
	const int r = (e<0 ? 0x0002 : 0xFFFE)<<(13-v); // generate regime bits
	m = ((m>>(v-(e<0)))+1+(e<-13)*0x2)>>1; // rounding: add 1 after truncated position; in case of lowest numbers, saturate
	return (b&0x80000000)>>16 | (e>-16)*((r+m)&0x7FFF) | (e>13)*0x7FFF; // sign | regime+mantissa ("+" handles rounding overflow) | saturate
}
float p160_to_float(const ushort x) {
	const uint sr = (x>>14)&1; // sign of regime
	ushort t = x<<2; // remove sign and first regime bit
	t = sr ? ~t : t; // positive regime r>=0 : negative regime r<0
	const int r = 142-(as_int((float)t)>>23); // evil log2 bit hack to count leading zeros for regime
	const uint m = (x<<(r+10))&0x007FFFFF; // extract mantissa and bit-shift it in place
	const int rs = sr ? r : -r-1; // negative regime r<0 : positive regime r>=0
	return as_float((x&0x8000)<<16 | (r!=158)*((rs+127)<<23 | m)); // sign | regime | mantissa
}



ushort float_to_p161(const float x) {
	const uint b = as_uint(x);
	const int e = ((b&0x7F800000)>>23)-127; // exponent-bias
	int m = (b&0x007FFFFF)>>10; // mantissa
	const int ae = abs(e);
	const int v = ae>>1; // shift, ">>1" is the same as "/2"
	const int e2 = ae&1; // "&1" is the same as "%2"
	const int r = ((e<0 ? 0x0002 : 0xFFFE<<e2)+e2)<<(13-v-e2); // generate regime bits, merge regime+exponent and shift in place
	m = ((m>>(v-(e<0)*(1-e2)))+(e>-28)+(e<-26)*0x3)>>1; // rounding: add 1 after truncated position; in case of lowest numbers, saturate
	return (b&0x80000000)>>16 | (e>-31)*((r+m)&0x7FFF) | (e>26)*0x7FFF; // sign | regime+exponent+mantissa ("+" handles rounding overflow) | saturate
}
float p161_to_float(const ushort x) {
	const uint sr = (x>>14)&1; // sign of regime
	ushort t = x<<2; // remove sign and first regime bit
	t = sr ? ~t : t; // positive regime r>=0 : negative regime r<0
	const int r = 142-(as_int((float)t)>>23); // evil log2 bit hack to count leading zeros for regime
	const uint e = (x>>(12-r))&1; // extract mantissa and bit-shift it in place
	const uint m = (x<<(r+11))&0x007FFFFF; // extract mantissa and bit-shift it in place
	const int rs = (sr ? r : -r-1)<<1; // negative regime r<0 : positive regime r>=0, "<<1" is the same as "*2"
	return as_float((x&0x8000)<<16 | (r!=158)*((rs+e+127)<<23 | m)); // sign | regime+exponent | mantissa
}



ushort float_to_p162C(const float x) {
	const int b = as_int(x);
	const int e = ((b&0x7F800000)>>23)-127; // exponent-bias
	int m = (b&0x007FFFFF)>>10; // mantissa
	const int ae = -e;
	const int v = ae>>2; // shift, ">>2" is the same as "/4"
	const int r = 0x4000>>v; // generate regime bits, merge regime+exponent and shift in place
	const int e4 = (3-(ae&3))<<(12-v); // generate exponent and shift in place
	m = ((m>>v)+(e>-54)+(e<-51)*0x3)>>1; // rounding: add 1 after truncated position; in case of lowest numbers, saturate
	return (b&0x80000000)>>16 | (e>-59)*((r+e4+m)&0x7FFF) | (e>0)*0x7FFF; // sign | regime+exponent+mantissa ("+" handles rounding overflow) | saturate
}
float p162C_to_float(const ushort x) {
	const int r = 158-(as_int((float)((uint)x<<17))>>23); // remove sign bit, evil log2 bit hack to count leading zeros for regime
	const int e = (x>>(12-r))&3; // extract mantissa and bit-shift it in place
	const int m = (x<<(r+11))&0x007FFFFF; // extract mantissa and bit-shift it in place
	return as_float((x&0x8000)<<16 | (r!=158)*((124-(r<<2)+e)<<23 | m)); // sign | regime+exponent | mantissa
}
\end{lstlisting}

\newpage
\subsection{LBM core of the \textsl{FluidX3D} OpenCL C implementation (D3Q19 SRT FP32/xx)}
\begin{lstlisting}[caption={LBM core of the \textsl{FluidX3D} OpenCL C implementation (D3Q19 SRT FP32/xx).}, captionpos=b, label={list:lbm-core}, numbers=left]
float __attribute__((always_inline)) sq(const float x) {
	return x*x;
}
uint3 __attribute__((always_inline)) coordinates(const uint n) { // disassemble 1D index to 3D coordinates (n -> x,y,z)
	const uint t = n%(def_sx*def_sy);
	return (uint3)(t%def_sx, t/def_sx, n/(def_sx*def_sy)); // n = x+(y+z*sy)*sx

uint __attribute__((always_inline)) f_index(const uint n, const uint i) { // 32-bit indexing (maximum box size for D3Q19: 608x608x608)
	return i*def_s+n; // SoA (229% faster on GPU compared to AoS)
}
void __attribute__((always_inline)) equilibrium(const float rho, float ux, float uy, float uz, float* feq) { // calculate f_equilibrium
	const float c3 = -3.0f*(sq(ux)+sq(uy)+sq(uz)), rhom1=rho-1.0f; // c3=-2*sq(u)/(2*sq(c))
	ux *= 3.0f;
	uy *= 3.0f;
	uz *= 3.0f;
	feq[ 0] = def_w0*fma(rho, 0.5f*c3, rhom1); // 000 (identical for all velocity sets)
	const float u0=ux+uy, u1=ux+uz, u2=uy+uz, u3=ux-uy, u4=ux-uz, u5=uy-uz;
	const float rhos=def_ws*rho, rhoe=def_we*rho, rhom1s=def_ws*rhom1, rhom1e=def_we*rhom1;
	feq[ 1] = fma(rhos, fma(0.5f, fma(ux, ux, c3), ux), rhom1s); feq[ 2] = fma(rhos, fma(0.5f, fma(ux, ux, c3), -ux), rhom1s); // +00 -00
	feq[ 3] = fma(rhos, fma(0.5f, fma(uy, uy, c3), uy), rhom1s); feq[ 4] = fma(rhos, fma(0.5f, fma(uy, uy, c3), -uy), rhom1s); // 0+0 0-0
	feq[ 5] = fma(rhos, fma(0.5f, fma(uz, uz, c3), uz), rhom1s); feq[ 6] = fma(rhos, fma(0.5f, fma(uz, uz, c3), -uz), rhom1s); // 00+ 00-
	feq[ 7] = fma(rhoe, fma(0.5f, fma(u0, u0, c3), u0), rhom1e); feq[ 8] = fma(rhoe, fma(0.5f, fma(u0, u0, c3), -u0), rhom1e); // ++0 --0
	feq[ 9] = fma(rhoe, fma(0.5f, fma(u1, u1, c3), u1), rhom1e); feq[10] = fma(rhoe, fma(0.5f, fma(u1, u1, c3), -u1), rhom1e); // +0+ -0-
	feq[11] = fma(rhoe, fma(0.5f, fma(u2, u2, c3), u2), rhom1e); feq[12] = fma(rhoe, fma(0.5f, fma(u2, u2, c3), -u2), rhom1e); // 0++ 0--
	feq[13] = fma(rhoe, fma(0.5f, fma(u3, u3, c3), u3), rhom1e); feq[14] = fma(rhoe, fma(0.5f, fma(u3, u3, c3), -u3), rhom1e); // +-0 -+0
	feq[15] = fma(rhoe, fma(0.5f, fma(u4, u4, c3), u4), rhom1e); feq[16] = fma(rhoe, fma(0.5f, fma(u4, u4, c3), -u4), rhom1e); // +0- -0+
	feq[17] = fma(rhoe, fma(0.5f, fma(u5, u5, c3), u5), rhom1e); feq[18] = fma(rhoe, fma(0.5f, fma(u5, u5, c3), -u5), rhom1e); // 0+- 0-+
}
void __attribute__((always_inline)) fields(const float* f, float* rhon, float* uxn, float* uyn, float* uzn) { // calculate density and velocity from fi
	float rho=f[0], ux, uy, uz;
	#pragma unroll
	for(uint i=1; i<def_set; i++) rho += f[i]; // calculate density from f
	rho += 1.0f; // add 1.0f last to avoid digit extinction effects when summing up f
	ux = f[ 1]-f[ 2]+f[ 7]-f[ 8]+f[ 9]-f[10]+f[13]-f[14]+f[15]-f[16]; // calculate velocity from fi (alternating + and - for best accuracy)
	uy = f[ 3]-f[ 4]+f[ 7]-f[ 8]+f[11]-f[12]+f[14]-f[13]+f[17]-f[18];
	uz = f[ 5]-f[ 6]+f[ 9]-f[10]+f[11]-f[12]+f[16]-f[15]+f[18]-f[17];
	*rhon = rho;
	*uxn = ux/rho;
	*uyn = uy/rho;
	*uzn = uz/rho;
}
void __attribute__((always_inline)) neighbors(const uint n, uint* j) { // calculate neighbor indices
	const uint3 xyz = coordinates(n);
	const uint x0 =   xyz.x; // pre-calculate indices (periodic boundary conditions on simulation box walls)
	const uint xp =  (xyz.x       +1)%def_sx;
	const uint xm =  (xyz.x+def_sx-1)%def_sx;
	const uint y0 =   xyz.y                  *def_sx;
	const uint yp = ((xyz.y       +1)%def_sy)*def_sx;
	const uint ym = ((xyz.y+def_sy-1)%def_sy)*def_sx;
	const uint z0 =   xyz.z                  *def_sy*def_sx;
	const uint zp = ((xyz.z       +1)%def_sz)*def_sy*def_sx;
	const uint zm = ((xyz.z+def_sz-1)%def_sz)*def_sy*def_sx;
	j[0] = n;
	j[ 1] = xp+y0+z0; j[ 2] = xm+y0+z0; // +00 -00
	j[ 3] = x0+yp+z0; j[ 4] = x0+ym+z0; // 0+0 0-0
	j[ 5] = x0+y0+zp; j[ 6] = x0+y0+zm; // 00+ 00-
	j[ 7] = xp+yp+z0; j[ 8] = xm+ym+z0; // ++0 --0
	j[ 9] = xp+y0+zp; j[10] = xm+y0+zm; // +0+ -0-
	j[11] = x0+yp+zp; j[12] = x0+ym+zm; // 0++ 0--
	j[13] = xp+ym+z0; j[14] = xm+yp+z0; // +-0 -+0
	j[15] = xp+y0+zm; j[16] = xm+y0+zp; // +0- -0+
	j[17] = x0+yp+zm; j[18] = x0+ym+zp; // 0+- 0-+
}
kernel void initialize(global fpXX* fc, global float* rho, global float* u) {
	const uint n = get_global_id(0); // n = x+(y+z*sy)*sx
	float feq[def_set]; // f_equilibrium
	equilibrium(rho[n], u[n], u[def_s+n], u[2*def_s+n], feq);
	#pragma unroll
	for(uint i=0; i<def_set; i++) store(fc, f_index(n,i), feq[i]); // write to fc
} // initialize()
kernel void stream_collide(const global fpXX* fc, global fpXX* fs, global float* rho, global float* u, global uchar* flags) {
	const uint n = get_global_id(0); // n = x+(y+z*sy)*sx
	const uchar flagsn = flags[n]; // cache flags[n] for multiple readings
	if(flagsn&TYPE_W) return; // if node is boundary node, just return (slight speed up)
	uint j[def_set]; // neighbor indices
	neighbors(n, j); // calculate neighbor indices
	uchar flagsj[def_set]; // cache neighbor flags for multiple readings
	flagsj[0] = flagsn;
	#pragma unroll
	for(uint i=1; i<def_set; i++) flagsj[i] = flags[j[i]];
	// read from fc in video memory and stream to fhn
	float fhn[def_set]; // cache f_half_step[n], do streaming step
	fhn[0] = fc[f_index(n, 0)]; // keep old center population
	#pragma unroll
	for(uint i=1; i<def_set; i+=2) { // perform streaming
		fhn[i  ] = load(fc, flagsj[i+1]&TYPE_W ? f_index(n, i+1) : f_index(j[i+1], i  )); // boundary : regular
		fhn[i+1] = load(fc, flagsj[i  ]&TYPE_W ? f_index(n, i  ) : f_index(j[i  ], i+1));
	}
	// collide fh
	float rhon, uxn, uyn, uzn; // cache density and velocity for multiple writings/readings
	fields(fhn, &rhon, &uxn, &uyn, &uzn); // calculate density and velocity fields from f
	uxn = clamp(uxn, -def_c, def_c); // limit velocity (for stability purposes)
	uyn = clamp(uyn, -def_c, def_c);
	uzn = clamp(uzn, -def_c, def_c);
	float feq[def_set]; // cache f_equilibrium[n]
	equilibrium(rhon, uxn, uyn, uzn, feq); // calculate equilibrium populations
	#pragma unroll
	for(uint i=0; i<def_set; i++) store(fs, f_index(n,i), fma(1.0f-def_w, fhn[i], def_w*feq[i])); // write to fs in video memory
} // stream_collide()
\end{lstlisting}

\end{document}